\UseRawInputEncoding
\documentclass[aip,jcp,reprint,twocolumn,superscriptaddress,showpacs,floatfix]{revtex4-1}
\usepackage{mathrsfs}
\usepackage{graphicx}
\usepackage{multirow}
\usepackage{color}
\usepackage{dcolumn}
\usepackage{bm}
\usepackage{float}
\usepackage{units}
\usepackage{booktabs}
\definecolor{red}{rgb}{1,0,0}

\begin{document}

\title{Benchmarking the Semi-Stochastic CC($\bm{P}$;$\bm{Q}$) Approach for 
Singlet--Triplet Gaps in Biradicals}

\author{Arnab Chakraborty}
\affiliation{Department of Chemistry,
Michigan State University, East Lansing, Michigan 48824, USA}

\author{Stephen H. Yuwono}
\affiliation{Department of Chemistry,
Michigan State University, East Lansing, Michigan 48824, USA}

\author{J. Emiliano Deustua}
\affiliation{Department of Chemistry,
Michigan State University, East Lansing, Michigan 48824, USA}

\author{Jun Shen}
\affiliation{Department of Chemistry,
Michigan State University, East Lansing, Michigan 48824, USA}

\author{Piotr Piecuch}
\thanks{Corresponding author}
\email[e-mail: ]{piecuch@chemistry.msu.edu.}
\affiliation{Department of Chemistry,
Michigan State University, East Lansing, Michigan 48824, USA}
\affiliation{Department of Physics and Astronomy,
Michigan State University, East Lansing, Michigan 48824, USA}

\date{\today}


\begin{abstract}
We recently
proposed
a semi-stochastic approach to converging high-level
coupled-cluster (CC) energetics, such as those obtained in the CC calculations
with singles, doubles, and triples (CCSDT), in which the deterministic
CC($P$;$Q$) framework is merged with the stochastic configuration
interaction Quantum Monte
Carlo propagations
[J. E. Deustua, J. Shen,
and P. Piecuch, {\it  Phys. Rev. Lett.} \textbf{119}, 223003 (2017)]. In this
work, we investigate the ability of the semi-stochastic CC($P$;$Q$) methodology
to recover the CCSDT energies of the lowest singlet and triplet states and the
corresponding singlet--triplet gaps of biradical systems using methylene,
${\rm (HFH)}^{-}$, cyclobutadiene, cyclopentadienyl cation, and
trimethylenemethane
as examples.
\end{abstract}

\maketitle


\section{Introduction}
\label{sec:intro}

There has been great progress in \textit{ab initio} computational quantum chemistry, but an
accurate description of multi-reference (MR) situations, such as
molecular potential energy surfaces along bond breaking coordinates,
electronic spectra of radical and biradical species, and excited states dominated
by two-electron transitions, continues to represent a major challenge. 
While a traditional way of addressing this challenge has been to use MR
approaches that rely on multi-dimensional model spaces spanned by multiple
reference determinants,\cite{Roos1987,ref:mrmpreview,chemrev-2012a,lindh-review-2012,sinha-review-2016,%
chemrev-2012b,succ5,evangelista-perspective-jcp-2018}
in this work we focus on methods based on the single-reference (SR) coupled-cluster (CC)
theory\cite{Coester:1958,Coester:1960,cizek1,cizek2,cizek4} (cf. Refs.\ \onlinecite{paldus-li,bartlett-musial2007}
for selected reviews), in which the
ground electronic state is expressed using the exponential wave function ansatz
$\left|\Psi\right\rangle = e^T\left|\Phi\right\rangle$, where
$T=\sum_{n=1}^{N}T_n$ is the cluster operator, $N$ is the total number of 
correlated electrons in the system, $T_n$ is the $n$-body ($n$-particle--$n$-hole or $n$-tuply excited)
component of $T$, and
$\left|\Phi\right\rangle$ is the reference determinant that serves as a Fermi vacuum.
By truncating $T$ at various excitation ranks, one obtains the well-known hierarchy of SRCC approximations,
including the CC method with singles and doubles (CCSD),\cite{ccsd,ccsd2,ccsdfritz,osaccsd} where 
$T$ is truncated at $T_2$, the CC method with singles, doubles, and triples
(CCSDT),\cite{ccsdt-hoffmann,ccfullt,ccfullt2,ccsdt-uhf} where $T$ is truncated at $T_3$,
the CC method with singles, doubles, triples, and quadruples (CCSDTQ),
\cite{ccsdtq0,ccsdtq1,ccsdtq2,ccsdtq3} where $T$ is truncated at $T_4$, and so on.
As long as the number of strongly correlated electrons is not too large, the CCSD, CCSDT, CCSDTQ, etc.
hierarchy
and its extensions to excited states and properties other than energy via the
equation-of-motion (EOM)\cite{emrich,eomcc1,eomcc3,eomccsdt1,eomccsdt2,eomccsdt3,kallaygauss,hirata1}
and linear response (LR)\cite{monk,monk2,mukherjee_lrcc,sekino-rjb-1984,lrcc3,lrcc4,jorgensen,kondo-1995,kondo-1996}
CC frameworks rapidly converge to the exact, full configuration interaction (FCI),
limit.\cite{bartlett-musial2007} As a result, the CCSDT, CCSDTQ, and similar methods and their
EOM and LR extensions are capable of accurately describing typical MR situations,
such as bond rearrangements in chemical reactions, singlet--triplet gaps in biradicals, and excited states
having substantial double excitation
character, via
particle--hole excitations from a single determinant.

The convergence of the CCSD, CCSDT, CCSDTQ, etc. hierarchy toward FCI in the majority of chemical
applications is fast,
but to achieve
a quantitative
description,
one
has to go beyond the basic
CCSD level and face
high, often prohibitive, computational
costs, such as the iterative $n_{o}^{3}n_{u}^{5}$ steps
of CCSDT or the even more 
demanding $n_{o}^{4}n_{u}^{6}$
steps of CCSDTQ
($n_{o}$ and $n_{u}$ are the numbers of correlated occupied and unoccupied orbitals,
respectively). Thus, one of the key challenges in the development of SRCC
methodologies
has been
the incorporation of many-electron correlation effects due to
higher--than--two-body components of the cluster operator $T$ without running into the
enormous computational costs of the high-level parent methods, such as CCSDT or CCSDTQ,
while avoiding the failures of perturbative approximations of the
CCSD(T)\cite{ccsdpt,watts-gauss-bartlett-1993} type when bond breaking, biradicals, and other MR
situations
are
examined.\cite{paldus-li,bartlett-musial2007,irpc,PP:TCA,piecuch-qtp}

One of the most promising ways of addressing the above challenge within the SRCC framework is offered
by the CC($P$;$Q$) formalism, which allows one to correct the
CC and EOMCC energies obtained using conventional as well as unconventional
truncations in the cluster and EOM excitation operators for the missing many-electron correlation effects
of interest
using
the
moment expansions developed in Refs.\
\onlinecite{jspp-chemphys2012,jspp-jcp2012,jspp-jctc2012,nbjspp-molphys2017}.
Focusing on the ground electronic states (or, in general, the lowest states of symmetries of interest, as in
the case of many lowest-energy singlet, doublet, or triplet states), the CC($P$;$Q$) methods based on
correcting the CC energies obtained with the conventional truncations in $T$ at a given excitation rank
reduce to
the left-eigenstate
completely renormalized (CR) CC approaches,
such as CR-CC(2,3),\cite{crccl_jcp,crccl_cpl,crccl_molphys,crccl_jpc,crccl_ijqc} in which one corrects
the CCSD energies for the leading $T_{3}$ correlations, and its higher-order extensions.
\cite{nbjspp-molphys2017,ccpq-be2-jpca-2018,ptcp2007,msg65,nuclei8} The CR-CC(2,3) method
improves poor performance of CCSD(T) in situations involving covalent bond breaking
\cite{jspp-chemphys2012,crccl_jcp,crccl_cpl,crccl_molphys,ptcp2007,crccl_jpc,ge1,ge2} and certain types
of noncovalent interactions,\cite{ccpq-be2-jpca-2018,ccpq-mg2-mp-2019}
while retaining the relatively inexpensive iterative $n_{o}^{2}n_{u}^{4}$ and noniterative
$n_{o}^{3}n_{u}^{4}$ steps of CCSD(T),
but neither CR-CC(2,3) nor its CR-CCSD(T)\cite{irpc,PP:TCA,leszcz,ren1} and locally
renormalized CCSD(T)\cite{ndcmmcc} predecessors, nor their CCSD(2)$_{\rm T}$,\cite{eomccpt,ccsdpt2,gwaltney1,gwaltney3}
CCSD(T)$_{\Lambda}$,\cite{stanton1997,crawford1998,ref:26} $\Lambda$-CCSD(T),\cite{bartlett2008a,bartlett2008b}
and CCSD(${\rm T}\mbox{--}n$) \cite{eriksen1,eriksen2}
counterparts, designed
to improve CCSD(T) as well,
are accurate enough when the coupling of the low-order $T_n$ components, such as $T_1$ and $T_2$,
and their higher-order counterparts with $n > 2$, e.g., $T_3$, becomes larger.\cite{jspp-jcp2012,jspp-jctc2012}
One can address this issue by solving for $T_1$ and $T_2$ clusters in the presence of the dominant
higher--than--doubly excited components of $T$ identified with the help of active orbitals, as in
the CCSDt,
CCSDtq, and similar schemes,\cite{ccsdtq3,eomccsdt1,eomccsdt2,semi0a,semi0b,%
semi2,semih2o,semi3c,semi4,piecuch-qtp} and account for the remaining correlation effects of interest
(e.g., the $T_{3}$ correlations not captured by CCSDt) using the CC($P$;$Q$) moment expansions, but
the resulting CC(t;3), CC(t,q;3), CC(t,q;3,4), etc. hierarchy,\cite{jspp-chemphys2012,jspp-jcp2012,jspp-jctc2012,%
nbjspp-molphys2017,ccpq-be2-jpca-2018,ccpq-mg2-mp-2019} while accurately reproducing the parent CCSDT and
CCSDTQ energetics at the fraction of the computational costs, requires choosing user- and system-dependent
active orbitals. The question arises if the CC($P$;$Q$) methodology can be similarly successful without having to define
active orbitals tuned to a given calculation.

To address this question, we have recently proposed an automated way of identifying the leading $T_{n}$
components with $n>2$ within the CC($P$;$Q$) framework by merging the deterministic CC($P$;$Q$) formalism
with the stochastic configuration interaction (CI) Quantum Monte Carlo (QMC)
\cite{Booth2009,Cleland2010,fciqmc-uga-2019,ghanem_alavi_fciqmc_jcp_2019,ghanem_alavi_fciqmc_2020}
and CC Monte Carlo (CCMC)\cite{Thom2010,Franklin2016,Spencer2016,Scott2017} propagations.
\cite{stochastic-ccpq-prl-2017,stochastic-ccpq-molphys-2020,jed-js-pp-jcp-2021}
As shown in Ref.\ \onlinecite{stochastic-ccpq-prl-2017}, the results of the
semi-stochastic CC($P$;$Q$) computations, in which the lists of determinants needed to define
higher--than--doubly excited cluster components are extracted from the FCIQMC and CCSDT-MC runs, rapidly
converge to the parent high-level CC energetics, represented in Ref.\ \onlinecite{stochastic-ccpq-prl-2017}
by CCSDT, out of the early stages of the underlying QMC propagations, even when electronic quasi-degeneracies
and the associated $T_{3}$ correlations become substantial.
In more recent studies,
it has been demonstrated that the semi-stochastic CC($P$;$Q$)
calculations exhibit a similarly fast convergence toward the parent CCSDT and CCSDTQ
energies when
the FCIQMC propagations are replaced by their less expensive truncated CISDT-MC and CISDTQ-MC
analogs\cite{jed-js-pp-jcp-2021} and that the CIQMC-driven CC($P$;$Q$) framework can be very effective
in converging the EOMCCSDT energies of excited states, including states of the same symmetry as the ground state,
states of different symmetries, and challenging excited states dominated by
two-electron transitions\cite{stochastic-ccpq-molphys-2020}
(cf., also, Ref.\ \onlinecite{eomccp-jcp-2019}). Our group has also explored
the deterministic alternative to the semi-stochastic CC($P$;$Q$) methodology
by replacing the CIQMC and CCMC propagations, needed to identify the dominant
higher--than--doubly excited components of $T$, by the sequences of Hamiltonian diagonalizations
defining the selected CI approach abbreviated as CIPSI,\cite{sci_3,cipsi_1,cipsi_2} obtaining encouraging
results.\cite{cipsi-ccpq-2021}

In the present work, we investigate the ability of the semi-stochastic CC($P$;$Q$) formalism, driven
by either FCIQMC or CISDTQ-MC propagations, to converge the energies of the lowest singlet and triplet
states and the corresponding singlet--triplet gaps, $\Delta E_{\rm S\mbox{-}T}$, resulting
from the high-level CCSDT calculations for the selected organic biradicals, including methylene,
cyclobutadiene, cyclopentadienyl cation, and trimethylenemethane, and a prototype magnetic system,
represented by the ${\rm (HFH)^-}$ ion. Understanding electronic structure and properties
of organic biradicals is of major significance in
areas such as chemical reaction dynamics,\cite{sjs-ref1} molecular magnetism,\cite{sjs-ref3}
spintronics,\cite{sjs-ref5,sjs-ref6} nonlinear optics,\cite{sjs-ref7} photochemical pathways,
\cite{zgierski1,zgierski2,jpclett-2021}
and photovoltaics.\cite{sjs-ref8,sjs-ref11,sjs-ref12} The linear ${\rm (HFH)}^{-}$ species, in
which one simultaneously stretches both H--F bonds,
is a model inorganic magnetic system, in which two paramagnetic centers carrying unpaired spins
associated with the terminal hydrogen atoms are coupled via a polarizable diamagnetic
bridge formed by ${\rm F^{-}}$.\cite{hfh-1992}
It is well established that an accurate determination of the ordering of the low-lying singlet and triplet
states and $\Delta E_{\rm S\mbox{-}T}$ values in biradicals,
which are among their most important physical characteristics, remains a significant challenge for
computational quantum chemistry, even when high-level
{\it ab initio} wave function methods are used in the calculations.
\cite{crccl_jpc,ch2_krylov,ch2-rmrccsdt,icmrcc6,BiradicalGeom,sjs-ref15,sjs-ref18,%
sjs-ref16,sjs-ref14,jspp-jctc2012,aj-js-pp-jpca2017,stoneburner-js-jcp2017,js-pp-dea2021,dipea7}
This is because one has to balance substantial nondynamical correlation effects needed
for an accurate description of the low-spin singlet state, which can be either of the
open-shell type involving the nearly degenerate singly occupied valence molecular orbitals (MOs) or of
the multi-configurational type mixing two closed-shell determinants involving the highest occupied and
lowest unoccupied MOs, HOMO and LUMO, respectively [the determinant in which HOMO is doubly occupied
and the doubly excited determinant of the ${\rm (HOMO)}^{2} \rightarrow {\rm (LUMO)}^{2}$ type],
with the dynamical correlations of the high-spin triplet state, which has a predominantly SR character.
If we limit ourselves to the black-box SRCC
methodologies, the
only methods that can provide a reliable and balanced description of the lowest singlet
and triplet states and singlet--triplet gaps in biradicals and magnetic systems are the high-level
CC approaches, beginning with CCSDT, and their particle-nonconserving double electron-attachment (DEA)
and double ionization potential (DIP) EOMCC counterparts,
\cite{dipea1,dipea2,dipea3,dip-stanton,dipea5,dipea6,kus-krylov-2011,kus-krylov-2012,%
jspp-dea-dip-2013,jspp-dea-dip-2014,aj-js-pp-jpca2017,dipea7,js-pp-dea2021}
especially those that incorporate the high-rank
4-particle--2-hole (4p2h) and 4-hole--2-particle (4h2p)
correlations on top of the CC treatment of the underlying closed-shell cores.
\cite{jspp-dea-dip-2013,jspp-dea-dip-2014,aj-js-pp-jpca2017,dipea7,js-pp-dea2021}
It is interesting to examine if our
semi-stochastic, CIQMC-driven, CC($P$;$Q$) methodology can be
as successful in recovering
the CCSDT results for the lowest
singlet and triplet states and $\Delta E_{\rm S\mbox{-}T}$
gaps of the biradical and magnetic systems listed above as in the previously reported benchmark
studies of bond dissociations,\cite{stochastic-ccpq-prl-2017,jed-js-pp-jcp-2021} chemical reaction pathways,
\cite{stochastic-ccpq-prl-2017,jed-js-pp-jcp-2021} and excited electronic states.\cite{stochastic-ccpq-molphys-2020}


\section{Theory}
\label{sec:theory}

We recall that the CC($P$;$Q$) formalism is a natural generalization of the 
biorthogonal moment expansions of Refs.\ \onlinecite{crccl_jcp,crccl_cpl,%
crccl_molphys}, which in the past resulted in the CR-CC/EOMCC triples corrections
to the CCSD/EOMCCSD energies, such as CR-CC(2,3),\cite{crccl_jcp,crccl_cpl,crccl_molphys,crccl_jpc,crccl_ijqc}
CR-EOMCC(2,3),\cite{crccl_molphys,crccl_ijqc2} and $\delta $-CR-EOMCC(2,3),\cite{7hq}
and their extensions to quadruples,\cite{nbjspp-molphys2017,ccpq-be2-jpca-2018,ptcp2007,msg65,nuclei8}
to unconventional truncations in the cluster and EOM excitation operators. In the case of the
ground electronic state, or the lowest-energy state of symmetry other than the ground state that can be
treated using the SRCC framework, the CC($P$;$Q$) calculation is
a two-step procedure. In the first step, abbreviated as CC($P$), we solve the CC amplitude equations
in the subspace of the many-electron Hilbert space, $\mathscr{H}^{(P)}$, referred to as the $P$ space,
spanned by the excited determinants that
together with the reference determinant $|\Phi\rangle$ dominate the target state $|\Psi\rangle$ of interest,
for the suitably truncated form of the cluster operator $T$ consistent with the content of the
$P$ space, designated as $T^{(P)}$, and the corresponding
energy $E^{(P)}$. In the second step, we correct the CC($P$) energy $E^{(P)}$ for the many-electron
correlation effects captured with the help of the excited determinants that span another
subspace of the Hilbert space, called the $Q$ space and denoted as $\mathscr{H}^{(Q)}$,
using the noniterative correction $\delta (P;Q)$ derived from the CC($P$;$Q$) moment expansion
introduced in Refs.\ \onlinecite{jspp-chemphys2012,jspp-jcp2012,jspp-jctc2012}. The final CC($P$;$Q$)
energy is determined as
\begin{equation}
E^{(P+Q)} = E^{(P)} + \delta (P;Q),
\label{eqn:ccpq}
\end{equation}
where the explicit formulas for the correction $\delta (P;Q)$ in terms of the
generalized moments\cite{moments,leszcz,ren1} of the CC($P$) equations, which correspond to
projections of these equations on the complementary $Q$-space determinants, and the left eigenstate
$\langle \Phi| (1 + \Lambda^{(P)})$
of the similarity-transformed Hamiltonian
$\overline{H}^{(P)} = e^{-T^{(P)}}He^{T^{(P)}}$ in the $P$ space, with $\Lambda^{(P)}$ representing the
relevant hole--particle deexcitation operator, can be found
in Refs.\ \onlinecite{jspp-chemphys2012,jspp-jcp2012,jspp-jctc2012,nbjspp-molphys2017} (see, also,
Refs.\ \onlinecite{ccpq-be2-jpca-2018,ccpq-mg2-mp-2019,%
stochastic-ccpq-prl-2017,stochastic-ccpq-molphys-2020,jed-js-pp-jcp-2021,cipsi-ccpq-2021}).
In the most complete variant of the CC($P$;$Q$) formalism adopted in this work, one uses the
Epstein--Nesbet denominators involving $\overline{H}^{(P)}$ in determining the $\delta (P;Q)$ corrections.
One could employ the M{\o}ller--Plesset denominators instead, but the
Epstein--Nesbet form has been shown to be generally more accurate (see, e.g., Refs.\
\onlinecite{jspp-jctc2012,nbjspp-molphys2017,stochastic-ccpq-prl-2017,jed-js-pp-jcp-2021}).

The main advantage of the CC($P$;$Q$) formalism compared to its CR-CC predecessors is the flexibility
in defining the underlying $P$ and $Q$ spaces that allows us to relax the lower-order $T_1$ and $T_2$ components
of the cluster operator $T$ in the presence of their higher-order counterparts, such as the leading $T_{3}$
contributions, which the CR-CC(2,3), CCSD(2)$_{\rm T}$, $\Lambda$-CCSD(T), and other noniterative triples
corrections to CCSD cannot do. As explained in the Introduction, one can incorporate the coupling
among the lower-rank $T_1$ and $T_2$ and higher-rank $T_{n}$ components with $n > 2$ within
the CC($P$;$Q$) framework by including
the dominant higher--than--doubly excited determinants identified with the help of active orbitals
or selected CI diagonalizations in the $P$ space, using the $\delta(P;Q)$ corrections
to capture the remaining correlations of interest, but this is not what we do in this work.
Here, we explore the alternative approach that utilizes the semi-stochastic CC($P$;$Q$)
methodology of Refs.\ \onlinecite{stochastic-ccpq-prl-2017,stochastic-ccpq-molphys-2020,jed-js-pp-jcp-2021},
in which the leading higher--than--doubly excited determinants included in the underlying $P$ spaces
are identified using CIQMC runs, whereas the corresponding $Q$ spaces, needed to determine corrections $\delta(P;Q)$,
are populated by the remaining determinants of interest not captured by CIQMC at a given propagation time $\tau$.
Given our interest in converging the CCSDT energies of the lowest singlet and triplet
states of biradical systems, we adopt in this study the following CC($P$;$Q$) algorithm:
\begin{itemize}
\item[1.]
Initiate the desired CIQMC (in this study, FCIQMC or CISDTQ-MC) propagation by placing a certain
number of walkers on the appropriate reference determinant $|\Phi\rangle$. In calculations for the lowest
singlet states of symmetries of interest, we used the relevant restricted Hartree--Fock (RHF) determinants
as reference functions. In the case of the lowest triplet states, we used the
restricted open-shell Hartree--Fock (ROHF) references.
\item[2.]
After a certain number of CIQMC time steps, referred to as MC iterations, or, equivalently, at some
propagation time $\tau > 0$, extract a list of triply excited determinants captured by CIQMC,
i.e., those triples that are populated by at least one walker.
\item[3.]
Solve the CC($P$)
amplitude equations and the associated left eigenstate problem involving $\overline{H}^{(P)}$
in the $P$ space spanned by all singly and doubly excited determinants
and the subset of triply excited determinants identified in step 2, using the same reference $|\Phi\rangle$
as that chosen to initiate the CIQMC run, to obtain the cluster operator $T^{(P)}$, defined
as $T_{1} + T_{2} + T_{3}^{\rm (MC)}$,
and its $\Lambda^{(P)} = \Lambda_{1} + \Lambda_{2} + \Lambda_{3}^{\rm (MC)}$  companion,
where the list of triples in $T_{3}^{\rm (MC)}$ and $\Lambda_{3}^{\rm (MC)}$
is extracted from the CIQMC propagation at time
$\tau$. Calculate the energy $E^{(P)}$.
\item[4.]
Correct the CC($P$) energy $E^{(P)}$ obtained in step 3 for the remaining $T_{3}$ correlations
using Eq.\ (\ref{eqn:ccpq}), where the $Q$ space needed to determine the
$\delta(P;Q)$ correction consists of the triply excited determinants not captured by the CIQMC propagation
at time $\tau$, to obtain the CC($P$;$Q$) energy $E^{(P+Q)}$.
\item[5.]
Repeat steps 2--4 at some later CIQMC propagation time $\tau^\prime > \tau$ to check the
convergence of the CC($P$;$Q$) energies $E^{(P+Q)}$. Stop the calculation if the
consecutive $E^{(P+Q)}$ values do not change within an {\it a priori} defined convergence threshold
or if the stochastically determined $P$ space captures a large enough fraction of the triply excited determinants
sufficient to achieve the desired accuracy relative to the parent CCSDT approach.
\end{itemize}

In this work, we explored two types of CIQMC propagations to carry out our semi-stochastic CC($P$;$Q$)
computations. For smaller systems, including methylene, ${\rm (HFH)^{-}}$, and cyclobutadiene, we used
FCIQMC, which allows the CIQMC algorithm to explore the entire many-electron Hilbert space. For the two
largest biradical species considered in our calculations, namely, cyclopentadienyl cation and
trimethylenemethane, we used the truncated CISDTQ-MC approach, in which spawning beyond quadruply
excited determinants relative to reference $|\Phi\rangle$ is disallowed, reducing computation costs
compared to FCIQMC, especially in later stages of CIQMC propagations.
As shown in Ref.\ \onlinecite{jed-js-pp-jcp-2021},
convergence of the CC($P$;$Q$) energies toward their high-level CC parents, such as CCSDT,
is not affected by the type of the CIQMC approach used to identify the relevant higher--than--doubly
excited determinants (in this work, the triply excited determinants), so choosing CISDTQ-MC as a substitute
for FCIQMC
has
no effect on our conclusions. Following our earlier studies
of the semi-stochastic CC($P$;$Q$) methodology
\cite{stochastic-ccpq-prl-2017,stochastic-ccpq-molphys-2020,jed-js-pp-jcp-2021} and related work,
\cite{eomccp-jcp-2019} in all of the CIQMC calculations reported in this article, we relied on the
initiator CIQMC ($i$-CIQMC) algorithm introduced in Ref.\ \onlinecite{Cleland2010}, based on integer walker
numbers, in which only the determinants populated by $n_a$ or more walkers are allowed to spawn new walkers
onto empty determinants, but the semi-stochastic CC($P$;$Q$) procedure summarized above is flexible and
could be merged with other CIQMC techniques developed in recent years, such as those described in Refs.\
\onlinecite{ghanem_alavi_fciqmc_jcp_2019,ghanem_alavi_fciqmc_2020}. While the convergence of the
CC($P$;$Q$) energies of the lowest singlet and triplet states of
biradicals
considered in this study
and of the corresponding singlet--triplet gaps obtained with the help of $i$-CIQMC propagations toward
the parent CCSDT
results
is
very fast, we will examine
the utility of the newer CIQMC algorithms in the future work.

As explained in Refs.\ \onlinecite{stochastic-ccpq-prl-2017,stochastic-ccpq-molphys-2020,jed-js-pp-jcp-2021}
(cf., also, Ref.\ \onlinecite{eomccp-jcp-2019}), the semi-stochastic CC($P$;$Q$) procedure, as summarized
above, offers considerable savings in the computational effort compared to its CCSDT parent. Indeed,
the computational times and walker populations characterizing the early stages of CIQMC propagations,
which are sufficient to produce enough information for the subsequent CC($P$;$Q$) calculations to recover the
target CCSDT energetics to within fractions of a millihartree, are very small compared to the converged
CIQMC runs. Next, the CC($P$) calculations using small fractions of the triply excited determinants captured
in the early stages of the CIQMC propagations are much faster than the parent CCSDT computations using all
triples. Finally, the computational times required to determine corrections
$\delta (P;Q)$ due to the remaining $T_{3}$ correlations not captured by the preceding
CC($P$) runs, which scale no worse than $\sim$$2 n_{o}^{3} n_{u}^{4}$, are much less than the $n_o^3n_u^5$
timings associated with full CCSDT
iterations.

Another interesting feature of the semi-stochastic CC($P$;$Q$) algorithm, as defined by the above steps 1--5,
is a systematic behavior of the $E^{(P+Q)}$ energies as $\tau$ varies from 0 to $\infty$. Indeed, the initial,
$\tau = 0$, CC($P$;$Q$) energy, where the $P$ space is spanned by all singly and doubly excited determinants and the $Q$ space
consists of all triples, is identical to that obtained using the CR-CC(2,3) correction to CCSD.
On the other hand, when $\tau = \infty$, so that the $P$ space is spanned by all singly, doubly, and triply excited
determinants and the $Q$ space is empty, the CC($P$;$Q$) energy $E^{(P+Q)}$ becomes equivalent to its CCSDT parent. Thus,
the CIQMC propagation time $\tau$ can be regarded as a continuation parameter connecting the approximate treatment of $T_{3}$
clusters, represented by CR-CC(2,3), with their complete description offered
by full CCSDT.
As $\tau \rightarrow \infty$,
the uncorrected CC($P$) energies $E^{(P)}$ converge to their CCSDT counterparts too,
but, as shown in Refs.\ \onlinecite{stochastic-ccpq-prl-2017,stochastic-ccpq-molphys-2020,jed-js-pp-jcp-2021}, and
as illustrated
via our calculations of the singlet--triplet gaps discussed in Section \ref{sec:results}, the
convergence toward CCSDT is in this case substantially slower,
since the initial, $\tau = 0$, CC($P$) energy is equivalent to
that of CCSD,
where $T_{3} = 0$.
The main
role of corrections $\delta(P;Q)$ in the
context of the semi-stochastic CC($P$;$Q$) algorithm considered in this study is to accelerate convergence of the
underlying CC($P$) energies toward the desired CCSDT limit.


\section{Results}
\label{sec:results}

As explained in the Introduction, in order to assess the performance of our semi-stochastic, CIQMC-driven,
CC($P$;$Q$) methodology
in converging the full CCSDT data for the singlet--triplet gaps and the corresponding singlet- and triplet-state 
energies of biradical systems, we applied it to methylene, ${\rm (HFH)^-}$, cyclobutadiene, cyclopentadienyl cation,
and trimethylenemethane. Following our earlier studies of the singlet--triplet gaps in the same systems using the
deterministic CC($P$;$Q$)\cite{jspp-jctc2012} and DEA/DIP-EOMCC\cite{jspp-dea-dip-2013,jspp-dea-dip-2014,%
aj-js-pp-jpca2017,stoneburner-js-jcp2017} approaches,
we used the aug-cc-pVTZ basis set\cite{ccpvnz,augccpvnz} for methylene, the 6-31G(d,p) basis\cite{631gs,ss} for
the ${\rm (HFH)^-}$ ion, and the cc-pVDZ basis set\cite{ccpvnz} for cyclobutadiene, cyclopentadienyl cation, and
trimethylenemethane. In the case of methylene and trimethylenemethane, we focused on the ability of the
semi-stochastic CC($P$;$Q$) algorithm to converge the adiabatic $\Delta E_{\rm S\mbox{-}T}$ data obtained with
CCSDT. When executing the semi-stochastic CC($P$;$Q$) calculations for ${\rm (HFH)^-}$, cyclobutadiene, and
cyclopentadienyl cation, we focused on recovering the CCSDT values of the vertical singlet--triplet gaps.
Throughout this work, we define $\Delta E_{\rm S\mbox{-}T}$ as $E_{\rm S} - E_{\rm T}$, where $E_{\rm S}$ and
$E_{\rm T}$ are the electronic energies of the corresponding singlet and triplet states, i.e., the positive
$\Delta E_{\rm S\mbox{-}T}$ value implies that triplet is lower in energy.

All of the CC calculations reported in this article were performed using our group's standalone codes,
interfaced with the RHF, ROHF, and integral transformation routines in the GAMESS package,\cite{gamess,gamess2020}
which were originally developed in Refs.\ \onlinecite{jspp-chemphys2012,jspp-jcp2012,jspp-jctc2012,nbjspp-molphys2017}
and extended to the stochastically generated $P$ spaces for the use in CC($P$) and CC($P$;$Q$) computations in Refs.\
\onlinecite{stochastic-ccpq-prl-2017,stochastic-ccpq-molphys-2020,jed-js-pp-jcp-2021,eomccp-jcp-2019}.
The $i$-FCIQMC [methylene, ${\rm (HFH)^-}$, and cyclobutadiene] and $i$-CISDTQ-MC (cyclopentadienyl
cation and trimethylenemethane) calculations, needed to generate the lists of triples for the semi-stochastic
CC($P$) and CC($P$;$Q$) runs, were carried out with the HANDE software.\cite{hande-jors-2015,hande-jctc-2019}
Each of the $i$-FCIQMC and $i$-CISDTQ-MC propagations was initiated by placing 1500 walkers on the
relevant reference determinant. The CIQMC time step $\delta \tau$ and the initiator parameter $n_a$
were set at 0.0001 a.u. and 3, respectively. In all post-Hartree--Fock calculations, the core MOs
correlating with the 1s orbitals of the C and F atoms were kept frozen. If the true point group of the biradical
system of interest was not Abelian, we used its largest Abelian subgroup, since our
CC codes interfaced with GAMESS and the CIQMC routines in HANDE cannot handle non-Abelian symmetries.


\subsection{Methylene}
\label{results:methylene}

We begin the discussion of our results by analyzing the performance of the semi-stochastic CC($P$;$Q$)
approach in converging the CCSDT energies of the ground (${\rm X \: ^3B_1}$) and first-excited (${\rm A \: ^1A_1}$)
states of the methylene/aug-cc-pVTZ system and the adiabatic gap between them. The $C_{2v}$-symmetric
geometries of ${\rm CH}_{2}$ in the two states, optimized using FCI and the [5s3p/3s] triple zeta
basis set of Dunning\cite{tz2p} augmented with two sets of polarization functions (TZ2P), were taken from
Ref.\ \onlinecite{ch2tz2p}. The electronically nondegenerate triplet ground state has a predominantly SR nature
dominated by the $(1a_1)^2(2a_1)^2(1b_2)^2(3a_1)^1(1b_1)^1$ configuration, whereas the first-excited singlet state
exhibits a significant MR character requiring a linear combination of the
$(1a_1)^2(2a_1)^2(1b_2)^2(3a_1)^2$ and doubly excited
$(1a_1)^2(2a_1)^2(1b_2)^2(1b_1)^2$ closed-shell determinants for a
proper zeroth-order description. Because of these fundamentally different characteristics of the
${\rm X \: ^3B_1}$ and ${\rm A \: ^1A_1}$ states, a well-balanced and accurate
treatment of dynamical and nondynamical correlation effects is
the key to a reliable description of the singlet--triplet gap in methylene.
It is, therefore, unsurprising that one usually resorts to methods of the
MRCI\cite{ch2dzp,ch2-bausch,ch2-mclean,ch2-shavitt,ch2-buenker,ch2tz2p} or
MRCC\cite{ch2_ppjp-dzp,ch2_ppjp,succ-balkova-ch2,rmrcc9} type, or to the
high-level SRCC theories that account for higher--than--doubly excited clusters
in an iterative manner, such as full CCSDT used in Refs.\ \onlinecite{ccsdt-uhf,jspp-jctc2012},
to accomplish this goal (for other examples of high-level \textit{ab initio}
calculations for the ${\rm X \: ^3B_1}$ and ${\rm A \: ^1A_1}$ states of methylene, see
Refs.\ \onlinecite{gour2010,jspp-dea-dip-2013,jspp-dea-dip-2014,aj-js-pp-jpca2017} and references
therein). The CCSDT results for the adiabatic singlet--triplet gap in methylene, which
are of interest in the present study, are indeed very accurate.
As shown, for example, in Ref.\ \onlinecite{jspp-jctc2012}, the difference
between the adiabatic $\Delta E_{\rm S\mbox{-}T}$ value obtained
in the CCSDT/TZ2P calculations
and the corresponding FCI result of 11.14 kcal/mol\cite{ch2tz2p}
is only 0.11 kcal/mol or 38 ${\rm cm^{-1}}$.
As demonstrated in Ref.\ \onlinecite{jspp-jctc2012} as well, the purely electronic ${\rm A\: ^1A_1-X\: ^3B_1}$
separation resulting from the CCSDT computations using the aug-cc-pVTZ basis employed in this work
is only about 0.15 kcal/mol ($\sim$50 ${\rm cm^{-1}}$) higher than the experimentally derived value of
9.37 kcal/mol reported in Ref.\ \onlinecite{ch2_ppjp}, obtained by correcting the vibrationless adiabatic
singlet--triplet gap determined in Ref.\ \onlinecite{methylene5} for the relativistic and nonadiabatic
(Born--Oppenheimer diagonal correction) effects estimated in Refs.\ \onlinecite{ch2-rel} and
\onlinecite{ch2-bodc}, respectively. It is, therefore,
interesting to examine if the semi-stochastic CC($P$;$Q$) approach advocated in this
work is capable of reproducing the high-quality CCSDT/aug-cc-pVTZ data for the ${\rm X \: ^3B_1}$
and ${\rm A \: ^1A_1}$ states of methylene and the adiabatic separation between them.

\begin{figure*}[h!]
\centering
    
\includegraphics[width=1.0\textwidth]{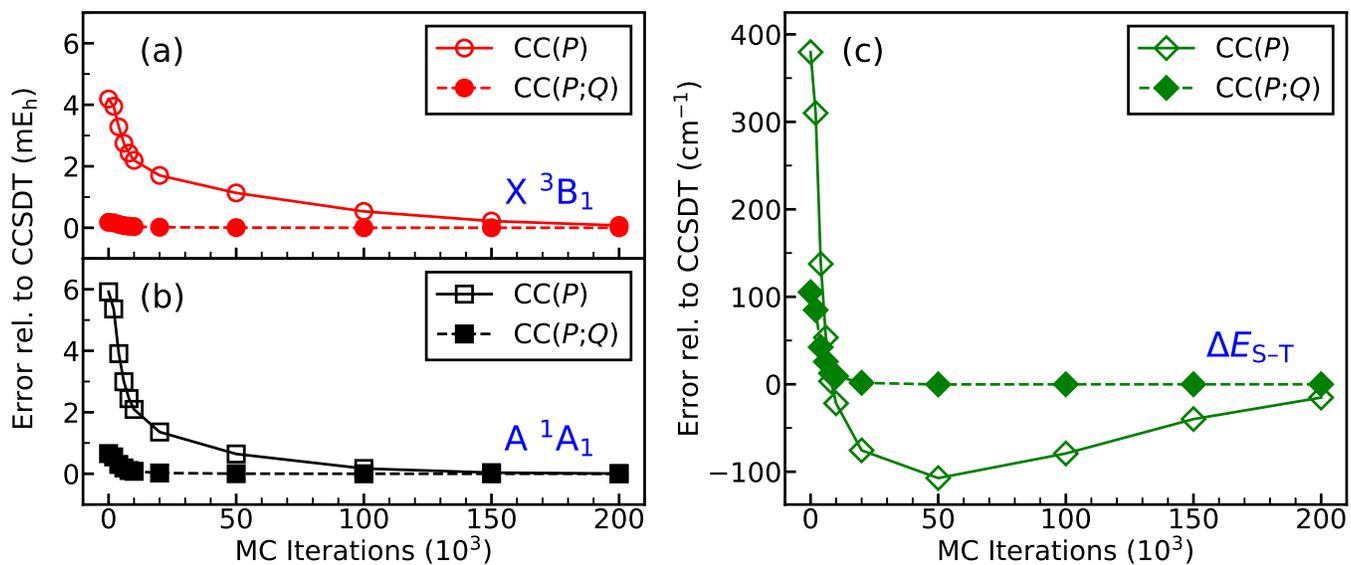}
\caption{\label{fig:figure1} 
Convergence of the CC($P$) and CC($P$;$Q$) energies of the ${\rm X \: ^3B_1}$ [panel (a)]
and ${\rm A \: ^1A_1}$ [panel (b)] states
of methylene, as described by the aug-cc-pVTZ basis set,
and of the corresponding adiabatic singlet--triplet gaps [panel (c)] toward their parent CCSDT values.
The geometries of the ${\rm X \: ^3B_1}$ and ${\rm A \: ^1A_1}$ states, optimized in the
FCI calculations using the TZ2P basis set, were taken from Ref.\ \onlinecite{ch2tz2p}.
The $P$ spaces consisted of all singles and doubles and subsets 
of triples identified during the $i$-FCIQMC propagations with $\delta \tau=0.0001$ a.u.\ and the $Q$ 
spaces consisted of the triples not captured by $i$-FCIQMC.}
\end{figure*}

The results of our FCIQMC-driven CC($P$;$Q$) calculations for the methylene/aug-cc-pVTZ system, reported
as errors relative to the parent CCSDT data, and their CC($P)$ counterparts
are shown in Table \ref{table:table1}
and Fig.\ \ref{fig:figure1}.
The reference determinants
$|\Phi\rangle$ used to initiate the $i$-FCIQMC propagations and to
carry out the CC($P)$, CC($P$;$Q$), CCSD, CR-CC(2,3), and CCSDT calculations were the ROHF determinant in
the case of the ${\rm X \: ^3B_1}$ state and the RHF determinant for the ${\rm A\: ^1A_1}$ state.
The subsets of triply excited determinants needed to construct the $P$ spaces used in the CC($P$) and CC($P$;$Q$)
computations at various propagation times $\tau$ were the $S_z = 1$ triples of the ${\rm B_1}$ symmetry
captured during the $i$-FCIQMC run for the ${\rm X\: ^3B_1}$ state and the $S_z = 0$
triples of the ${\rm A_1}$ symmetry captured during the analogous run for the ${\rm A\: ^1A_1}$ state.
Following the semi-stochastic CC($P$;$Q$) algorithm described in Section \ref{sec:theory},
the $Q$ spaces needed to determine
corrections $\delta(P;Q)$
were defined as the remaining triples not captured by $i$-FCIQMC.

Let us start our analysis by examining the CC($P$) and CC($P$;$Q$) data at
$\tau = 0$, where the $P$ spaces do not contain any triply excited
determinants. As shown in Table \ref{table:table1}, the CC($P$) energies
of the ${\rm X\: ^3B_1}$ and ${\rm A\: ^1A_1}$ states at $\tau = 0$, which
are equivalent to those obtained using conventional CCSD, are above their CCSDT
[i.e., $\tau = \infty$ CC($P$)] counterparts by 4.187 and 5.918 millihartree,
respectively. This translates into a 380 ${\rm cm^{-1}}$ or $\sim$11\%
error in the adiabatic $\Delta E_{\rm S\mbox{-}T}$ value when compared to the
3328 ${\rm cm^{-1}}$ singlet--triplet gap obtained with CCSDT. The situation improves
when the CC($P$;$Q$) corrections $\delta (P;Q)$ due to $T_{3}$ correlation effects,
calculated by placing all triply excited determinants in the respective $Q$ spaces,
are added to the CC($P$) energies. The $\tau = 0$ CC($P$;$Q$) or CR-CC(2,3) energy 
characterizing the ${\rm X\: ^3B_1}$ state is only 0.177 millihartree above
the parent CCSDT value, which is an error reduction relative to CCSDT compared to 
the underlying CC($P$) result by a factor of $\sim $24. The $\delta (P;Q)$ correction
improves the $\tau = 0$ CC($P$) energy of the more challenging ${\rm A\: ^1A_1}$ state as well,
although
the difference between the resulting CR-CC(2,3) energy and its CCSDT
counterpart, of 0.656 millihartree, is almost 4 times larger than the analogous difference
obtained for the ${\rm X\: ^3B_1}$ state. As a result, the 105 ${\rm cm^{-1}}$ error relative
to CCSDT characterizing the adiabatic ${\rm A\: ^1A_1 - X\: ^3B_1}$ separation obtained in the
$\tau = 0$ CC($P$;$Q$) or CR-CC(2,3) calculations, while considerably smaller than the
380 ${\rm cm^{-1}}$ obtained in the underlying CC($P$) (i.e., CCSD) runs, leaves room
for further
improvements. One can improve the CR-CC(2,3) energies of the ${\rm X\: ^3B_1}$ and ${\rm A\: ^1A_1}$ states
and the gap between them
by enriching the $P$ spaces defining the CC($P$) calculations with the leading
triply excited determinants identified using active orbitals and correcting the resulting CCSDt energies
for the remaining $T_3$ correlations that have not been captured
by CCSDt,\cite{jspp-jctc2012} but our
objective here is to examine
if one can accomplish the same, or improve the CC(t;3) results reported in Ref.\ \onlinecite{jspp-jctc2012}
even further, by turning to the more black-box semi-stochastic CC($P$;$Q$) methodology, in which the dominant
triply excited determinants are identified with CIQMC.

The results in Table \ref{table:table1} and Fig. \ref{fig:figure1}
show that when the $\tau = 0$ $P$ spaces are augmented with the subsets
of triply excited determinants captured in the $i$-FCIQMC runs at $\tau > 0$ and,
following the CC($P$;$Q$) recipe, the resulting CC($P$) energies
are corrected for the remaining $T_{3}$ correlations, the convergence of the total electronic energies
of the ${\rm X\: ^3B_1} $ and ${\rm A\: ^1A_1}$ states and the adiabatic separation between them toward
their CCSDT parents is rapid. We can see this already in the early stages of the $i$-FCIQMC propagations.
For example, at
$\tau = 0.8$ a.u., i.e., after only 8000 $\delta \tau = 0.0001$ a.u. MC iterations, the errors in the CC($P$;$Q$)
energies of the ${\rm X\: ^3B_1}$ and ${\rm A\: ^1A_1}$ states and the corresponding $\Delta E_{\rm S\mbox{-}T}$
value relative to CCSDT are 0.049 millihartree, 0.106 millihartree, and 13 ${\rm cm^{-1}}$, respectively,
substantially improving the CR-CC(2,3) [i.e., $\tau = 0$ CC($P$;$Q$)] calculations,
which give 0.177 millihartree for the ${\rm X\: ^3B_1}$ state,
0.656 millihartree for the ${\rm A\: ^1A_1}$ state, and 105 ${\rm cm^{-1}}$ for $\Delta E_{\rm S\mbox{-}T}$.
This confirms our expectation that the main source of errors in the CR-CC(2,3) computations,
especially in the case of the more MR ${\rm A\: ^1A_1}$ state, which is characterized by larger $T_{3}$ effects,
is the use of the unrelaxed $T_{1}$ and $T_{2}$ amplitudes obtained with CCSD in constructing the
correction due to triples. The FCIQMC-based CC($P$;$Q$)
calculations at $\tau = 0.8$ a.u., which use as little as 16\% of all triply excited determinants to define the
$P$ space for the ${\rm X\: ^3B_1}$ state and only 25\% of all triples in the $P$ space for the ${\rm A\: ^1A_1}$
state, are also more accurate than the CC(t;3) computations reported in Ref.\ \onlinecite{jspp-jctc2012}, which
produced the 0.130 millihartree, 0.409 millihartree, and 61 ${\rm cm^{-1}}$ errors relative to CCSDT for the
${\rm X\: ^3B_1}$ and ${\rm A\: ^1A_1}$ energies and $\Delta E_{\rm S\mbox{-}T}$, respectively.
This is all very promising, especially if we realize that the $i$-FCIQMC propagations used to
generate the lists of triples for our semi-stochastic CC($P$;$Q$) runs, which work so well,
are very far from convergence when $\tau = 0.8$ a.u.
Indeed, as seen in Table S.1 of the supplementary material, the total
numbers of walkers at 8000
$\delta \tau = 0.0001$ a.u.\ MC iterations, which are 132689 in the case of the ${\rm X\: ^3B_1}$ state
and 165564 for the ${\rm A\: ^1A_1}$ state, represent tiny fractions, 2.17\% and 1.11\%, respectively,
of the total walker populations at $\tau = 20.0$ a.u., where we stopped our $i$-FCIQMC propagations.

As demonstrated in Table \ref{table:table1} and Fig. \ref{fig:figure1},
the convergence of the energies of the ${\rm X\: ^3B_1} $ and ${\rm A\: ^1A_1}$ states and the gap
between them resulting from the FCIQMC-driven CC($P$;$Q$) calculations remains fast at the larger propagation
times $\tau$ as well. For example, if we allow $i$-FCIQMC to populate the respective $P$ spaces with about
26--38\% of all triples, which happens after 20000 $\delta \tau = 0.0001$ a.u.\ MC iterations,
the CC($P$;$Q$) energies of the ${\rm X\: ^3B_1}$ and ${\rm A\: ^1A_1}$ states and the
resulting singlet--triplet gap become practically indistinguishable from the parent CCSDT data,
with errors in the total electronic energies and $\Delta E_{\rm S\mbox{-}T}$ being only $\sim$20 microhartree
and 2 ${\rm cm^{-1}}$, respectively.
As shown in Table S.1 of the supplementary material,
walker populations characterizing the ${\rm X\: ^3B_1}$ and ${\rm A\: ^1A_1}$ states produced by $i$-FCIQMC
at 20000 $\delta \tau = 0.0001$ a.u.\ MC time steps are still very
small compared to the total numbers of walkers at $\tau = 20.0$ a.u., where we terminated
our $i$-FCIQMC propagations (4.11\% for the ${\rm X\: ^3B_1}$ state and 2.17\% in the case
of the ${\rm A\: ^1A_1}$ state). It is also interesting to note that the more MR ${\rm A\: ^1A_1}$ state
requires a higher fraction of triply excited determinants in the $P$ space than its
SR ${\rm X\: ^3B_1}$ counterpart to achieve similar accuracy levels in the semi-stochastic
CC($P$;$Q$) computations for both states. For example, the $i$-FCIQMC propagation has to capture about 25\% of
all triples, for the inclusion in the $P$ space, if we are to reduce errors relative to CCSDT in the
CC($P$;$Q$) calculations for the ${\rm A\: ^1A_1}$ state to $\sim$0.1 millihartree. In the case of the
${\rm X\: ^3B_1}$ state, the analogous fraction of triples is about 10\% (cf. Table \ref{table:table1}).
This highlights
the importance of balancing
the SR
triplet state with the more MR
singlet state in obtaining accurate $\Delta E_{\rm S\mbox{-}T}$ estimates, which
is not a problem for the semi-stochastic CC($P$;$Q$) methodology because the
underlying $i$-FCIQMC wave function sampling is very effective in identifying
the dominant
higher--than--doubly excited determinants, to be included in the relevant $P$ spaces, and
the $\delta (P;Q)$ corrections to the CC($P$) energies take care of the remaining correlation effects of
interest.

Before concluding this subsection and discussing other molecular examples,
we would like to comment on the effectiveness of the
noniterative corrections $\delta (P;Q)$, adopted in the CC($P$;$Q$) formalism, in accelerating
convergence of the underlying CC($P$) calculations toward CCSDT. The CC($P$) and CC($P$;$Q$)
error curves shown in Fig. \ref{fig:figure1} illustrate this best. It is clear from this figure that
the CC($P$;$Q$) energies of the ${\rm X\: ^3B_1}$ and ${\rm A\: ^1A_1}$ states [Fig. \ref{fig:figure1} (a) and (b)]
and the corresponding $\Delta E_{\rm S\mbox{-}T}$ values [Fig. \ref{fig:figure1} (c)] converge to the
parent CCSDT data much faster than in the case of the uncorrected CC($P$) computations. One can see
the same by inspecting the numerical data shown in Table \ref{table:table1}. In this context, it is worth
commenting on the CC($P$) and CC($P$;$Q$) results obtained after 8000 MC iterations.
In that case, the CC($P$;$Q$) calculations
reduce the $\sim$2.4 millihartree errors relative to CCSDT characterizing the CC($P$) energies of the
${\rm X\: ^3B_1}$ and ${\rm A\: ^1A_1}$ states to 0.1 millihartree or less, which is a desired behavior,
but the CC($P$;$Q$) $\Delta E_{\rm S\mbox{-}T}$ value is less accurate than that obtained with the uncorrected
CC($P$). One should not read too much into this though. The fact that the CC($P$;$Q$) calculations at
8000 MC iterations increase the very small 3 ${\rm cm^{-1}}$ error obtained with CC($P$) to 13 ${\rm cm^{-1}}$
is a coincidence arising from the accidental cancellation of errors in the CC($P$) total electronic energies
obtained at this particular propagation time.
Indeed, when the later stages of the $i$-FCIQMC propagations are considered, the differences between the CC($P$)
and CCSDT values of $\Delta E_{\rm S\mbox{-}T}$ become increasingly negative, reaching $-107$ ${\rm cm^{-1}}$
at 50000 MC iterations, before eventually converging to the CCSDT limit, whereas the corresponding
CC($P$;$Q$) results display a smooth behavior, rapidly approaching CCSDT.
In particular, they reduce the relatively large
negative error value obtained for $\Delta E_{\rm S\mbox{-}T}$ in the CC($P$) calculations at
50000 MC iterations to a numerical 0 ${\rm cm^{-1}}$. This highlights, once again, the ability of the
CC($P$;$Q$) corrections $\delta (P;Q)$ to offer a well-balanced description of the lowest singlet and
triplet states in methylene, in addition to improving the individual state energies.


\subsection{$\bm{\mathrm{(HFH)^{-}}}$}
\label{results:hfhminus}

Our next example is the linear, $D_{\infty h}$-symmetric, ${\rm (HFH)^{-}}$ anion, a prototype
magnetic system in which unpaired spins of terminal hydrogen atoms couple to singlet and triplet
states via a polarizable diamagnetic bridge
of
${\rm F}^{-}$.\cite{hfh-1992}
The energies of the lowest two electronic states of the ${\rm (HFH)^{-}}$ system, including the
singlet ground state ${\rm X\: ^1\Sigma}_{g}^{+}$ and the first-excited triplet state
${\rm A\: ^3\Sigma}_{u}^{+}$, and the vertical gap between them,
which is proportional to the magnetic exchange coupling constant $J$ and which should
approach zero as both H--F bonds are stretched to infinity, were used in the past to test
various quantum chemistry approaches.
\cite{hfh-1992,crccl_jpc,crccl_ijqc,h3,jspp-jctc2012,jspp-dea-dip-2013,jspp-dea-dip-2014,%
nbjspp-molphys2017,hfh-2020-stanton} Among them
were methods developed in
our group, including CR-CC(2,3),\cite{crccl_jpc,crccl_ijqc} CR-CC(2,4),\cite{nbjspp-molphys2017} the
DIP-EOMCC approaches with full and active-space treatments of
4h2p
correlations of top of CCSD,\cite{jspp-dea-dip-2013,jspp-dea-dip-2014}
and the active-orbital-based CC(t;3), CC(t,q;3), and CC(t,q;3,4) hierarchy.\cite{jspp-jctc2012,nbjspp-molphys2017}
Here, we test the alternative
to CC(t;3) offered
by the semi-stochastic, FCIQMC-driven, CC($P$;$Q$)
algorithm aimed at the CCSDT energetics. As in our
previous studies,\cite{crccl_jpc,crccl_ijqc,jspp-jctc2012,jspp-dea-dip-2013,jspp-dea-dip-2014,%
nbjspp-molphys2017} we used the 6-31G(d,p) basis set and several stretches of both H--F bonds,
including $R_{\rm H\mbox{-}F} = 1.50$, 1.75, 2.00, 2.50, and 4.00 {\AA}, where
$R_{\rm H\mbox{-}F}$ is the
distance between the hydrogen and fluorine nuclei.

An accurate computation of the singlet--triplet gap in the ${\rm (HFH)^{-}}$ system is complicated by the fact
that, unlike the ${\rm A\: ^3\Sigma}_{u}^{+}$ state, which is weakly correlated and well represented by
a single ROHF determinant, its ground-state counterpart ${\rm X\: ^1\Sigma}_{g}^{+}$ displays a
substantial MR character that includes a significant contribution from the doubly excited
${\rm (HOMO)^2 \rightarrow (LUMO)^2}$ determinant, in addition to the RHF configuration.
The MR character of the ${\rm X\: ^1\Sigma}_{g}^{+}$ state, which is already
noticeable at shorter H--F separations and which substantially strengthens as $R_{\rm H\mbox{-}F}$ increases,
can be illustrated by the ratio of the FCI expansion coefficients at the ${\rm (HOMO)^2 \rightarrow (LUMO)^2}$
and RHF determinants or the equivalent $T_{2}$ cluster amplitude extracted from
FCI, which increases, in absolute value, from 0.38 at $R_{\rm H\mbox{-}F} = 1.50$ {\AA} to
1.17 at $R_{\rm H\mbox{-}F} = 4.00$ {\AA}, when the 6-31G(d,p) basis is employed\cite{crccl_jpc,crccl_ijqc}
(the HOMO and LUMO have
different symmetries, $\sigma_{g}$ and $\sigma_{u}$, respectively, so that the ${\rm HOMO \rightarrow LUMO}$
$T_{1}$ amplitude is zero).
As a result of all this, it is difficult to balance the lowest two states of the ${\rm (HFH)^{-}}$ system in a
single quantum chemistry calculation, especially when the SRCC framework using the RHF reference determinant
for the ${\rm X\: ^1\Sigma}_{g}^{+}$ state and the ROHF reference for the ${\rm A\: ^3\Sigma}_{u}^{+}$ state
is employed. Indeed, as shown in
Refs.\ \onlinecite{crccl_jpc,crccl_ijqc,jspp-jctc2012}, the differences between
the energies obtained in the CCSD/6-31G(d,p) computations and their FCI counterparts at $R_{\rm H\mbox{-}F} = 1.50$
{\AA} are 12.674 millihartree for the ${\rm X\: ^1\Sigma}_{g}^{+}$ state and only 2.628 millihartree when
the ${\rm A\: ^3\Sigma}_{u}^{+}$ state is considered. The analogous differences at
$R_{\rm H\mbox{-}F} = 2.00$ {\AA} are 19.398 and 2.068 millihartree, respectively.
The observed large discrepancies between the errors in the CCSD energies for the ${\rm X\: ^1\Sigma}_{g}^{+}$
and ${\rm A\: ^3\Sigma}_{u}^{+}$ states translate into a poor description of the singlet--triplet gaps.
One can see this by comparing the $\Delta E_{\rm S\mbox{-}T}$ values resulting from the RHF/ROHF-based
CCSD/6-31G(d,p) computations at $R_{\rm H\mbox{-}F} = 1.50$, 1.75, 2.00, 2.50,
and 4.00 {\AA}
with the
corresponding FCI data. CCSD/6-31G(d,p)
gives $-7320$, $-1838$, 1656, 3605, and 230 ${\rm cm^{-1}}$,
respectively, as opposed to $-9525$, $-4911$, $-2147$, $-277$, and 0 ${\rm cm^{-1}}$ obtained with FCI.
\cite{crccl_jpc,crccl_ijqc,jspp-jctc2012}
If we are to improve the CCSD results within the SRCC framework, we must turn to higher-level theories,
such as the CCSDT approach that interests us in
this
study,\cite{jspp-jctc2012,nbjspp-molphys2017,h3}
CCSDTQ,\cite{nbjspp-molphys2017} or the DIP-EOMCC methodology, especially after incorporating
4h2p correlations.\cite{jspp-dea-dip-2013,jspp-dea-dip-2014}
The CCSDT method is indeed very accurate, reducing the 2205, 3073, 3804, 3882, and 230 ${\rm cm^{-1}}$ errors
relative to FCI in the $\Delta E_{\rm S\mbox{-}T}$ values obtained with CCSD/6-31G(d,p) at
$R_{\rm H\mbox{-}F} = 1.50$, 1.75, 2.00, 2.50, and 4.00 {\AA} to
198, 270, 341, 420, and 58 ${\rm cm^{-1}}$, respectively.\cite{jspp-jctc2012,nbjspp-molphys2017,h3} It also
greatly improves the total electronic energies.
Indeed, the differences
between the CCSDT and FCI energies of the ${\rm X\: ^1\Sigma}_{g}^{+}$ and ${\rm A\: ^3\Sigma}_{u}^{+}$
states in the entire $R_{\rm H\mbox{-}F} = 1.50 - 4.00$ {\AA} region obtained using the 6-31G(d,p) basis set do not
exceed 2.276 and 0.389 millihartree,
respectively.\cite{jspp-jctc2012,nbjspp-molphys2017}
The analogous differences between the CCSD and FCI energies
are as large as 20.546 millihartree for the former state and 2.628 millihartree when the latter state is considered.
One can reduce the remaining small errors in the CCSDT results even further
or practically eliminate them
by using CCSDTQ\cite{nbjspp-molphys2017} or the DIP-EOMCC approaches with
4h2p
contributions,
\cite{jspp-dea-dip-2013,jspp-dea-dip-2014} but the objective of this study is to assess the performance of our
semi-stochastic CC($P$;$Q$) methodology in converging
the CCSDT data.

\begin{figure*}[!htbp]
\centering

\includegraphics[width=0.8\textwidth]{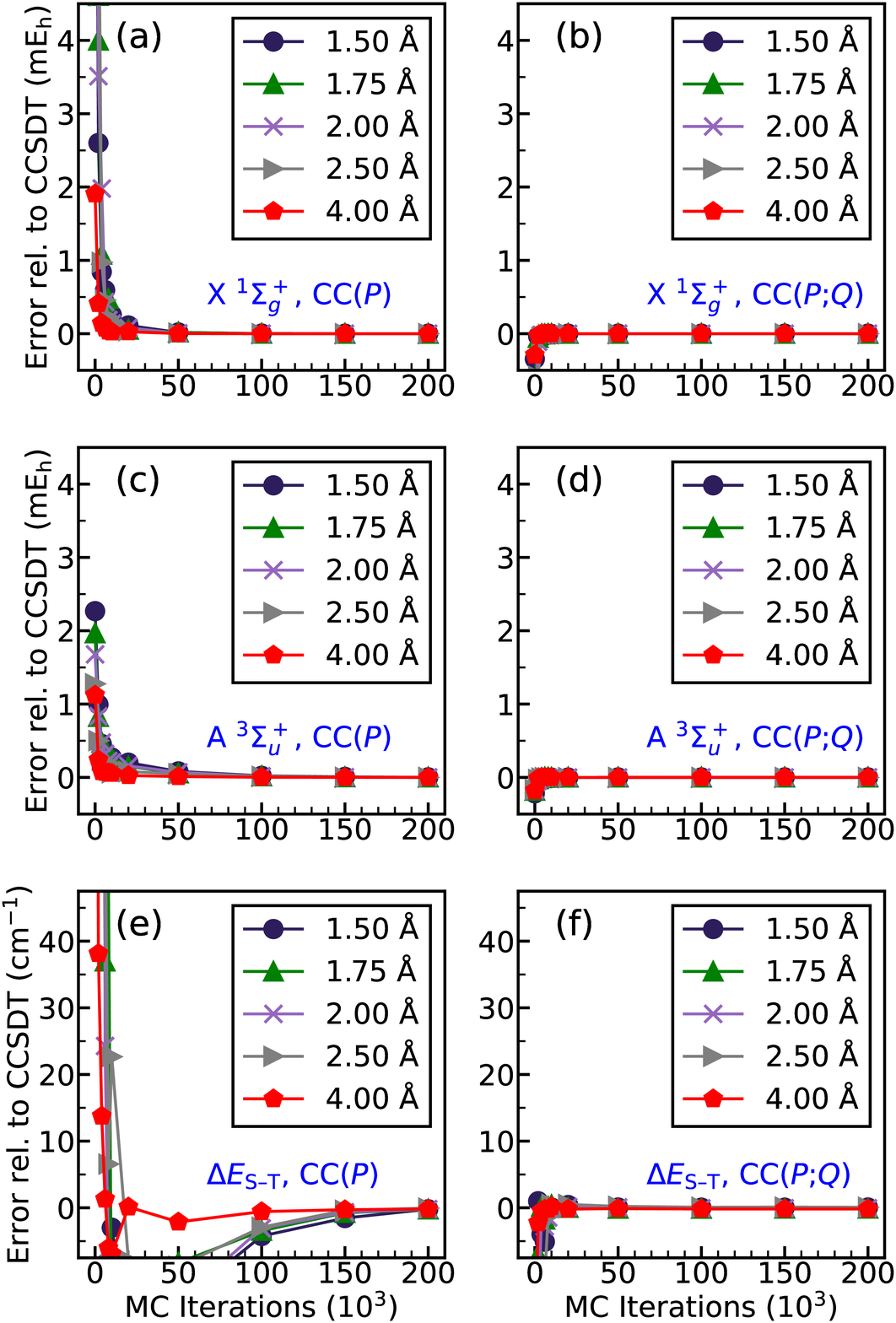}
\caption{ \label{fig:figure2}
Convergence of the CC($P$) and CC($P$;$Q$) energies of the ${\rm X\: ^1\Sigma}^+_g$ [panels (a) and (b)]
and ${\rm A\: ^3\Sigma}^+_u$ [panels (c) and (d)] states of ${\rm (HFH)^-}$, as described by the
6-31G(d,p) basis set, and of the corresponding singlet--triplet gaps [panels (e) and (f)]
toward their parent CCSDT values. The H--F distances $R_{\rm H\mbox{-}F}$ used are 1.50 \AA,
1.75 \AA, 2.00 \AA, 2.50 \AA, and 4.00 \AA. The $P$ spaces consisted of all singles and doubles and subsets
of triples identified during $i$-FCIQMC propagations with $\delta \tau=0.0001$ a.u.\ and the $Q$
spaces consisted of the triples not captured by $i$-FCIQMC.}
\end{figure*}

The results of our FCIQMC-driven CC($P$;$Q$)/6-31G(d,p) computations for the ${\rm X\: ^1 \Sigma}_{g}^{+}$ and
${\rm A\: ^3 \Sigma}_{u}^{+}$ states of the linear ${\rm (HFH)^{-}}$ system
at the H--F distances $R_{\rm H\mbox{-}F} = 1.50$, 1.75, 2.00, 2.50, and 4.00 {\AA} and the corresponding
$\Delta E_{\rm S\mbox{-}T}$ values, along with the underlying CC($P$) data, are reported in Tables
\ref{table:table3}--\ref{table:table5}
and Fig.\ \ref{fig:figure2}.
In all of our
CC($P$) and CC($P$;$Q$) computations and the
underlying $i$-FCIQMC runs for the $D_{\infty h}$-symmetric ${\rm (HFH)^{-}}$ system, we used the $D_{2h}$
Abelian subgroup of $D_{\infty h}$. In particular, the $i$-FCIQMC calculations for the ${\rm X\: ^1\Sigma}_{g}^{+}$
and ${\rm A\: ^3\Sigma}_{u}^{+}$ states
were set up to converge the lowest-energy states of the ${\rm ^{1}A}_g (D_{2h})$ and
${\rm ^{3}B}_{1u} (D_{2h})$ symmetries.
As a result, the subsets of triply excited determinants used to construct the $P$ spaces for the
subsequent CC($P$) and CC($P$;$Q$) computations for the ${\rm X\: ^1\Sigma}_{g}^{+}$ state at
the various $R_{\rm H\mbox{-}F}$ and $\tau$ values considered in this work were defined
as the $S_{z} = 0$ triples of the ${\rm A}_{g} (D_{2h})$ symmetry captured in the underlying
$i$-FCIQMC propagations. Similarly, the subsets of triply excited determinants used to
design the $P$ spaces for the CC($P$) and CC($P$;$Q$) calculations for the ${\rm A\: ^3 \Sigma}_{u}^{+}$\
state were the $S_{z} = 1$ triples of the ${\rm B}_{1u} (D_{2h})$ symmetry extracted from $i$-FCIQMC.
In analogy to all other CC($P$;$Q$) computations performed in this work, the $Q$ spaces used to determine
the $\delta (P;Q)$ corrections to the CC($P$) energies were defined as the remaining triples not captured
by the respective $i$-FCIQMC runs.

As shown in Table \ref{table:table3}, and in line with our earlier CC($P$;$Q$) work\cite{jspp-jctc2012} and the
above remarks,
the CC($P$) energies of the ${\rm X\: ^1\Sigma}_{g}^{+}$ state of ${\rm (HFH)^{-}}$
obtained at $\tau=0$, which are identical to those resulting from the conventional CCSD calculations
reported in Refs.\ \onlinecite{crccl_jpc,crccl_ijqc,jspp-jctc2012}, are
characterized by large errors relative to their $\tau = \infty$, i.e., CCSDT, parents. Indeed, the
differences between the $\tau=0$ and $\tau = \infty$ CC($P$) energies for the ${\rm X\: ^1\Sigma}_{g}^{+}$ state
increase from 11.412 millihartree at ${R_{\rm H\mbox{-}F}}=1.50$ {\AA} to more than 17 millihartree at
${R_{\rm H\mbox{-}F}}=2.00$ and 2.50 {\AA}. These differences become smaller at large H--F separations, represented
in our calculations by ${R_{\rm H\mbox{-}F}}=4.00$ {\AA}, where the $D_{\infty h}$-symmetric
${\rm (HFH)^{-}}$ system is essentially dissociated into the stretched hydrogen molecule,
which has only two electrons, so that CCSD becomes exact, and the closed-shell fluoride ion, which has
the electronic structure of the neon atom and which is characterized by small $T_{n}$ correlations with
$n > 2$, but they remain large when the ${R_{\rm H\mbox{-}F}}$ values are smaller. This should be contrasted
with the small, $\sim$1--2 millihartree, differences between the $\tau=0$ and $\tau = \infty$ CC($P$)
energies obtained at all values of ${R_{\rm H\mbox{-}F}}$ for the predominantly SR ${\rm A\: ^3\Sigma}_{u}^{+}$
state (see Table \ref{table:table4}).
As already alluded to above, and as shown in Table \ref{table:table5}, this imbalance in the description of the
${\rm X\: ^1\Sigma}_{g}^{+}$ and ${\rm A\: ^3\Sigma}_{u}^{+}$ states by the CCSD, i.e., $\tau=0$ CC($P$),
calculations gives rise to large errors in the resulting $\Delta E_{\rm S\mbox{-}T}$ values relative
to their $\tau = \infty$ (CCSDT) counterparts,
which range from 2007 ${\rm cm^{-1}}$ to 3462 ${\rm cm^{-1}}$ in the
$R_{\rm H\mbox{-}F} = 1.50\mbox{--}2.50$ {\AA} region.
Once again, these errors become small at large H--F separations,
such as ${R_{\rm H\mbox{-}F}}=4.00$ {\AA} used in this work, where ${\rm (HFH)^{-}}$ is more or less
equivalent to the stretched ${\rm H}_{2}$ and ${\rm F}^{-}$, resulting in the nearly degenerate
singlet and triplet states and the 172 ${\rm cm^{-1}}$ difference between the $\tau=0$ and $\tau = \infty$ CC($P$)
values of $\Delta E_{\rm S\mbox{-}T}$, but at shorter H--F distances they
are large
and comparable
to or even larger than the
singlet--triplet gap values provided by CCSDT or FCI.

The situation dramatically changes, when the $\tau=0$ CC($P$) or CCSD energies are corrected for
$T_{3}$ correlations with the help of the noniterative correction $\delta (P;Q)$, as in the
$\tau=0$ CC($P$;$Q$) calculations, which are equivalent to the purely deterministic CR-CC(2,3) runs
reported in Refs.\ \onlinecite{crccl_jpc,crccl_ijqc,jspp-jctc2012,nbjspp-molphys2017}.
As shown in Tables \ref{table:table3}--\ref{table:table5}, the $\tau=0$ CC($P$;$Q$), i.e., CR-CC(2,3),
energies of the ${\rm X\: ^1\Sigma}_{g}^{+}$ and ${\rm A\: ^3\Sigma}_{u}^{+}$ states at the various
H--F distances considered in this study and the gaps between them are substantially more accurate than
their uncorrected CC($P$) (i.e., CCSD) counterparts. For example, the CR-CC(2,3) approach
reduces the large, more than 17 millihartree, errors in the CCSD energies of the
${\rm X\: ^1\Sigma}_{g}^{+}$ state relative to their CCSDT [$\tau = \infty$ CC($P$) or CC($P$;$Q$)]
parents at $R_{\rm H\mbox{-}F} = 2.00$ and 2.50 {\AA} to $\sim$1--3 millihartree. We see
similarly significant improvements in the CCSD energies of the ${\rm X\: ^1\Sigma}_{g}^{+}$ state
by CR-CC(2,3) at other H--F distances, even at the ``easiest'' $R_{\rm H\mbox{-}F} = 4.00$ {\AA} value,
where the triples correction $\delta (P;Q)$ is capable of reducing the already small, 1.907 millihartree,
difference between the CCSD and CCSDT energies to the much smaller (in absolute value) 0.291 millihartree
(see Table \ref{table:table3}). Consistent with our earlier studies,
\cite{crccl_jpc,crccl_ijqc,jspp-jctc2012,nbjspp-molphys2017} 
the CR-CC(2,3) method performs even better when the weakly correlated ${\rm A\: ^3\Sigma}_{u}^{+}$ state
is examined, reducing the $\sim$1--2 millihartree errors in the underlying CCSD energetics
relative to our CCSDT target to about 0.2 millihartree (see Table \ref{table:table4}).
As a result of all of these accuracy improvements, the singlet--triplet gap
values obtained using CR-CC(2,3) are much closer to their CCSDT parents than
their CCSD counterparts, reducing the 2007, 2803, 3462, 3462, and 172 ${\rm cm^{-1}}$ errors
relative to CCSDT obtained with CCSD
at $R_{\rm H\mbox{-}F} = 1.50$, 1.75, 2.00, 2.50, and 4.00 {\AA}, respectively, by factors ranging from
6 at $R_{\rm H\mbox{-}F} = 2.50$ {\AA} to 72 at $R_{\rm H\mbox{-}F} = 1.50$ {\AA}, but, as shown
in Table \ref{table:table5} [see, also, Ref.\ \onlinecite{jspp-jctc2012}, where one can find a comparison
of the CCSD, CR-CC(2,3), and CCSDT $\Delta E_{\rm S\mbox{-}T}$ data for additional H--F distances],
the differences on the order of $(-600)$--$(-300)$ ${\rm cm^{-1}}$
between the CR-CC(2,3) and CCSDT singlet--triplet separations
in the intermediate $R_{\rm H\mbox{-}F} = 2.00\mbox{--}3.00$ {\AA} region
remain. The question arises if one can refine the CR-CC(2,3)
results
by enriching
the $P$ spaces used in the CC($P$;$Q$) calculations, which in CR-CC(2,3) consist of only
singles and doubles, with the subsets of triply excited determinants identified by $i$-FCIQMC propagations.

As shown in Tables \ref{table:table3}--\ref{table:table5} and Fig.\ \ref{fig:figure2}, once
the leading triply excited determinants, captured using $i$-FCIQMC at $\tau > 0$, are included
in the respective $P$ spaces and the $\delta(P;Q)$ corrections due to the remaining $T_{3}$
correlations are added to the energies obtained in the CC($P$) calculations, the resulting CC($P$;$Q$)
values of the ${\rm X\: ^1\Sigma}_{g}^{+}$ and ${\rm A\: ^3\Sigma}_{u}^{+}$ energies and vertical gaps
between them display very fast convergence toward their CCSDT counterparts. This is already observed
when the $i$-FCIQMC propagation times are short, engaging tiny walker populations that are orders of
magnitude smaller than those required to converge the $i$-FCIQMC runs, and the fractions of
the triply excited determinants captured by $i$-FCIQMC are small. For example, after
as few as 2000 $\delta \tau = 0.0001$ a.u. MC iterations, where $\tau$ is only 0.2 a.u. and where, as shown
in Table S.2 of the supplementary material,
the total walker populations characterizing the underlying $i$-FCIQMC runs are
0.01--0.11\% of the respective numbers of walkers at $\tau = 20.0$ a.u. [the termination time for our $i$-FCIQMC
propagations for ${\rm (HFH)}^{-}$], the differences between the CC($P$;$Q$) and CCSDT
energies obtained for the strongly correlated
${\rm X\: ^1\Sigma}_{g}^{+}$ state are $-0.035$ millihartree for $R_{\rm H\mbox{-}F} = 1.50$ {\AA},
$-0.056$ millihartree for $R_{\rm H\mbox{-}F} = 1.75$ {\AA},
$-0.110$ millihartree for $R_{\rm H\mbox{-}F} = 2.00$ {\AA},
$-0.583$ millihartree for $R_{\rm H\mbox{-}F} = 2.50$ {\AA},
and
$-0.025$ millihartree for $R_{\rm H\mbox{-}F} = 4.00$ {\AA}. In spite of using only about 10--30\% of all triply
excited determinants in the underlying $P$ spaces,
the FCIQMC-based CC($P$;$Q$) energies of the ${\rm X\: ^1\Sigma}_{g}^{+}$
state obtained after 2000 MC iterations reduce the errors relative to CCSDT characterizing the CR-CC(2,3)
[i.e., $\tau = 0$ CC($P$;$Q$)] computations in the $R_{\rm H\mbox{-}F} = 1.50\mbox{--}4.00$ {\AA} region
by factors ranging from 5 to 13 (see Table \ref{table:table3}). In fact, with an exception
of $R_{\rm H\mbox{-}F} = 2.00$ and 2.50 {\AA}, they are much more accurate than the results
produced by the purely deterministic CC(t;3) analog of the semi-stochastic CC($P$;$Q$)
methodology,
reported in Refs.\ \onlinecite{jspp-jctc2012,nbjspp-molphys2017}.
One can observe even more dramatic improvements over CR-CC(2,3) offered by the FCIQMC-driven CC($P$;$Q$)
approach, when the propagation time $\tau$ increases.
For example, after 4000 MC iterations, where the $i$-FCIQMC propagations are still far from being converged
(cf. the total walker populations used by our $i$-FCIQMC runs relative to the termination time $\tau = 20.0$ a.u.
in Table S.2 of the supplementary material)
and the fractions of triples included in the stochastically determined $P$ spaces, which range from 12\% at
$R_{\rm H\mbox{-}F} = 4.00$ {\AA} to 56\% at $R_{\rm H\mbox{-}F} = 1.50$ {\AA}, remain relatively small,
the differences between the CC($P$;$Q$) and CCSDT energies obtained for the ${\rm X\: ^1\Sigma}_{g}^{+}$ state
at $R_{\rm H\mbox{-}F} = 1.50$, 1.75, 2.00, 2.50, and 4.00 {\AA} are $-28$, $-9$, $-17$, $-50$, and $-4$
microhartree, respectively, reducing the errors relative to CCSDT that characterize the corresponding CR-CC(2,3)
calculations by factors ranging from 12 to 86 [2 to 61 when compared to the CC(t;3) results
reported in Refs.\ \onlinecite{jspp-jctc2012,nbjspp-molphys2017}]. As shown in Table \ref{table:table4},
the performance of the FCIQMC-driven CC($P$;$Q$) approach becomes even more
impressive when the ${\rm A\: ^3\Sigma}_{u}^{+}$ state, which has a SR character, is examined.
After 2000 $\delta \tau = 0.0001$ a.u. MC time steps, the errors in the CC($P$;$Q$) energies relative to
their CCSDT parents obtained for the ${\rm A\: ^3\Sigma}_{u}^{+}$ state
at $R_{\rm H\mbox{-}F} = 1.50$, 1.75, 2.00, 2.50, and 4.00 {\AA} are only
$-40$, $-24$, $-38$, $-29$, and $-14$ microhartree, respectively. After 4000 MC iterations, they become
$-10$, $-10$, $-12$, $-9$, and $-2$ microhartree, respectively. Once again, these are considerable
improvements compared to CR-CC(2,3) and CC(t;3) that both give
errors on the order of $-0.2$ millihartree,\cite{jspp-jctc2012,nbjspp-molphys2017} especially if we realize that 
the fractions of triples captured by the $i$-FCIQMC runs after 2000 and 4000
MC iterations are relatively
small (5--28\% and 5--49\%, respectively) and,
as shown in Table S.2 of the supplementary material,
the corresponding numbers of walkers
represent only about 1--2\% of the total numbers of walkers at $\tau = 20.0$ a.u., where
we stopped our $i$-FCIQMC propagations.

As a consequence of the small errors in the CC($P$;$Q$) total energies characterizing the ${\rm X\: ^1\Sigma}_{g}^{+}$
and ${\rm A\: ^3\Sigma}_{u}^{+}$ states in the early stages of the $i$-FCIQMC propagations, the
resulting singlet--triplet gap values are very accurate as well. This is demonstrated in Table \ref{table:table5},
where one can see that after 2000 $\delta \tau = 0.0001$ a.u. MC iterations, which is, as already explained,
a very short propagation time engaging tiny walker populations and small fractions of triples,
most of the differences between the CC($P$;$Q$) and CCSDT $\Delta E_{\rm S\mbox{-}T}$ values in the
$R_{\rm H\mbox{-}F} = 1.50\mbox{--}4.00$ {\AA} region
are on the order of a few reciprocal centimeter. The only exception is the semi-stochastic CC($P$;$Q$)
run at $R_{\rm H\mbox{-}F} = 2.50$ {\AA}, where the $-122$ ${\rm cm^{-1}}$ error relative to CCSDT
characterizing the singlet--triplet gap obtained after 2000 MC time steps, while representing a five-fold
error reduction compared to CR-CC(2,3), is comparable, in magnitude,
to the CCSDT value of $\Delta E_{\rm S\mbox{-}T}$.
This happens because the CC($P$;$Q$) energy of the strongly correlated ${\rm X\: ^1\Sigma}_{g}^{+}$ state obtained
after 2000 MC iterations at $R_{\rm H\mbox{-}F} = 2.50$ {\AA} differs from its CCSDT parent by $-0.583$ millihartree,
whereas the analogous difference between the CC($P$;$Q$) and CCSDT energies for its weakly correlated
${\rm A\: ^3\Sigma}_{u}^{+}$ companion is only $-29$ microhartree. This is not a problem though, since
by running $i$-FCIQMC a little longer and capturing about 20\% of all triply excited determinants in the
relevant $P$ spaces, as is the case when 4000 $\delta \tau = 0.0001$ a.u. MC time steps are considered, one reduces
the differences between the CC($P$;$Q$) and CCSDT energies of the ${\rm X\: ^1\Sigma}_{g}^{+}$
and ${\rm A\: ^3\Sigma}_{u}^{+}$ states to $-50$ and $-9$
microhartree, respectively (cf. Tables \ref{table:table3} and \ref{table:table4}), so that
the 122 ${\rm cm^{-1}}$ unsigned error in the CC($P$;$Q$) value of $\Delta E_{\rm S\mbox{-}T}$ relative to CCSDT
obtained after 2000 MC iterations decreases to less than 10 ${\rm cm^{-1}}$. This is yet another illustration of the
ability of the semi-stochastic CC($P$;$Q$) methodology pursued in this work to balance the more MR singlet and
weakly correlated triplet states of biradical systems in a single computation at the fraction of the cost of
the parent high-level CC calculations. As shown in Table \ref{table:table5}, at $\tau = 0.4$ a.u., where the
$i$-FCIQMC propagations are still far from being converged, the FCIQMC-driven CC($P$;$Q$) calculations recover the
CCSDT values of the singlet--triplet gaps in ${\rm (HFH)}^{-}$ at all H--F distances considered in this study
to within a few reciprocal centimeter, reaching a 1--2 ${\rm cm^{-1}}$ or better accuracy after 6000 MC
iterations.

Last, but not least, the results reported in Tables \ref{table:table3}--\ref{table:table5} and Fig.\
\ref{fig:figure2} also demonstrate
the remarkable efficiency of the $\delta(P;Q)$ corrections in accelerating the convergence of the CC($P$)
energies of the ${\rm X\: ^1\Sigma}_{g}^{+}$ and ${\rm A\: ^3\Sigma}_{u}^{+}$ states and the vertical gaps
between them toward CCSDT, independent of the H--F distance considered.
Let us, for example, compare the uncorrected CC($P$) and
corrected CC($P$;$Q$) energies of the ${\rm X\: ^1\Sigma}_{g}^{+}$ and ${\rm A\: ^3\Sigma}_{u}^{+}$ states
of ${\rm (HFH)}^{-}$ at the five H--F separations considered in this work obtained after 2000 MC iterations.
In the case of the former, more MR, state, the CC($P$;$Q$) corrections $\delta(P;Q)$ reduce the positive
2.601, 3.998, 3.511, 6.586, and 0.412 millihartree errors relative to CCSDT resulting from the CC($P$)
computations at $R_{\rm H\mbox{-}F} = 1.50$, 1.75, 2.00, 2.50, and 4.00 {\AA} to the much smaller
negative error values of $-0.035$, $-0.056$, $-0.110$, $-0.583$, and $-0.025$ millihartree, respectively.
When the latter state, which is characterized by much weaker correlations,
is considered,
the CC($P$;$Q$) approach reduces
the 0.995, 0.826, 0.834, 0.502, and 0.239
millihartree errors obtained with CC($P$)
to $-40$, $-24$, $-38$, $-29$, and $-14$ microhartree, respectively. It is
interesting to notice that while the errors
characterizing the CC($P$) calculations for the ${\rm A\: ^3\Sigma}_{u}^{+}$ state are generally much smaller
than their ${\rm X\: ^1\Sigma}_{g}^{+}$ counterparts, and the two states have a substantially different
character, the error reductions offered by the CC($P$;$Q$) corrections $\delta(P;Q)$,
by at least one order of magnitude, apply to both states. As already alluded to above, and as shown in
Table \ref{table:table5} and Fig.\ \ref{fig:figure2} (e) and (f), where we examine the convergence of the
CC($P$) and CC($P$;$Q$) $\Delta E_{\rm S\mbox{-}T}$ values toward their CCSDT parents, the noniterative
corrections $\delta(P;Q)$ are also very effective in improving the balance in the description of the
${\rm X\: ^1\Sigma}_{g}^{+}$ and ${\rm A\: ^3\Sigma}_{u}^{+}$ states by the CC($P$) approach and
smoothing the convergence of the resulting singlet--triplet gaps toward their CCSDT limits. This can be
illustrated by comparing the behavior of the error values relative to CCSDT characterizing the CC($P$) calculations
of $\Delta E_{\rm S\mbox{-}T}$ at $R_{\rm H\mbox{-}F} = 2.00$ {\AA} with their CC($P$;$Q$) counterparts, shown
in Table \ref{table:table5}. In the former case, the 3462 ${\rm cm^{-1}}$ error at $\tau = 0$ decreases,
in absolute value, to 8 ${\rm cm^{-1}}$ at $\tau = 0.8$ a.u. (8000 MC iterations), to increase to 17 ${\rm cm^{-1}}$
at $\tau = 2.0$ a.u. (20000 MC iterations), to decrease again to a numerical 0 ${\rm cm^{-1}}$ at
$\tau = 20.0$ a.u. (200000 MC iterations). Once the CC($P$) energies of the ${\rm X\: ^1\Sigma}_{g}^{+}$ and
${\rm A\: ^3\Sigma}_{u}^{+}$ states are corrected using the $\delta(P;Q)$ corrections, the unsigned errors in the
resulting CC($P$;$Q$) values of $\Delta E_{\rm S\mbox{-}T}$ relative to their CCSDT parent monotonically and rapidly
decrease, from 282 ${\rm cm^{-1}}$ at $\tau = 0$ to a numerical 0 ${\rm cm^{-1}}$ at $\tau \geq 0.8$ a.u.
It is clear from Tables \ref{table:table3}--\ref{table:table5} and Fig.\ \ref{fig:figure2} that while
both the CC($P$) and CC($P$;$Q$) energies converge to the parent CCSDT limit, the latter energies
and the gaps between them converge to CCSDT a lot faster.


\subsection{Cyclobutadiene and cyclopentadienyl cation}
\label{results:cbdcpc}

We now proceed to the examination of the performance of the semi-stochastic CC($P$;$Q$) algorithm
in calculations involving medium-sized organic biradicals, starting from two prototypical anti-aromatic systems,
cyclobutadiene and cyclopentadienyl cation, both described using the cc-pVDZ basis set. As in the
rest of the present study, we are mainly interested in how efficient the CIQMC-driven CC($P$;$Q$)
methodology is in recovering the CCSDT energies of the lowest singlet and triplet states and gaps between them.
In the case of cyclobutadiene and cyclopentadienyl cation discussed in this subsection, we focus on examining
vertical singlet--triplet gaps.

We begin with the FCIQMC-driven CC($P$;$Q$) calculations for cyclobutadiene, in which we adopted
the $D_{4h}$-symmetric geometry that represents the transition state for the
automerization of cyclobutadiene proceeding on the lowest singlet
potential, optimized
with the MR average-quadratic CC (MR-AQCC) approach\cite{mraqcc1,mraqcc2} using the cc-pVDZ basis in
Ref.\ \onlinecite{MR-AQCC}. We employed this geometry for two reasons. One of them is the fact that we used the
same geometry in our earlier CIQMC- and CCMC-based,\cite{stochastic-ccpq-prl-2017,jed-js-pp-jcp-2021}
CIPSI-driven,\cite{cipsi-ccpq-2021} and active-orbital-based\cite{jspp-jcp2012} CC($P$;$Q$) calculations
for cyclobutadiene, when examining its automerization. Because of this, we could verify the correctness of
our FCIQMC-driven CC($P$;$Q$) calculations for the lowest-energy singlet state, which is also the
ground state of the system. Another is the observation that the $D_{4h}$-symmetric transition-state
structure characterizing the automerization of cyclobutadiene is practically identical to the
$D_{4h}$-symmetric minimum on the lowest triplet surface.
Indeed, the MR-AQCC/cc-pVDZ C--C and C--H bond lengths
defining the transition state on the ground-state
singlet potential differ from those
characterizing the triplet minimum optimized using unrestricted CCSD (UCCSD)
in Ref.\ \onlinecite{BiradicalGeom} by less than
0.009 and 0.001 {\AA}, respectively.

At the $D_{4h}$-symmetric geometry used in our calculations, cyclobutadiene is characterized by the
delocalization of four $\pi$ electrons over four $\pi$ MOs, which gives rise to the close-lying
singlet and triplet states that require a highly accurate treatment of electron correlation effects
if we are to obtain a well-balanced description of the two states and the small energy separation between them.
One can understand this by examining the valence $\pi$ network of the $D_{4h}$-symmetric cyclobutadiene species,
which consists of the doubly occupied nondegenerate $a_{2u}$ orbital, the doubly degenerate $e_{g}$ level,
in which each component MO is occupied by a single electron, and the nondegenerate $b_{1u}$ orbital,
which in the zeroth-order description of the lowest singlet and triplet states remains empty.
The two valence electrons in the degenerate $e_{g}$ shell can couple to a singlet or a triplet,
resulting in the open-shell singlet ground state, ${\rm X\: ^1B}_{1g}$, which has a substantial MR character,
and the first excited triplet state, ${\rm A\: ^3A}_{2g}$, which is predominantly SR in nature.
In order to balance the substantial nondynamical correlation effects, needed for an accurate description of
the low-spin ${\rm X\: ^1B}_{1g}$ state, with the dynamical correlations dominating its high-spin triplet
${\rm A\: ^3A}_{2g}$ companion within a conventional, particle-conserving, SRCC framework and produce
reliable $\Delta E_{\rm S\mbox{-}T}$ values for cyclobutadiene, which could compete with the high-accuracy
{\it ab initio} data reported in Refs.\
\onlinecite{balkova1994,krylov2004,MR-AQCC,BiradicalGeom,stoneburner-js-jcp2017,aj-js-pp-jpca2017,%
zimmerman2017,mazziotti2021,loos2022}, one has to consider robust treatments of the connected triply
excited clusters, such as that offered by CCSDT.\cite{balkova1994,loos2022}
Indeed, full CCSDT, which
is the target of this investigation,
produces high-quality results for the lowest singlet and triplet states of the $D_{4h}$-symmetric
cyclobutadiene system and the energy separation between them. For example, the $\Delta E_{\rm S\mbox{-}T}$
value obtained in the CCSDT/cc-pVDZ calculations at the transition-state geometry used
in the present study, of $-4.8$ kcal/mol, is practically identical to the results of the state-of-the-art
DEA-EOMCC computations including the high-rank
4p2h
correlations on top of CCSD, reported
in Refs.\ \onlinecite{stoneburner-js-jcp2017,aj-js-pp-jpca2017}, which give $-5.0$ kcal/mol
when the cc-pVDZ basis set is employed (for similar recent observations regarding the reliability
of full CCSDT in generating virtually exact singlet--triplet gap values for cyclobutadiene, see
Ref.\ \onlinecite{loos2022}). It is, therefore, interesting to explore if the semi-stochastic CC($P$;$Q$)
methodology investigated in this work is capable of converging the CCSDT results for the ${\rm X\: ^1B}_{1g}$ and
${\rm A\: ^3A}_{2g}$ states of cyclobutadiene and vertical gap between them out of the early stages of
CIQMC propagations.

\begin{figure*}[!htbp]
\centering

\includegraphics[width=1.0\textwidth]{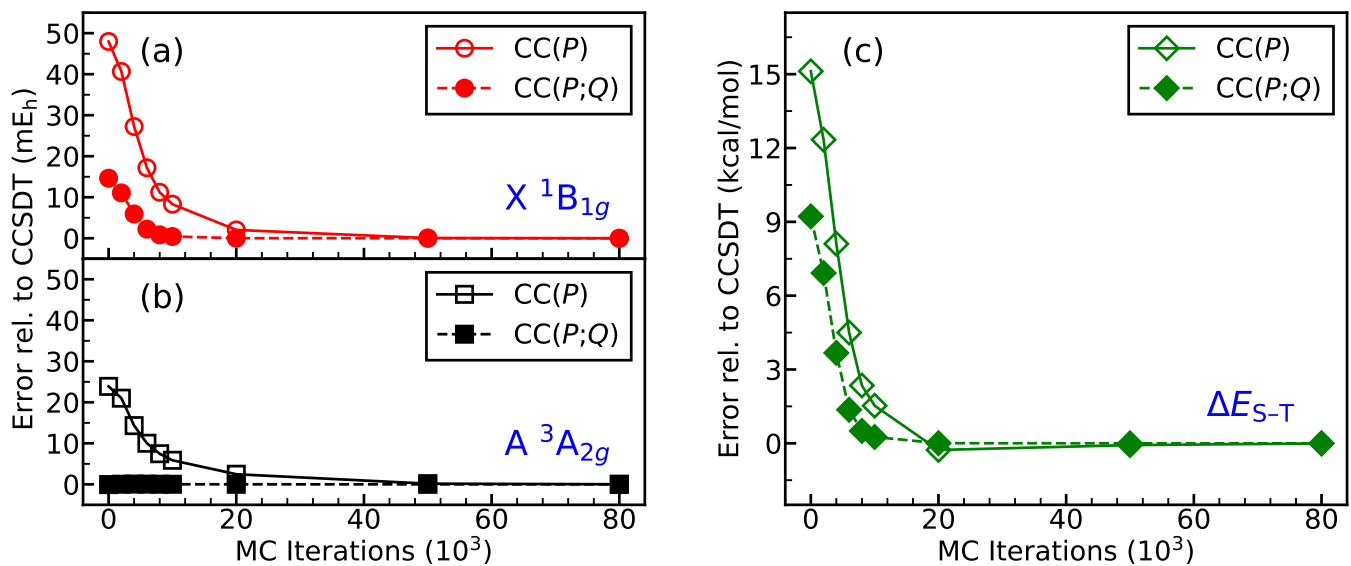}
\caption{ \label{fig:figure3}
Convergence of the CC($P$) and CC($P$;$Q$) energies of the ${\rm X \: ^1B}_{1g}$ [panel (a)]
and ${\rm A \: ^3A}_{2g}$ [panel (b)] states
of cyclobutadiene, as described by the cc-pVDZ basis set,
and of the corresponding vertical singlet--triplet gaps [panel (c)] toward their parent CCSDT values.
All calculations were performed at the $D_{4h}$-symmetric transition-state geometry of the ${\rm X\: ^1B}_{1g}$ state
optimized in the MR-AQCC calculations in Ref.\ \onlinecite{MR-AQCC}.
The $P$ spaces consisted of all singles and doubles and subsets
of triples identified during the $i$-FCIQMC propagations with $\delta \tau=0.0001$ a.u.\ and the $Q$
spaces consisted of the triples not captured by $i$-FCIQMC.}
\end{figure*}

The results of our FCIQMC-driven CC($P$) and CC($P$;$Q$) computations for cyclobutadiene
are summarized in Table \ref{table:table7}
and Fig.\ \ref{fig:figure3}.
In all of our
calculations, starting with the stochastic $i$-FCIQMC steps and ending with the
deterministic CC($P$;$Q$) and CCSDT runs, we used the $D_{2h}$ Abelian subgroup of the $D_{4h}$
point group characterizing the cyclobutadiene's geometry adopted in this work. Consequently, the
$i$-FCIQMC propagations for the ${\rm X\: ^1B}_{1g}$ and ${\rm A\: ^3A}_{2g}$ states
were set up to converge the lowest states of the
${\rm ^1A}_{g} (D_{2h})$ and ${\rm ^3B}_{1g} (D_{2h})$ symmetries. Consistent with
the CC($P$) and CC($P$;$Q$)
runs that follow the $i$-FCIQMC steps
and the accompanying CCSD, CR-CC(2,3), and CCSDT computations,
the reference determinants
used to initiate our $i$-FCIQMC propagations were the closed-shell, $D_{2h}$-adapted, RHF function
obtained by placing two electrons on one of the $e_{g}$ valence orbitals for the lowest-energy
${\rm ^1A}_{g} (D_{2h})$ state and the high-spin ROHF determinant,
adapted to $D_{2h}$ as well,
for the lowest ${\rm ^3B}_{1g} (D_{2h})$ state.
As a result, the lists of
triply excited determinants extracted from the $i$-FCIQMC runs at the various propagation times $\tau > 0$,
needed to define the $P$ spaces
for the CC($P$) and CC($P$;$Q$)
computations, consisted of the
$S_z=0$ triples of the ${\rm A}_{g} (D_{2h})$ symmetry for the ${\rm X\: ^1B}_{1g}$ state and the $S_z=1$ triples
of the ${\rm B}_{1g} (D_{2h})$ symmetry in the case of the ${\rm A\: ^3A}_{2g}$ state. Given our interest
in converging the CCSDT energetics, the $Q$ spaces
used to construct the
$\delta(P;Q)$ corrections consisted of the remaining triply excited determinants, absent
in the $i$-FCIQMC wave functions of the ${\rm X\: ^1B}_{1g}$ and ${\rm A\: ^3A}_{2g}$ states
at a given $\tau$.

The results shown in Table \ref{table:table7} and Fig.\ \ref{fig:figure3} display several similarities
with the previously discussed methylene and ${\rm (HFH)}^{-}$ cases. One cannot, for example, obtain
an accurate description of the more MR singlet ground state and the energy separation between
the ${\rm X\: ^1B}_{1g}$ and ${\rm A\: ^3A}_{2g}$ states without incorporating the leading triply excited
determinants in the $P$ space. Indeed, when the $P$ space consists of only singly and doubly excited determinants,
as in the $\tau = 0$ CC($P$) (i.e., CCSD) and CC($P$;$Q$) [i.e., CR-CC(2,3)] calculations, one ends up with
the enormous errors in the energies of the ${\rm X\: ^1B}_{1g}$ state relative to their CCSDT parent, which are
47.979 millihartree in the former case and 14.636 millihartree when the latter computation is considered.
The $\tau = 0$ CC($P$;$Q$) energy of the ${\rm A\: ^3A}_{2g}$ state is a lot more accurate, reducing the
large, 23.884 millihartree, error relative to CCSDT obtained in the underlying CC($P$) calculation
to $-60$ microhartree, but this does not help too much. The corresponding CR-CC(2,3) triples correction to CCSD,
which neglects the coupling of the low-order $T_{1}$ and $T_{2}$ clusters with their higher-order $T_{3}$
counterpart, is incapable of offering a balanced description of the ${\rm X\: ^1B}_{1g}$ and ${\rm A\: ^3A}_{2g}$
states, so that the resulting singlet--triplet gap is very poor. The 9.2 kcal/mol
difference between the $\Delta E_{\rm S\mbox{-}T}$ values obtained in the $\tau = 0$ CC($P$;$Q$)
or CR-CC(2,3) and CCSDT calculations is so large that the ${\rm X\: ^1B}_{1g}-{\rm A\: ^3A}_{2g}$
separation predicted by CR-CC(2,3) has a wrong sign compared to its $-4.8$ kcal/mol CCSDT counterpart,
while being nearly identical in magnitude. This difference becomes even larger when the uncorrected
$\tau = 0$ CC($P$), meaning CCSD, calculations are
considered (15.1 kcal/mol).

As shown in Table \ref{table:table7} and Fig.\ \ref{fig:figure3}, the situation dramatically changes when
the $P$ spaces used in the CC($P$) and CC($P$;$Q$) calculations are enriched with the subsets of
triply excited determinants captured by the $i$-FCIQMC propagations. The convergence of the CC($P$;$Q$)
energies of the ${\rm X\: ^1B}_{1g}$ and ${\rm A\: ^3A}_{2g}$ states, especially the former ones,
and the vertical separations between them is particularly impressive. For example, after as few
as 6000 $\delta \tau = 0.0001$ a.u. MC time steps and $i$-FCIQMC capturing less than 30\% of all
triples in the $P$ space, where,
as demonstrated in Table S.3 of the supplementary material,
the walker population
characterizing the $i$-FCIQMC run for the ${\rm X\: ^1B}_{1g}$ state is only 0.02\% of the total number
of walkers at $\tau = 8.0$ a.u. (the termination time for our $i$-FCIQMC propagations for cyclobutadiene),
the CC($P$;$Q$) approach reduces the 14.636 millihartree difference between the CR-CC(2,3) and CCSDT
energies of the strongly correlated singlet ground state to 2.223 millihartree. While the CR-CC(2,3)
description of the ${\rm A\: ^3A}_{2g}$ state, which has a largely SR character,
is already excellent, the CC($P$;$Q$) calculation
performed after 6000 MC iterations, which uses only 26\% of triples in the $P$ space and a tiny
walker population that amounts to 0.04\% of all walkers at $\tau = 8.0$ a.u. in the underlying $i$-FCIQMC propagation,
improves it too, reducing the small, $60$ microhartree, unsigned difference between the CR-CC(2,3) and CCSDT energies
to an even smaller 51 microhartree. As a consequence of the above improvements, especially for the
${\rm X\: ^1B}_{1g}$ state, the error relative to CCSDT characterizing the
$\Delta E_{\rm S\mbox{-}T}$ value obtained in the FCIQMC-driven CC($P$;$Q$) calculations after
6000 $\delta \tau = 0.0001$ a.u. MC time steps, where the underlying $i$-FCIQMC propagations are
still in their early stages, is only 1.4 kcal/mol, as opposed to 9.2 kcal/mol obtained at $\tau = 0$
with CR-CC(2,3). The resulting ${\rm X\: ^1B}_{1g}-{\rm A\: ^3A}_{2g}$ energy separation, of $-3.4$ kcal/mol,
has not only the correct sign, but is also very close to the $-4.8$ kcal/mol value obtained with CCSDT.
If we wait a little longer, by executing the extra 2000 MC iterations, so that the $i$-FCIQMC propagations can
capture 34\%--39\% of all triply excited determinants, we can reduce the already small 2.223 millihartree, 51
microhartree, and 1.4 kcal/mol errors in the CC($P$;$Q$) energies of the ${\rm X\: ^1B}_{1g}$ and ${\rm A\: ^3A}_{2g}$
states and separation between them relative to CCSDT, obtained after 6000 MC time steps, to
0.835 millihartree, 31 microhartree, and 0.5 kcal/mol, respectively. It is clear from
Table \ref{table:table7} and Fig.\ \ref{fig:figure3} that the convergence of the semi-stochastic CC($P$;$Q$) results
for the lowest-energy singlet and triplet states of cyclobutadiene, especially the ${\rm X\: ^1B}_{1g}$ energies
and the ${\rm X\: ^1B}_{1g}-{\rm A\: ^3A}_{2g}$ gap values, which the $\tau = 0$ CC($P$;$Q$) or CR-CC(2,3)
calculations describe poorly, toward CCSDT is very fast, even when the underlying $i$-FCIQMC propagations are far from
convergence. It is also apparent from our calculations that the noniterative corrections $\delta(P;Q)$ play a
significant role in accelerating convergence of the corresponding CC($P$) energetics toward CCSDT.
As shown, for example, in Table
\ref{table:table7}, the relatively large differences between the uncorrected CC($P$) energies of the
${\rm X\: ^1B}_{1g}$ and ${\rm A\: ^3A}_{2g}$ states and vertical gap between them obtained at $\tau = 0.8$ a.u.,
i.e., after 8000 $\delta \tau = 0.0001$ a.u. MC iterations,
and the corresponding CCSDT data, which exceed 11 and 7 millihartree and 3 kcal/mol, respectively,
are reduced to 0.835 millihartree, 31 microhartree, and 0.5 kcal/mol, when the CC($P$;$Q$) approach is
employed. We can see similar improvements in the CC($P$) energies at other $\tau$ values.

Most of the observations regarding the performance of the semi-stochastic CC($P$;$Q$) methodology
and its CC($P$) counterpart remain valid when the larger cyclopentadienyl cation, which is also
the largest molecular system considered in our CC($P$)/CC($P$;$Q$) work to date, is examined. 
Following our previous DEA-EOMCC studies of cyclopentadienyl cation,
\cite{stoneburner-js-jcp2017,aj-js-pp-jpca2017}
where we investigated the effect of high-order
4p2h
correlations on its singlet--triplet gap,
we used the $D_{5h}$-symmetric geometry corresponding to a minimum on the lowest triplet surface
obtained in the UCCSD/cc-pVDZ optimization in Ref.\ \onlinecite{BiradicalGeom}. At this geometry,
cyclopentadienyl cation is characterized by the delocalization of four $\pi$ electrons over five $\pi$
MOs, resulting in the doubly occupied nondegenerate $a_{2}^{\prime \prime}$ orbital,
the doubly degenerate $e_{1}^{\prime \prime}$ shell, in which each component MO is occupied by
a single electron, and the doubly degenerate $e_{2}^{\prime \prime}$ shell,
which in the zeroth-order description of the lowest-energy singlet and triplet states remains empty.
In analogy to the previously discussed cyclobutadiene system, the two electrons in the degenerate
$e_{1}^{\prime \prime}$ MOs can couple to a singlet or triplet, but compared to cyclobutadiene,
where the lowest-energy singlet state is also a ground state, the state ordering in cyclopentadienyl
cation is reversed, so that the lowest triplet, designated as ${\rm X\: ^3A_{2}^{\prime}}$, is the ground state
and the lowest-energy singlet, denoted as ${\rm A\: ^1E_{2}^{\prime}}$, is the first excited state.
Similar to all other examples considered in this work,
in order to obtain a well-balanced description of the ${\rm X\: ^3A_{2}^{\prime}}$ state, which has a
SR character dominated by dynamical correlations, and its ${\rm A\: ^1E_{2}^{\prime}}$ companion, which is
an open-shell singlet characterized by significant MR correlations, and obtain an accurate value of
$\Delta E_{\rm S\mbox{-}T}$ within a conventional SRCC framework, one must turn to higher-level
theories that can offer a robust treatment of $T_{n}$ clusters with $n > 2$. Otherwise, as
shown in Ref.\ \onlinecite{BiradicalGeom}, and as confirmed in our calculations, the results can be very poor.
For example, the ${\rm A\: ^1E_{2}^{\prime}}-{\rm X\: ^3A_{2}^{\prime}}$ separation in cyclopentadienyl cation
resulting from the restricted CCSD calculations using the cc-pVDZ basis, which are equivalent to our
$\tau = 0$ CC($P$) computations, is about 23 kcal/mol. This is in large disagreement with the
most accurate {\it ab initio} calculations of the singlet--triplet gap in cyclopentadienyl cation
performed to date using the DEA-EOMCC formalism including
3p1h as well as 4p2h
correlations on top of the CCSD treatment of the
underlying closed-shell core, which give about 14 kcal/mol when the cc-pVDZ basis set is employed
\cite{stoneburner-js-jcp2017,aj-js-pp-jpca2017} (for the examples of other high-level SRCC and MRCC
calculations of the singlet--triplet gap in cyclopentadienyl cation, see Ref.\ \onlinecite{BiradicalGeom};
Ref.\ \onlinecite{stoneburner-js-jcp2017} also provides the well-converged MR perturbation theory data, which
agree with the state-of-the-art DEA-EOMCC computations reported in Refs.\
\onlinecite{stoneburner-js-jcp2017,aj-js-pp-jpca2017}). The restricted CCSDT approach, which is the target SRCC
method in this study, provides a much better description, reducing the approximately 9 kcal/mol
error relative to the most accurate DEA-EOMCC calculations with up to
4p2h
excitations
reported in Refs.\ \onlinecite{stoneburner-js-jcp2017,aj-js-pp-jpca2017} obtained with restricted CCSD
to less than 3 kcal/mol, when the cc-pVDZ basis set is employed. It would certainly be interesting to
examine if the inclusion of higher--than--triply excited clusters, such as $T_{4}$, could further improve
the CCSDT description of the singlet--triplet gap in cyclopentadienyl cation, but in this work we focus
on the ability of the semi-stochastic, CIQMC-based, CC($P$;$Q$) methodology to improve the CR-CC(2,3)
$\Delta E_{\rm S\mbox{-}T}$ values and converge the results
of CCSDT computations. We hope to return to the topic of the role of $T_{4}$ clusters in
describing the singlet--triplet gap in cyclopentadienyl cation in one of our future studies.
It may be worth pointing out
that the ${\rm A\: ^1E_{2}^{\prime}}-{\rm X\: ^3A_{2}^{\prime}}$ gap obtained in
the restricted CCSDT/cc-pVDZ calculations, which give $\Delta E_{\rm S\mbox{-}T} = 16.7$ kcal/mol,
is in very good agreement with the 16.1 kcal/mol resulting from the DEA-EOMCC/cc-pVDZ computations
truncated at
3p1h excitations.\cite{stoneburner-js-jcp2017,aj-js-pp-jpca2017}

\begin{figure*}[!htbp]
\centering

\includegraphics[width=1.0\textwidth]{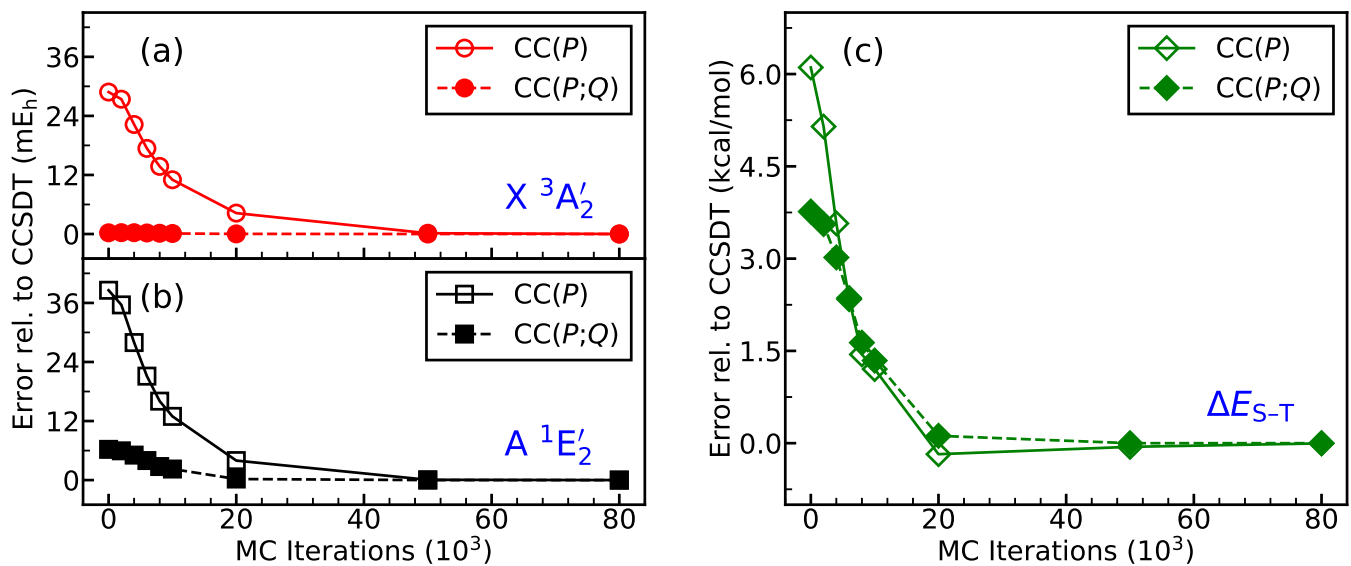}
\caption{ \label{fig:figure4} 
Convergence of the CC($P$) and CC($P$;$Q$) energies of the ${\rm X \: ^3A^\prime_2}$ [panel (a)]
and ${\rm A \: ^1E^\prime_2}$ [panel (b)] states
of cyclopentadienyl cation, as described by the cc-pVDZ basis set,
and of the corresponding vertical singlet--triplet gaps [panel (c)] toward their parent CCSDT values.
All calculations were performed at the $D_{5h}$-symmetric geometry
of the ${\rm X\: ^3A^\prime_2}$ state optimized using the unrestricted CCSD/cc-pVDZ approach in
Ref.\ \onlinecite{BiradicalGeom}.
The $P$ spaces consisted of all singles and doubles and subsets
of triples identified during the $i$-CISDTQ-MC propagations with $\delta \tau=0.0001$ a.u.\ and the $Q$
spaces consisted of the triples not captured by $i$-CISDTQ-MC.}
\end{figure*}

The results of our CIQMC-driven CC($P$) and CC($P$;$Q$) computations for cyclopentadienyl cation
are reported in Table \ref{table:table9} and Fig.\ \ref{fig:figure4}.
As already alluded to above,
to reduce the computational
costs of the CIQMC propagations preceding the CC($P$) and CC($P$;$Q$) steps, especially in the later stages
of the CIQMC runs that are included in Table \ref{table:table9} and Fig.\ \ref{fig:figure4} for the
completeness of our presentation, we replaced the $i$-FCIQMC algorithm, which we
exploited in our calculations for methylene,
${\rm (HFH)}^{-}$, and cyclobutadiene, by its truncated $i$-CISDTQ-MC counterpart. It has been established
in Ref.\ \onlinecite{jed-js-pp-jcp-2021} that the replacement of $i$-FCIQMC by $i$-CISDTQ-MC, when identifying
the leading higher--than--doubly excited
determinants
for the
inclusion in the $P$ spaces used in the semi-stochastic CC($P$) and CC($P$;$Q$) runs, has
virtually no effect on the rate at which these runs converge the parent
SRCC energetics.
In analogy to
cyclobutadiene, all of our 
$i$-CISDTQ-MC, semi-stochastic CC($P$) and CC($P$;$Q$), and deterministic CCSD, CR-CC(2,3), and
CCSDT computations utilized the largest Abelian subgroup of the $D_{5h}$ point group characterizing
the cyclopentadienyl cation's structure examined in the present study, which is $C_{2v}$. This
means that in setting up our calculations for the ${\rm X\: ^3A_{2}^{\prime}}$ state, we treated it
as the lowest state of the ${\rm ^3B_2}(C_{2v})$ symmetry, whereas the doubly degenerate
${\rm A\: ^1E_{2}^{\prime}}$ state was represented by its ${\rm ^1A_1}(C_{2v})$ component.
Similar to cyclobutadiene, and to remain consistent with the CC($P$), CC($P$;$Q$), and other
SRCC runs for cyclopentadienyl cation carried out in this study, the reference determinant
used to initiate the $i$-CISDTQ-MC propagation for the lowest-energy ${\rm ^3B_2}(C_{2v})$ state
was the triplet ROHF determinant. In the case of the
${\rm ^1A_1}(C_{2v})$ component of the ${\rm A\: ^1E_{2}^{\prime}}$ state,
we used the
RHF determinant obtained by pairing the two valence electrons in one of the $e_{1}^{\prime\prime}$
MOs
to initiate the corresponding $i$-CISDTQ-MC run.
Consistent with the above description, the subsets of triply excited determinants
used to construct the $P$ spaces for the semi-stochastic CC($P$) and CC($P$;$Q$) computations
for the ${\rm X\: ^3A_{2}^{\prime}}$ state were the $S_z=1$ triples of the ${\rm B_2}(C_{2v})$ symmetry
captured by $i$-CISDTQ-MC. In the case of the ${\rm A\: ^1E_{2}^{\prime}}$ state, represented,
as explained above, by its ${\rm ^1A_1}(C_{2v})$ component, we used the $S_z=0$ triples of the
${\rm A_1}(C_{2v})$ symmetry identified by the $i$-CISDTQ-MC propagation set up to
converge the lowest ${\rm A_1}(C_{2v})$ state. As usual, the corresponding $Q$ spaces were
spanned by the remaining triply excited determinants that were not captured by the $i$-CISDTQ-MC runs
when the lists of $P$-space triples were created.

Our calculations for cyclopentadienyl cation, summarized in Table \ref{table:table9} and Fig.\ \ref{fig:figure4},
demonstrate that the CC($P$;$Q$) energies of the ${\rm X\: ^3A_{2}^{\prime}}$ and ${\rm A\: ^1E_{2}^{\prime}}$
states and vertical gaps between them display fast convergence toward the respective CCSDT values with the
propagation time $\tau$. This is particularly apparent in the case of the CC($P$;$Q$) energies of the more MR
${\rm A\: ^1E_{2}^{\prime}}$ state and the ${\rm A\: ^1E_{2}^{\prime}}-{\rm X\: ^3A_{2}^{\prime}}$ separation,
which cannot be accurately described if the underlying $P$ spaces contain only singly and doubly excited
determinants. Indeed, the CR-CC(2,3) energy of the ${\rm A\: ^1E_{2}^{\prime}}$ state, which is
equivalent to the $\tau = 0$ CC($P$;$Q$) value, is much more accurate than the result of the associated
CC($P$) or CCSD calculation, which produces the enormous error relative to CCSDT exceeding 38
millihartree, but the substantial, $>6$ millihartree, difference with the CCSDT energy remains.
The situation for the SR ${\rm X\: ^3A_{2}^{\prime}}$ state, where the CR-CC(2,3) approach reduces the
nearly 29 millihartree error relative to CCSDT obtained in the CCSD calculations to $\sim$0.2 millihartree,
is a lot better, but this does not help the resulting $\Delta E_{\rm S\mbox{-}T}$ value, which differs from
its CCSDT counterpart by almost 4 kcal/mol (almost a quarter of the CCSDT value of $\Delta E_{\rm S\mbox{-}T}$).
The discrepancy between the errors in the CR-CC(2,3) energies of the ${\rm X\: ^3A_{2}^{\prime}}$ and
${\rm A\: ^1E_{2}^{\prime}}$ states is simply too large. Clearly, one needs to incorporate some triples in the
corresponding $P$ spaces, especially in the case of the more challenging ${\rm A\: ^1E_{2}^{\prime}}$ state.

Once the $\tau = 0$ $P$ spaces are augmented with the leading triply excited determinants identified
by the $i$-CISDTQ-MC propagations and the noniterative corrections $\delta(P;Q)$ are added to the
CC($P$) energies to estimate the effects of the remaining $T_{3}$ correlations, we observe smooth
convergence of the resulting CC($P$;$Q$) energetics toward their respective CCSDT limits. This includes
significant improvements in the poor description of the ${\rm A\: ^1E_{2}^{\prime}}$ state and the
${\rm A\: ^1E_{2}^{\prime}}-{\rm X\: ^3A_{2}^{\prime}}$ separation by CR-CC(2,3).
As shown in Table \ref{table:table9},
already after 10000 $\delta \tau = 0.0001$ a.u. MC time steps, where the $i$-CISDTQ-MC propagations
are still in their infancy, capturing only 25--30\% of all triples and using tiny walker populations,
on the order of 0.1--0.2\% of the total numbers of walkers at $\tau = 8.0$ a.u.
(see Table S.4 of the supplementary material),
the 6.245 millihartree and 3.8 kcal/mol errors in the energy of the ${\rm A\: ^1E_{2}^{\prime}}$ state and
the $\Delta E_{\rm S\mbox{-}T}$ value relative to CCSDT obtained with CR-CC(2,3) reduce
in the CC($P$;$Q$) calculations to 2.248 millihartree
and 1.3 kcal/mol, respectively. By running $i$-CISDTQ-MC a little longer and capturing about 50--60\%
of all triples in the relevant $P$ spaces, as is the case after 20000 MC iterations,
where the walker populations compared to $\tau = 8.0$ a.u. are still tiny, the errors
in the CC($P$;$Q$) values of the ${\rm A\: ^1E_{2}^{\prime}}$ energy and $\Delta E_{\rm S\mbox{-}T}$
relative to their CCSDT parents drop down by an order of magnitude compared to 10000 MC iterations,
to 0.217 millihartree and 0.1 kcal/mol, respectively. Although the excellent description of the
predominantly SR ${\rm X\: ^3A_{2}^{\prime}}$ state by the CR-CC(2,3) approach hardly needs any improvement,
the $i$-CISDTQ-MC-driven CC($P$;$Q$) calculations are helping here too, reducing the 0.245
millihartree difference between the CR-CC(2,3) and CCSDT energies to 0.108 millihartree after 10000 MC iterations
(26 microhartree when the number of MC iterations is increased to 20000).
As anticipated, the uncorrected CC($P$) energies
of the ${\rm X\: ^3A_{2}^{\prime}}$ and ${\rm A\: ^1E_{2}^{\prime}}$ states converge to the
respective CCSDT limits too,
but they do it at a much slower pace than their CC($P$;$Q$) counterparts. A comparison of the results of
the CC($P$) and CC($P$;$Q$) calculations for the ${\rm A\: ^1E_{2}^{\prime}}-{\rm X\: ^3A_{2}^{\prime}}$ gap
shown in Table \ref{table:table9} and Fig. \ref{fig:figure4} (c) may create an impression as if
the noniterative corrections $\delta (P;Q)$ offer very little, but this would be misleading.
The relatively fast convergence of the CC($P$) values of $\Delta E_{\rm S\mbox{-}T}$ toward
their CCSDT parent in the early stages of the underlying $i$-CISDTQ-MC propagations, which compares well
with that observed in the corresponding CC($P$;$Q$) computations, is a result of the fortuitous
cancellation of large errors characterizing the CC($P$) energies of the ${\rm X\: ^3A_{2}^{\prime}}$ and
${\rm A\: ^1E_{2}^{\prime}}$ states.
Since no other system examined in this study displays similar error cancellations,
and since costs of
computing
corrections
$\delta (P;Q)$, which offer major error reductions in the
individual CC($P$) energies, while accelerating their convergence
toward the SRCC target,
are low, we recommend using the $\delta (P;Q)$-corrected CC($P$;$Q$) energetics.

\subsection{Trimethylenemethane}
\label{results:tmm}

Our final example is trimethylenemethane, a fascinating non-Kekul\'{e} hydrocarbon examined as early as in
1948\cite{TMM-1} and 1950,\cite{TMM-2} in which four valence $\pi$ electrons are delocalized over four
closely spaced $\pi$-type orbitals. Assuming the $D_{3h}$ symmetry, which is the symmetry of the minimum-energy
structure on the ground-state triplet surface of trimethylenemethane,
the four MOs of this system's valence $\pi$ network consist of the nondegenerate
$1a_{2}^{\prime \prime}$ orbital, the doubly degenerate $1e^{\prime\prime}$ shell, and the nondegenerate
$2a_{2}^{\prime \prime}$ orbital. If one adopts the $C_{2v}$ symmetry, relevant to the
low-lying singlet states, which is also the largest Abelian subgroup of
$D_{3h}$ exploited in our CCSD, CR-CC(2,3), CCSDT, and CIQMC-driven CC($P$) and CC($P$;$Q$) computations,
the nondegenerate $1a_{2}^{\prime \prime}$ and $2a_{2}^{\prime \prime}$ orbitals in a $D_{3h}$ description
become the $1b_1$ and $3b_1$ MOs, respectively, whereas the degenerate $1e^{\prime\prime}$ shell splits into
the $1a_2$ and $2b_1$ components. 

The first experimental identification of trimethylenemethane dates back to 1966,\cite{TMM-32} a
definitive experimental verification, using electron paramagnetic resonance, of its triplet ground state
was accomplished already in 1976,\cite{TMM-33} and the electronic structure of trimethylenemethane
has been well understood for decades (cf., e.g., Ref.\ \onlinecite{TMM-8} and references therein),
but an accurate characterization of its triplet ground state and low-lying singlet states and
energy separations between them continues to present a significant challenge to quantum chemistry
approaches.\cite{jspp-jctc2012,jspp-dea-dip-2013,jspp-dea-dip-2014,aj-js-pp-jpca2017,%
TMM-14,TMM-15,TMM-16,TMM-17,TMM-18,TMM-19,%
TMM-20,TMM-21,TMM-22,TMM-23,TMM-24,TMM-25,TMM-26,TMM-27,TMM-28,TMM-29,TMM-30,%
ch2_krylov,TMM-krylov,krylov-little-t,TMM-BWCC,BCCC5,ch2-rmrccsdt,TMM-EXTRA-1,ssmrpt2019,ssmrpt2021}
The $D_{3h}$-symmetric triplet ground state, designated as ${\rm X\: ^3A}_{2}^{\prime}$
(in a $C_{2v}$ description adopted in this study, ${\rm X\: {^{3}}B}_{2}$), which is dominated by the
$\vert \lbrace{{\rm core}\rbrace}(1a_{2}^{\prime\prime})^2 (1e_{1}^{\prime\prime})^1 (1e_{2}^{\prime\prime})^1 \vert$
configuration (in $C_{2v}$, $\vert \lbrace{{\rm core}\rbrace} (1b_1)^2 (1a_2)^1 (2b_1)^1 \vert$), is relatively
easy to describe, but the next two states, which are the nearly degenerate singlets stabilized by the
Jahn--Teller distortion that lifts their exact degeneracy in a $D_{3h}$ description, are not. The lower of the
two singlets, which is characterized by a $C_s$-symmetric minimum that can be approximated by a twisted $C_{2v}$
structure and which is, therefore, usually designated as the ${\rm A\: ^1B_1}$ state, is
an open-shell singlet that emerges from the $\vert \lbrace{{\rm core}\rbrace} (1b_1)^2 (1a_2)^1 (2b_1)^1 \vert$
configuration. The second singlet, labeled as the ${\rm B\: ^1A_1}$ state, is a $C_{2v}$-symmetric
multi-configurational state dominated by the $\vert \lbrace{{\rm core}\rbrace} (1b_1)^2 (1a_2)^2 \vert$
and $\vert \lbrace{{\rm core}\rbrace} (1b_1)^2 (2b_1)^2 \vert$ closed-shell determinants. The ${\rm A\: ^1B_1}$ state,
although lower in energy compared to its ${\rm B\: ^1A_1}$ counterpart, has not been observed experimentally
due to unfavorable Frank--Condon factors,\cite{TMM-Expt-1,TMM-29} so we do not consider it in this work.
However, the second singlet, ${\rm B\: ^1A_1}$, has been detected in photoelectron spectroscopy experiments
reported in Refs.\  \onlinecite{TMM-Expt-1,TMM-Expt-2}, which located it at $16.1 \pm 0.1$ kcal/mol above the
${\rm X\: ^3A}_{2}^{\prime}$ ground state. Thus, following our previous deterministic, active-orbital-based,
CC($P$;$Q$) work\cite{jspp-jctc2012} and the state-of-the-art DEA- and DIP-EOMCC computations with up to
4p2h and 4h2p
excitations reported in Refs.\
\onlinecite{jspp-dea-dip-2013,jspp-dea-dip-2014,aj-js-pp-jpca2017}, in carrying out the CIQMC-driven
CC($P$) and CC($P$;$Q$) calculations discussed in this subsection and executing the accompanying
CCSD, CR-CC(2,3), and CCSDT runs, we focused on the $D_{3h}$-symmetric triplet ground state,
${\rm X\: ^3A}_{2}^{\prime}$, the $C_{2v}$-symmetric ${\rm B\: ^1A_1}$ singlet, and the adiabatic gap between them,
adopting the geometries of the two states optimized using the spin-flip density functional theory (SF-DFT)
and the 6-31G(d) basis in Ref.\ \onlinecite{ch2_krylov}.
In analogy to other organic biradicals discussed in this article, we employed the cc-pVDZ basis set,
so that the parent CCSDT computations, needed to judge the performance of our semi-stochastic
CC($P$) and CC($P$;$Q$) methods, and the more expensive CC($P$) and CC($P$;$Q$) calculations employing
large, near-100\%, fractions of triples in the relevant $P$ spaces (captured in the later stages of the
underlying CIQMC propagations) were not too difficult to execute on the computers available to us.
As shown in our earlier deterministic CC($P$;$Q$) work,\cite{jspp-jctc2012}
in which we tested the active-orbital-based CC(t;3) method, which recovers the CCSDT energetics
to within small fractions of kilocalorie per mole, and as confirmed by the authors of
Ref.\ \onlinecite{TMM-EXTRA-1}, who managed to perform the CCSDT/cc-pVTZ calculations,
the use of a larger cc-pVTZ basis changes the adiabatic ${\rm B\: ^1A_1}-{\rm X\: ^3A}_{2}^{\prime}$ gap
by about 0.5--1 kcal/mol, i.e., the use the cc-pVDZ basis is sufficient to draw meaningful
conclusions regarding the performance of the semi-stochastic CC($P$) and CC($P$;$Q$) approaches.

While the main goal of this study is to examine the efficiency of the CIQMC-driven CC($P$;$Q$) approaches
in converging the CCSDT energetics,
it is worth pointing out that the parent CCSDT calculations using the ROHF reference determinant for the
${\rm X\: ^3A}_{2}^{\prime}$ state and the RHF reference for the more strongly correlated
${\rm B\: ^1A_1}$ state, in spite of their
SR character, are capable of producing a reasonably accurate description of the adiabatic
${\rm B\: ^1A_1}-{\rm X\: ^3A}_{2}^{\prime}$ separation in trimethylenemethane. Indeed, the purely
electronic ${\rm B\: ^1A_1}-{\rm X\: ^3A}_{2}^{\prime}$ gap, designated, in analogy to other
singlet--triplet gaps considered in this work, as $\Delta E_{\rm S\mbox{-}T}$, resulting from the
ROHF/RHF-based CCSDT/cc-pVDZ computations using the SF-DFT/6-31G(d) geometries of the ${\rm X\: ^3A}_{2}^{\prime}$
and ${\rm B\: ^1A_1}$ states optimized in Ref.\ \onlinecite{ch2_krylov} is 21.7 kcal/mol\cite{jspp-jctc2012}
(cf. Table \ref{table:table11}). The corresponding experimentally derived result, obtained by subtracting
the zero-point vibrational energy correction $\Delta$ZPVE resulting from the SF-DFT/6-31G(d) calculations
reported in Ref.\ \onlinecite{ch2_krylov} from the experimental ${\rm B\: ^1A_1}-{\rm X\: ^3A}_{2}^{\prime}$ gap
determined in Refs.\ \onlinecite{TMM-Expt-1,TMM-Expt-2}, is 18.1 kcal/mol. The CCSDT/cc-pVDZ value of
$\Delta E_{\rm S\mbox{-}T}$ is not as accurate as the electronic ${\rm B\: ^1A_1}-{\rm X\: ^3A}_{2}^{\prime}$ gaps
generated in the high-level DEA- and DIP-EOMCC calculations with the explicit inclusion of
4p2h and 4h2p
correlations on top of CCSD, which produce 18--19 kcal/mol,
\cite{jspp-dea-dip-2013,jspp-dea-dip-2014,aj-js-pp-jpca2017} but it is certainly much better than
46.1, 24.4, and 29.8 kcal/mol obtained with the ROHF/RHF-based CCSD, CCSD(T), and CR-CC(2,3) methods,
respectively, when the cc-pVDZ basis set is employed
\cite{jspp-jctc2012} [as demonstrated in Ref.\ \onlinecite{jspp-jctc2012}, the use of a larger cc-pVTZ basis
makes the CCSD, CCSD(T), and CR-CC(2,3) results even worse; the CCSD/cc-pVDZ and CR-CC(2,3)/cc-pVDZ
values of $\Delta E_{\rm S\mbox{-}T}$ are included in Table \ref{table:table11} as the $\tau = 0$ CC($P$) and
CC($P$;$Q$) data, respectively]. While much of the
3.6 kcal/mol difference between the electronic ${\rm B\: ^1A_1}-{\rm X\: ^3A}_{2}^{\prime}$ separation
obtained in the ROHF/RHF-based CCSDT/cc-pVDZ calculations and its experimentally derived estimate
of 18.1 kcal/mol determined in Ref.\ \onlinecite{ch2_krylov} is, most likely, a consequence of the neglect
of $T_{4}$ clusters in the CCSDT approach, we should keep in mind that the latter estimate depends on the source
of the information about the $\Delta$ZPVE correction. For example, if one replaces the $\Delta$ZPVE value
obtained in the SF-DFT/6-31G(d) calculations reported in Ref.\ \onlinecite{ch2_krylov} by its
CCSD(T)/6-311++G(2d,2p) estimate and accounts for the core polarization effects determined
with the help of the CCSD(T)/cc-pCVQZ computations, combining the resulting information with the
experimental ${\rm B\: ^1A_1}-{\rm X\: ^3A}_{2}^{\prime}$ separation
determined in Refs.\ \onlinecite{TMM-Expt-1,TMM-Expt-2},
the purely electronic, experimentally derived, adiabatic $\Delta E_{\rm S\mbox{-}T}$ gap increases to
19.4 kcal/mol,\cite{TMM-EXTRA-1} which differs from our CCSDT/cc-pVDZ result by 2.3 kcal/mol.\cite{}
On the other hand, as shown in Ref.\ \onlinecite{TMM-EXTRA-1}, the CCSDT value of the adiabatic
${\rm B\: ^1A_1}-{\rm X\: ^3A}_{2}^{\prime}$ gap increases with the basis set too, to 23.1 kcal/mol when the
cc-pVTZ basis is employed, which reinforces our view that without accounting for $T_{4}$ correlations one cannot
bring the results of conventional SRCC computations to a close agreement with the experimentally derived data.
While the examination of the role of $T_{4}$ clusters, basis set, geometries of the ${\rm X\: ^3A}_{2}^{\prime}$
and ${\rm B\: ^1A_1}$ states employed in the calculations, $\Delta$ZPVE corrections, etc. would certainly be
interesting, it would also be outside the scope of the present study. Thus, in the remainder of this
subsection, we return to the analysis of the performance of the CIQMC-driven CC($P$;$Q$) approach
and its CC($P$) counterpart, especially their ability to converge the parent CCSDT energetics when the
cc-pVDZ basis is employed.

\begin{figure*}[!htbp]
\centering

\includegraphics[width=1.0\textwidth]{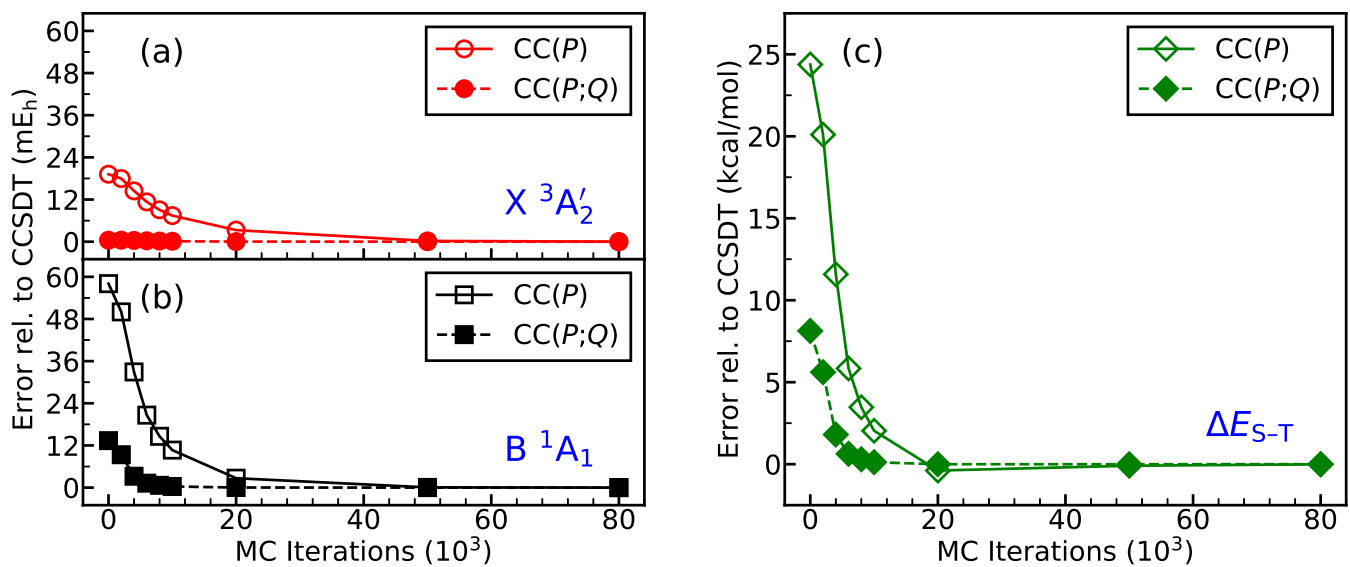}
\caption{\label{fig:figure5}
Convergence of the CC($P$) and CC($P$;$Q$) energies of the ${\rm X \: ^3A^\prime_2}$ [panel (a)]
and ${\rm B \: ^1A_1}$ [panel (b)] states
of trimethylenemethane, as described by the cc-pVDZ basis set,
and of the corresponding adiabatic singlet--triplet gaps [panel (c)] toward their parent CCSDT values.
The geometries of the ${\rm X \: ^3A^\prime_2}$ and ${\rm B \: ^1A_1}$ states, optimized in the
SF-DFT/6-31G(d) calculations, were taken from Ref.\ \onlinecite{ch2_krylov}.
The $P$ spaces consisted of all singles and doubles and subsets
of triples identified during the $i$-CISDTQ-MC propagations with $\delta \tau=0.0001$ a.u.\ and the $Q$
spaces consisted of the triples not captured by $i$-CISDTQ-MC.}
\end{figure*}

The results of our semi-stochastic CC($P$)/cc-pVDZ and CC($P$;$Q$)/cc-pVDZ computations for the
${\rm X\: ^3A}_{2}^{\prime}$ and ${\rm B\: ^1A_1}$ states of trimethylenemethane and the adiabatic gap between
them, along with the associated CCSD, CR-CC(2,3), and CCSDT data, are summarized in Table \ref{table:table11}
and Fig.\ \ref{fig:figure5}. As in the case of cyclopentadienyl cation, to reduce the computational
costs of the underlying CIQMC propagations, especially in their later stages, we resorted to the
truncated $i$-CISDTQ-MC approach. In analogy to cyclobutadiene and cyclopentadienyl cation, we terminated
our $i$-CISDTQ-MC propagations after 80000 $\delta \tau = 0.0001$ a.u. MC time steps,
where the differences between the CC($P$;$Q$) and CCSDT energies of the
${\rm X\: ^3A}_{2}^{\prime}$ and ${\rm B\: ^1A_1}$ states fall
below 1 microhartree.
Consistent with the
CC($P$), CC($P$;$Q$), and other SRCC calculations for trimethylenemethane
reported in Table \ref{table:table11} and Fig.\ \ref{fig:figure5}, we used the ROHF
determinant to initiate the $i$-CISDTQ-MC propagation for the $D_{3h}$-symmetric ${\rm X\: ^3A}_{2}^{\prime}$
(in $C_{2v}$, ${\rm X\: {^{3}}B}_{2}$) state and the RHF determinant to initiate the $i$-CISDTQ-MC run for the
$C_{2v}$-symmetric ${\rm B\: ^1A_1}$ state.
The lists of triply excited determinants captured by the $i$-CISDTQ-MC runs at the various times $\tau > 0$,
needed to construct the $P$ spaces for the CC($P$) and CC($P$;$Q$) computations, were the $S_z=1$ triples
of the ${\rm B_2}(C_{2v})$ symmetry in the case of the ${\rm X\: ^3A}_{2}^{\prime}$ state and the
$S_z=0$ triples of the ${\rm A_1}(C_{2v})$ symmetry when considering the ${\rm B\: ^1A_1}$ state. 
The remaining triples not captured by $i$-CISDTQ-MC defined the corresponding $Q$ spaces.

It is clear from the results presented in Table \ref{table:table11} and Fig.\ \ref{fig:figure5}
that the semi-stochastic CC($P$;$Q$) approach is very effective in converging the parent CCSDT
energetics characterizing the ${\rm X\: ^3A}_{2}^{\prime}$ and ${\rm B\: ^1A_1}$ states of
trimethylenemethane and the adiabatic gap between them. It offers substantial improvements in the results
of the CR-CC(2,3) calculations in the early stages of the underlying $i$-CISDTQ-MC propagations, especially
when the multi-configurational ${\rm B\: ^1A_1}$ state and the adiabatic ${\rm B\: ^1A_1}-{\rm X\: ^3A}_{2}^{\prime}$
separation $\Delta E_{\rm S\mbox{-}T}$, which are poorly described by CR-CC(2,3), are examined, while greatly
accelerating the convergence of the CC($P$) energies toward CCSDT. Indeed, after 6000 $\delta \tau = 0.0001$ a.u.
MC iterations, which is a very short propagation time engaging only $\sim$0.1\% of the total walker populations
at $\tau = 8.0$ a.u., where we terminated our $i$-CISDTQ-MC runs
(cf. Table S.5 of the supplementary material),
and $i$-CISDTQ-MC
capturing as little as 14--17\% of all triply excited determinants, the semi-stochastic CC($P$;$Q$) methodology
reduces the 13.370 millihartree difference between the CR-CC(2,3) and CCSDT energies of the ${\rm B\: ^1A_1}$ state
and the 8.1 kcal/mol error in the CR-CC(2,3) value of the ${\rm B\: ^1A_1}-{\rm X\: ^3A}_{2}^{\prime}$ gap
relative to CCSDT to 1.260 millihartree and 0.6 kcal/mol, respectively, which is a chemical accuracy regime.
Interestingly, the $i$-CISDTQ-MC-based CC($P$;$Q$) value of $\Delta E_{\rm S\mbox{-}T}$ obtained after 6000
MC iterations matches the quality of the ${\rm B\: ^1A_1}-{\rm X\: ^3A}_{2}^{\prime}$ gap resulting from
the fully deterministic CC($P$;$Q$) calculations using the CC(t;3) approach, which give a 0.5 kcal/mol error
relative to CCSDT when the cc-pVDZ basis set is employed.\cite{jspp-jctc2012}
After the additional 4000 MC time steps, where the $i$-CISDTQ-MC propagations for the ${\rm X\: ^3A}_{2}^{\prime}$
and ${\rm B\: ^1A_1}$ states are still very far from convergence and where the fractions of triples captured by
$i$-CISDTQ-MC increase to about 30\%, the small errors relative to CCSDT characterizing the $i$-CISDTQ-MC-based
CC($P$;$Q$) values of the energy of the ${\rm B\: ^1A_1}$ state and $\Delta E_{\rm S\mbox{-}T}$ at $\tau = 0.6$ a.u.
drop down by factors of 4--6, to 0.314 millihartree and 0.1 kcal/mol, respectively, illustrating how rapid the
convergence of the CIQMC-driven CC($P$;$Q$) calculations toward the parent SRCC data can be. While the
CR-CC(2,3) description of the ${\rm X\: ^3A}_{2}^{\prime}$ state, which has a SR character, is much better
than in the case of its strongly correlated ${\rm B\: ^1A_1}$ counterpart, the semi-stochastic CC($P$;$Q$)
computations offer great improvements in this case too. They are, for example, capable of reducing the
$\sim$0.4 millihartree difference between the CR-CC(2,3) and CCSDT energies to a 0.1 millihartree level after
10000 MC iterations and $i$-CISDTQ-MC capturing less than 30\% of all triples. In analogy to all other molecular
examples considered in this article, the uncorrected CC($P$) values of the energies of the
${\rm X\: ^3A}_{2}^{\prime}$ and ${\rm B\: ^1A_1}$ states and separation between them converge to their
CCSDT limits too, but it is clear from Table \ref{table:table11} and Fig.\ \ref{fig:figure5} that
they do it at a much slower rate than their CC($P$;$Q$) counterparts. This can be illustrated by comparing
the errors relative to CCSDT characterizing the CC($P$) and CC($P$;$Q$) energies of the ${\rm X\: ^3A}_{2}^{\prime}$
and ${\rm B\: ^1A_1}$ states and separation between them obtained after 6000 MC iterations. They are
more than 11 millihartree, about 21 millihartree, and almost 6 kcal/mol, respectively, in the former case
and only 0.253 millihartree, 1.260 millihartree, and 0.6 kcal/mol, when the CC($P$) energies are corrected
for the remaining $T_{3}$ correlations using the CC($P$;$Q$) approach. As explained in Section \ref{sec:theory},
the CC($P$) energies converge to CCSDT more slowly than their $\delta(P;Q)$-corrected CC($P$;$Q$) counterparts,
since the initial, $\tau = 0$, CC($P$) calculation for a given electronic state is equivalent to CCSD, where
$T_{3} = 0$. The CIQMC-driven CC($P$;$Q$) calculations start from CR-CC(2,3), which provides information
about $T_{3}$ clusters via noniterative corrections to CCSD. This once again emphasizes the benefits
of using corrections $\delta (P;Q)$ in the context of the semi-stochastic CC($P$;$Q$) work.

\section{Conclusions}
\label{sec:conclusions}

One of the most promising ideas in the area of converging high-level SRCC energetics without having to
resort to the very expensive methods such as CCSDT or CCSDTQ has been
the CC($P$;$Q$) formalism, originally introduced
in Refs.\ \onlinecite{jspp-chemphys2012,jspp-jcp2012,jspp-jctc2012}, in which one solves the CC
amplitude equations in a suitably defined subspace of the many-electron Hilbert space, called the $P$ space,
and then improves the resulting CC($P$) energies using the {\it a posteriori} corrections $\delta(P;Q)$
determined with the help of another subspace of the Hilbert space,
referred to as the $Q$ space.
In addition to conventional choices of the $P$ and $Q$ spaces, which result in methods
such as CR-CC(2,3),\cite{crccl_jcp,crccl_cpl,crccl_molphys,crccl_jpc,crccl_ijqc}
one can consider various unconventional ways of setting up these spaces
in the CC($P$;$Q$) calculations.\cite{jspp-chemphys2012,jspp-jcp2012,jspp-jctc2012,%
nbjspp-molphys2017,ccpq-be2-jpca-2018,ccpq-mg2-mp-2019,stochastic-ccpq-prl-2017,stochastic-ccpq-molphys-2020,%
jed-js-pp-jcp-2021,cipsi-ccpq-2021} In this work, we have focused on the semi-stochastic
formulation of the CC($P$;$Q$) formalism, introduced in Ref.\ \onlinecite{stochastic-ccpq-prl-2017} and
further developed in Refs.\ \onlinecite{stochastic-ccpq-molphys-2020,jed-js-pp-jcp-2021}, in which the
leading higher--than--doubly excited determinants incorporated in the $P$ space, needed to solve the CC($P$)
equations, are identified, in an automated fashion, by the stochastic wave function propagations employing
the CIQMC methodology of Refs. \onlinecite{Booth2009,Cleland2010,fciqmc-uga-2019,ghanem_alavi_fciqmc_jcp_2019,%
ghanem_alavi_fciqmc_2020}.

The specific objective of this study has been the investigation of the effectiveness of the semi-stochastic
CC($P$;$Q$) approaches driven by $i$-FCIQMC and $i$-CISDTQ-MC propagations, which were used to capture
the leading triply excited determinants for the inclusion in the respective $P$ spaces, in converging the
singlet--triplet gaps and the underlying total electronic energies resulting from the high-level CCSDT
computations. Molecular systems that have been used to examine the ability of the semi-stochastic CC($P$;$Q$)
methodology to recover the CCSDT energetics of the low-lying singlet and triplet states included the
methylene, cyclobutadiene, cyclopentadienyl cation, and trimethylenemethane biradicals and a prototype
magnetic system represented by the ${\rm (HFH)^-}$ ion. Cyclopentadienyl cation and trimethylenemethane
are the largest polyatomic species that have been used in the semi-stochastic CC($P$;$Q$) calculations reported
to date. 

An accurate determination of the singlet--triplet gaps
in the above five systems requires a well-balanced treatment of substantial nondynamical correlation effects,
needed for a reliable description of the low-spin singlet states that have a manifestly MR character, and
dynamical correlations of the high-spin triplet states, which are SR in nature. Within the conventional,
particle-conserving, black-box SRCC framework, the only methods that can do this in a robust manner are the
high-level CC approaches with a full treatment of $T_{n}$ correlations with $n > 2$, beginning with CCSDT.
The basic CCSD approximation, which ignores the $T_{n}$ cluster
components with $n > 2$ altogether, and the noniterative $T_{3}$ corrections to CCSD, including even CR-CC(2,3),
which can handle selected MR situations, such as single bond breaking, fail. Putting aside the quasi-perturbative
character of the majority of the existing triples corrections to CCSD, a significant part of the problem is their
inability to capture the coupling between the low-rank $T_{1}$ and $T_{2}$ clusters with their higher-rank $T_{3}$
counterpart, which cannot be ignored when the low-lying singlet states of biradicals are considered. We have
shown in this study that by relaxing $T_{1}$ and $T_{2}$ amplitudes in the presence of the $T_{3}$ component
defined using the list of the leading triply excited determinants identified by the CIQMC propagations and using
the $\delta(P;Q)$ corrections to account for the remaining $T_{3}$ correlations that are not described by
the underlying CC($P$) calculations,
the semi-stochastic CC($P$;$Q$) methodology provides an efficient mechanism
for incorporating the coupling
among the $T_{1}$, $T_{2}$, and $T_{3}$ clusters relevant to a reliable description of the singlet--triplet
gaps in biradicals, while capturing the vast majority of the many-electron correlation effects included in the
parent CCSDT computations. In particular, we have demonstrated that the semi-stochastic CC($P$;$Q$) calculations
are capable of reaching millihartree or submillihartree accuracy levels relative to the parent CCSDT results,
including total electronic energies of the lowest singlet and triplet states and adiabatic as well as vertical
gaps between them, with small (typically about 20--30\%; sometimes even less) fractions of triply excited determinants
captured in the early stages of the underlying CIQMC propagations and with tiny walker populations that are orders
of magnitude smaller than the total numbers of walkers required to converge them. We have also demonstrated
the vital role of the noniterative corrections $\delta(P;Q)$ in accelerating and, in the case of the
singlet--triplet gap values, smoothing convergence of the corresponding CC($P$) energetics toward CCSDT.

The numerical results and analyses reported in this article encourage us to pursue the semi-stochastic
CC($P$;$Q$) methodology even further. Putting aside the need for improving the computational efficiency
of our CIQMC-driven CC($P$;$Q$) codes, so that we could consider larger molecular problems than those
considered in this study, it would be interesting to examine if our semi-stochastic CC($P$;$Q$) computations,
including those discussed in this work, could take advantage of the recent advances in CIQMC, such as the
adaptive CIQMC algorithm of Refs.\ \onlinecite{ghanem_alavi_fciqmc_jcp_2019,ghanem_alavi_fciqmc_2020} that
might replace the $i$-CIQMC approaches that we have used so far. As implied by the remarks made in
Section \ref{sec:results}, some of the biradicals considered in this work might benefit from an accurate
description of $T_{3}$ as well as $T_{4}$ correlations. It would, therefore, be interesting to
examine if one could improve the results presented in this work by using the semi-stochastic CC($P$;$Q$)
approaches that aim at converging the CCSDTQ energetics, such as those discussed in Ref.\
\onlinecite{jed-js-pp-jcp-2021}. Last, but not least, we have started developing extensions of
the semi-stochastic CC($P$) and CC($P$;$Q$) approaches, pursued in Refs.\
\onlinecite{stochastic-ccpq-prl-2017,jed-js-pp-jcp-2021} and this study, and their EOMCC analogs
\cite{eomccp-jcp-2019,stochastic-ccpq-molphys-2020} to the particle-nonconserving EOMCC models, especially
EA/IP-EOMCC and DEA/DIP-EOMCC. As pointed out in this article, and as shown in Refs.\
\onlinecite{jspp-dea-dip-2013,jspp-dea-dip-2014,stoneburner-js-jcp2017,aj-js-pp-jpca2017,js-pp-dea2021},
the deterministic DEA/DIP-EOMCC approaches with the explicit inclusion of
4p2h and 4h2p
components of the corresponding electron attaching and ionizing operators on top of CCSD can provide a nearly exact
description of the singlet--triplet in biradical species, so it will be interesting if their semi-stochastic
counterparts can do the same.

\section*{Supplementary Material}
See the supplementary material for the information about the total numbers of walkers characterizing the
$i$-FCIQMC [methylene, ${\rm (HFH)^-}$, cyclobutadiene] and $i$-CISDTQ-MC (cyclopentadienyl cation,
trimethylenemethane) propagations carried out in this study.

\begin{acknowledgments}
This work has been supported by the Chemical Sciences, Geosciences and
Biosciences Division, Office of Basic Energy Sciences, Office of Science, U.S.
Department of Energy (Grant No. DE-FG02-01ER15228 to P.P.) and the National
Science Foundation (Grant No. CHE-1763371 to P.P.). The computational resources
provided by the Institute for Cyber-Enabled Research at Michigan State University are
gratefully acknowledged too.
\end{acknowledgments}

\section*{Data Availability}
The data that support the findings of this study are available within the article
and its supplementary material.


\begin{thebibliography}{203}%
\makeatletter
\providecommand \@ifxundefined [1]{%
 \@ifx{#1\undefined}
}%
\providecommand \@ifnum [1]{%
 \ifnum #1\expandafter \@firstoftwo
 \else \expandafter \@secondoftwo
 \fi
}%
\providecommand \@ifx [1]{%
 \ifx #1\expandafter \@firstoftwo
 \else \expandafter \@secondoftwo
 \fi
}%
\providecommand \natexlab [1]{#1}%
\providecommand \enquote  [1]{``#1''}%
\providecommand \bibnamefont  [1]{#1}%
\providecommand \bibfnamefont [1]{#1}%
\providecommand \citenamefont [1]{#1}%
\providecommand \href@noop [0]{\@secondoftwo}%
\providecommand \href [0]{\begingroup \@sanitize@url \@href}%
\providecommand \@href[1]{\@@startlink{#1}\@@href}%
\providecommand \@@href[1]{\endgroup#1\@@endlink}%
\providecommand \@sanitize@url [0]{\catcode `\\12\catcode `\$12\catcode
  `\&12\catcode `\#12\catcode `\^12\catcode `\_12\catcode `\%12\relax}%
\providecommand \@@startlink[1]{}%
\providecommand \@@endlink[0]{}%
\providecommand \url  [0]{\begingroup\@sanitize@url \@url }%
\providecommand \@url [1]{\endgroup\@href {#1}{\urlprefix }}%
\providecommand \urlprefix  [0]{URL }%
\providecommand \Eprint [0]{\href }%
\providecommand \doibase [0]{http://dx.doi.org/}%
\providecommand \selectlanguage [0]{\@gobble}%
\providecommand \bibinfo  [0]{\@secondoftwo}%
\providecommand \bibfield  [0]{\@secondoftwo}%
\providecommand \translation [1]{[#1]}%
\providecommand \BibitemOpen [0]{}%
\providecommand \bibitemStop [0]{}%
\providecommand \bibitemNoStop [0]{.\EOS\space}%
\providecommand \EOS [0]{\spacefactor3000\relax}%
\providecommand \BibitemShut  [1]{\csname bibitem#1\endcsname}%
\let\auto@bib@innerbib\@empty
\bibitem [{\citenamefont {Roos}(1987)}]{Roos1987}%
  \BibitemOpen
  \bibfield  {author} {\bibinfo {author} {\bibfnamefont {B.~O.}\ \bibnamefont
  {Roos}},\ }\href {\doibase 10.1002/9780470142943.ch7} {\bibfield  {journal}
  {\bibinfo  {journal} {Adv. Chem. Phys.}\ }\textbf {\bibinfo {volume} {69}},\
  \bibinfo {pages} {399} (\bibinfo {year} {1987})}\BibitemShut {NoStop}%
\bibitem [{\citenamefont {Schmidt}\ and\ \citenamefont
  {Gordon}(1998)}]{ref:mrmpreview}%
  \BibitemOpen
  \bibfield  {author} {\bibinfo {author} {\bibfnamefont {M.~W.}\ \bibnamefont
  {Schmidt}}\ and\ \bibinfo {author} {\bibfnamefont {M.~S.}\ \bibnamefont
  {Gordon}},\ }\href@noop {} {\bibfield  {journal} {\bibinfo  {journal} {Annu.
  Rev. Phys. Chem.}\ }\textbf {\bibinfo {volume} {49}},\ \bibinfo {pages} {233}
  (\bibinfo {year} {1998})}\BibitemShut {NoStop}%
\bibitem [{\citenamefont {Szalay}\ \emph {et~al.}(2012)\citenamefont {Szalay},
  \citenamefont {M{\" u}ller}, \citenamefont {Gidofalvi}, \citenamefont
  {Lischka},\ and\ \citenamefont {Shepard}}]{chemrev-2012a}%
  \BibitemOpen
  \bibfield  {author} {\bibinfo {author} {\bibfnamefont {P.~G.}\ \bibnamefont
  {Szalay}}, \bibinfo {author} {\bibfnamefont {T.}~\bibnamefont {M{\" u}ller}},
  \bibinfo {author} {\bibfnamefont {G.}~\bibnamefont {Gidofalvi}}, \bibinfo
  {author} {\bibfnamefont {H.}~\bibnamefont {Lischka}}, \ and\ \bibinfo
  {author} {\bibfnamefont {R.}~\bibnamefont {Shepard}},\ }\href@noop {}
  {\bibfield  {journal} {\bibinfo  {journal} {Chem. Rev.}\ }\textbf {\bibinfo
  {volume} {112}},\ \bibinfo {pages} {108} (\bibinfo {year}
  {2012})}\BibitemShut {NoStop}%
\bibitem [{\citenamefont {Roca-Sanju{\' a}n}, \citenamefont {Aquilante},\ and\
  \citenamefont {Lindh}(2012)}]{lindh-review-2012}%
  \BibitemOpen
  \bibfield  {author} {\bibinfo {author} {\bibfnamefont {D.}~\bibnamefont
  {Roca-Sanju{\' a}n}}, \bibinfo {author} {\bibfnamefont {F.}~\bibnamefont
  {Aquilante}}, \ and\ \bibinfo {author} {\bibfnamefont {R.}~\bibnamefont
  {Lindh}},\ }\href {\doibase https://doi.org/10.1002/wcms.97} {\bibfield
  {journal} {\bibinfo  {journal} {WIREs Comput. Mol. Sci.}\ }\textbf {\bibinfo
  {volume} {2}},\ \bibinfo {pages} {585} (\bibinfo {year} {2012})}\BibitemShut
  {NoStop}%
\bibitem [{\citenamefont {Chattopadhyay}\ \emph {et~al.}(2016)\citenamefont
  {Chattopadhyay}, \citenamefont {Chaudhuri}, \citenamefont {Mahapatra},
  \citenamefont {Ghosh},\ and\ \citenamefont {Ray}}]{sinha-review-2016}%
  \BibitemOpen
  \bibfield  {author} {\bibinfo {author} {\bibfnamefont {S.}~\bibnamefont
  {Chattopadhyay}}, \bibinfo {author} {\bibfnamefont {R.~K.}\ \bibnamefont
  {Chaudhuri}}, \bibinfo {author} {\bibfnamefont {U.~S.}\ \bibnamefont
  {Mahapatra}}, \bibinfo {author} {\bibfnamefont {A.}~\bibnamefont {Ghosh}}, \
  and\ \bibinfo {author} {\bibfnamefont {S.~S.}\ \bibnamefont {Ray}},\ }\href
  {\doibase https://doi.org/10.1002/wcms.1248} {\bibfield  {journal} {\bibinfo
  {journal} {WIREs Comput. Mol. Sci.}\ }\textbf {\bibinfo {volume} {6}},\
  \bibinfo {pages} {266} (\bibinfo {year} {2016})}\BibitemShut {NoStop}%
\bibitem [{\citenamefont {Lyakh}\ \emph {et~al.}(2012)\citenamefont {Lyakh},
  \citenamefont {Musia{\l}}, \citenamefont {Lotrich},\ and\ \citenamefont
  {Bartlett}}]{chemrev-2012b}%
  \BibitemOpen
  \bibfield  {author} {\bibinfo {author} {\bibfnamefont {D.~I.}\ \bibnamefont
  {Lyakh}}, \bibinfo {author} {\bibfnamefont {M.}~\bibnamefont {Musia{\l}}},
  \bibinfo {author} {\bibfnamefont {V.~F.}\ \bibnamefont {Lotrich}}, \ and\
  \bibinfo {author} {\bibfnamefont {R.~J.}\ \bibnamefont {Bartlett}},\
  }\href@noop {} {\bibfield  {journal} {\bibinfo  {journal} {Chem. Rev.}\
  }\textbf {\bibinfo {volume} {112}},\ \bibinfo {pages} {182} (\bibinfo {year}
  {2012})}\BibitemShut {NoStop}%
\bibitem [{\citenamefont {Piecuch}\ and\ \citenamefont
  {Kowalski}(2002)}]{succ5}%
  \BibitemOpen
  \bibfield  {author} {\bibinfo {author} {\bibfnamefont {P.}~\bibnamefont
  {Piecuch}}\ and\ \bibinfo {author} {\bibfnamefont {K.}~\bibnamefont
  {Kowalski}},\ }\href@noop {} {\bibfield  {journal} {\bibinfo  {journal} {Int.
  J. Mol. Sci.}\ }\textbf {\bibinfo {volume} {3}},\ \bibinfo {pages} {676}
  (\bibinfo {year} {2002})}\BibitemShut {NoStop}%
\bibitem [{\citenamefont
  {Evangelista}(2018)}]{evangelista-perspective-jcp-2018}%
  \BibitemOpen
  \bibfield  {author} {\bibinfo {author} {\bibfnamefont {F.~A.}\ \bibnamefont
  {Evangelista}},\ }\href@noop {} {\bibfield  {journal} {\bibinfo  {journal}
  {J. Chem. Phys.}\ }\textbf {\bibinfo {volume} {149}},\ \bibinfo {pages}
  {030901} (\bibinfo {year} {2018})}\BibitemShut {NoStop}%
\bibitem [{\citenamefont {Coester}(1958)}]{Coester:1958}%
  \BibitemOpen
  \bibfield  {author} {\bibinfo {author} {\bibfnamefont {F.}~\bibnamefont
  {Coester}},\ }\href@noop {} {\bibfield  {journal} {\bibinfo  {journal} {Nucl.
  Phys.}\ }\textbf {\bibinfo {volume} {7}},\ \bibinfo {pages} {421} (\bibinfo
  {year} {1958})}\BibitemShut {NoStop}%
\bibitem [{\citenamefont {Coester}\ and\ \citenamefont {K{\"
  u}mmel}(1960)}]{Coester:1960}%
  \BibitemOpen
  \bibfield  {author} {\bibinfo {author} {\bibfnamefont {F.}~\bibnamefont
  {Coester}}\ and\ \bibinfo {author} {\bibfnamefont {H.}~\bibnamefont {K{\"
  u}mmel}},\ }\href@noop {} {\bibfield  {journal} {\bibinfo  {journal} {Nucl.
  Phys.}\ }\textbf {\bibinfo {volume} {17}},\ \bibinfo {pages} {477} (\bibinfo
  {year} {1960})}\BibitemShut {NoStop}%
\bibitem [{\citenamefont {{\v C}{\'\i}{\v z}ek}(1966)}]{cizek1}%
  \BibitemOpen
  \bibfield  {author} {\bibinfo {author} {\bibfnamefont {J.}~\bibnamefont {{\v
  C}{\'\i}{\v z}ek}},\ }\href@noop {} {\bibfield  {journal} {\bibinfo
  {journal} {J. Chem. Phys.}\ }\textbf {\bibinfo {volume} {45}},\ \bibinfo
  {pages} {4256} (\bibinfo {year} {1966})}\BibitemShut {NoStop}%
\bibitem [{\citenamefont {{\v C}{\'\i}{\v z}ek}(1969)}]{cizek2}%
  \BibitemOpen
  \bibfield  {author} {\bibinfo {author} {\bibfnamefont {J.}~\bibnamefont {{\v
  C}{\'\i}{\v z}ek}},\ }\href@noop {} {\bibfield  {journal} {\bibinfo
  {journal} {Adv. Chem. Phys.}\ }\textbf {\bibinfo {volume} {14}},\ \bibinfo
  {pages} {35} (\bibinfo {year} {1969})}\BibitemShut {NoStop}%
\bibitem [{\citenamefont {Paldus}, \citenamefont {{\v C}{\'\i}{\v z}ek},\ and\
  \citenamefont {Shavitt}(1972)}]{cizek4}%
  \BibitemOpen
  \bibfield  {author} {\bibinfo {author} {\bibfnamefont {J.}~\bibnamefont
  {Paldus}}, \bibinfo {author} {\bibfnamefont {J.}~\bibnamefont {{\v
  C}{\'\i}{\v z}ek}}, \ and\ \bibinfo {author} {\bibfnamefont {I.}~\bibnamefont
  {Shavitt}},\ }\href@noop {} {\bibfield  {journal} {\bibinfo  {journal} {Phys.
  Rev. A}\ }\textbf {\bibinfo {volume} {5}},\ \bibinfo {pages} {50} (\bibinfo
  {year} {1972})}\BibitemShut {NoStop}%
\bibitem [{\citenamefont {Paldus}\ and\ \citenamefont {Li}(1999)}]{paldus-li}%
  \BibitemOpen
  \bibfield  {author} {\bibinfo {author} {\bibfnamefont {J.}~\bibnamefont
  {Paldus}}\ and\ \bibinfo {author} {\bibfnamefont {X.}~\bibnamefont {Li}},\
  }\href@noop {} {\bibfield  {journal} {\bibinfo  {journal} {Adv. Chem. Phys.}\
  }\textbf {\bibinfo {volume} {110}},\ \bibinfo {pages} {1} (\bibinfo {year}
  {1999})}\BibitemShut {NoStop}%
\bibitem [{\citenamefont {Bartlett}\ and\ \citenamefont
  {Musia{\l}}(2007)}]{bartlett-musial2007}%
  \BibitemOpen
  \bibfield  {author} {\bibinfo {author} {\bibfnamefont {R.~J.}\ \bibnamefont
  {Bartlett}}\ and\ \bibinfo {author} {\bibfnamefont {M.}~\bibnamefont
  {Musia{\l}}},\ }\href@noop {} {\bibfield  {journal} {\bibinfo  {journal}
  {Rev. Mod. Phys.}\ }\textbf {\bibinfo {volume} {79}},\ \bibinfo {pages} {291}
  (\bibinfo {year} {2007})}\BibitemShut {NoStop}%
\bibitem [{\citenamefont {Purvis}\ and\ \citenamefont {Bartlett}(1982)}]{ccsd}%
  \BibitemOpen
  \bibfield  {author} {\bibinfo {author} {\bibfnamefont {G.~D.}\ \bibnamefont
  {Purvis}, \bibfnamefont {III}}\ and\ \bibinfo {author} {\bibfnamefont
  {R.~J.}\ \bibnamefont {Bartlett}},\ }\href@noop {} {\bibfield  {journal}
  {\bibinfo  {journal} {J. Chem. Phys.}\ }\textbf {\bibinfo {volume} {76}},\
  \bibinfo {pages} {1910} (\bibinfo {year} {1982})}\BibitemShut {NoStop}%
\bibitem [{\citenamefont {Cullen}\ and\ \citenamefont {Zerner}(1982)}]{ccsd2}%
  \BibitemOpen
  \bibfield  {author} {\bibinfo {author} {\bibfnamefont {J.~M.}\ \bibnamefont
  {Cullen}}\ and\ \bibinfo {author} {\bibfnamefont {M.~C.}\ \bibnamefont
  {Zerner}},\ }\href@noop {} {\bibfield  {journal} {\bibinfo  {journal} {J.
  Chem. Phys.}\ }\textbf {\bibinfo {volume} {77}},\ \bibinfo {pages} {4088}
  (\bibinfo {year} {1982})}\BibitemShut {NoStop}%
\bibitem [{\citenamefont {Scuseria}\ \emph {et~al.}(1987)\citenamefont
  {Scuseria}, \citenamefont {Scheiner}, \citenamefont {Lee}, \citenamefont
  {Rice},\ and\ \citenamefont {Schaefer}}]{ccsdfritz}%
  \BibitemOpen
  \bibfield  {author} {\bibinfo {author} {\bibfnamefont {G.~E.}\ \bibnamefont
  {Scuseria}}, \bibinfo {author} {\bibfnamefont {A.~C.}\ \bibnamefont
  {Scheiner}}, \bibinfo {author} {\bibfnamefont {T.~J.}\ \bibnamefont {Lee}},
  \bibinfo {author} {\bibfnamefont {J.~E.}\ \bibnamefont {Rice}}, \ and\
  \bibinfo {author} {\bibfnamefont {H.~F.}\ \bibnamefont {Schaefer},
  \bibfnamefont {III}},\ }\href@noop {} {\bibfield  {journal} {\bibinfo
  {journal} {J. Chem. Phys.}\ }\textbf {\bibinfo {volume} {86}},\ \bibinfo
  {pages} {2881} (\bibinfo {year} {1987})}\BibitemShut {NoStop}%
\bibitem [{\citenamefont {Piecuch}\ and\ \citenamefont
  {Paldus}(1989)}]{osaccsd}%
  \BibitemOpen
  \bibfield  {author} {\bibinfo {author} {\bibfnamefont {P.}~\bibnamefont
  {Piecuch}}\ and\ \bibinfo {author} {\bibfnamefont {J.}~\bibnamefont
  {Paldus}},\ }\href@noop {} {\bibfield  {journal} {\bibinfo  {journal} {Int.
  J. Quantum Chem.}\ }\textbf {\bibinfo {volume} {36}},\ \bibinfo {pages} {429}
  (\bibinfo {year} {1989})}\BibitemShut {NoStop}%
\bibitem [{\citenamefont {Hoffmann}\ and\ \citenamefont
  {Schaefer}(1986)}]{ccsdt-hoffmann}%
  \BibitemOpen
  \bibfield  {author} {\bibinfo {author} {\bibfnamefont {M.~R.}\ \bibnamefont
  {Hoffmann}}\ and\ \bibinfo {author} {\bibfnamefont {H.~F.}\ \bibnamefont
  {Schaefer}, \bibfnamefont {III}},\ }\href@noop {} {\bibfield  {journal}
  {\bibinfo  {journal} {Adv. Quantum Chem.}\ }\textbf {\bibinfo {volume}
  {18}},\ \bibinfo {pages} {207} (\bibinfo {year} {1986})}\BibitemShut
  {NoStop}%
\bibitem [{\citenamefont {Noga}\ and\ \citenamefont
  {Bartlett}(1987)}]{ccfullt}%
  \BibitemOpen
  \bibfield  {author} {\bibinfo {author} {\bibfnamefont {J.}~\bibnamefont
  {Noga}}\ and\ \bibinfo {author} {\bibfnamefont {R.~J.}\ \bibnamefont
  {Bartlett}},\ }\href@noop {} {\bibfield  {journal} {\bibinfo  {journal} {J.
  Chem. Phys.}\ }\textbf {\bibinfo {volume} {86}},\ \bibinfo {pages} {7041}
  (\bibinfo {year} {1987})},\ \bibinfo {note} {{\bf 89}, 3401 (1988)
  [Erratum]}\BibitemShut {NoStop}%
\bibitem [{\citenamefont {Scuseria}\ and\ \citenamefont
  {Schaefer}(1988)}]{ccfullt2}%
  \BibitemOpen
  \bibfield  {author} {\bibinfo {author} {\bibfnamefont {G.~E.}\ \bibnamefont
  {Scuseria}}\ and\ \bibinfo {author} {\bibfnamefont {H.~F.}\ \bibnamefont
  {Schaefer}, \bibfnamefont {III}},\ }\href@noop {} {\bibfield  {journal}
  {\bibinfo  {journal} {Chem. Phys. Lett.}\ }\textbf {\bibinfo {volume}
  {152}},\ \bibinfo {pages} {382} (\bibinfo {year} {1988})}\BibitemShut
  {NoStop}%
\bibitem [{\citenamefont {Watts}\ and\ \citenamefont
  {Bartlett}(1990)}]{ccsdt-uhf}%
  \BibitemOpen
  \bibfield  {author} {\bibinfo {author} {\bibfnamefont {J.~D.}\ \bibnamefont
  {Watts}}\ and\ \bibinfo {author} {\bibfnamefont {R.~J.}\ \bibnamefont
  {Bartlett}},\ }\href@noop {} {\bibfield  {journal} {\bibinfo  {journal} {J.
  Chem. Phys.}\ }\textbf {\bibinfo {volume} {93}},\ \bibinfo {pages} {6104}
  (\bibinfo {year} {1990})}\BibitemShut {NoStop}%
\bibitem [{\citenamefont {Oliphant}\ and\ \citenamefont
  {Adamowicz}(1991{\natexlab{a}})}]{ccsdtq0}%
  \BibitemOpen
  \bibfield  {author} {\bibinfo {author} {\bibfnamefont {N.}~\bibnamefont
  {Oliphant}}\ and\ \bibinfo {author} {\bibfnamefont {L.}~\bibnamefont
  {Adamowicz}},\ }\href@noop {} {\bibfield  {journal} {\bibinfo  {journal} {J.
  Chem. Phys.}\ }\textbf {\bibinfo {volume} {95}},\ \bibinfo {pages} {6645}
  (\bibinfo {year} {1991}{\natexlab{a}})}\BibitemShut {NoStop}%
\bibitem [{\citenamefont {Kucharski}\ and\ \citenamefont
  {Bartlett}(1991)}]{ccsdtq1}%
  \BibitemOpen
  \bibfield  {author} {\bibinfo {author} {\bibfnamefont {S.~A.}\ \bibnamefont
  {Kucharski}}\ and\ \bibinfo {author} {\bibfnamefont {R.~J.}\ \bibnamefont
  {Bartlett}},\ }\href@noop {} {\bibfield  {journal} {\bibinfo  {journal}
  {Theor. Chim. Acta}\ }\textbf {\bibinfo {volume} {80}},\ \bibinfo {pages}
  {387} (\bibinfo {year} {1991})}\BibitemShut {NoStop}%
\bibitem [{\citenamefont {Kucharski}\ and\ \citenamefont
  {Bartlett}(1992)}]{ccsdtq2}%
  \BibitemOpen
  \bibfield  {author} {\bibinfo {author} {\bibfnamefont {S.~A.}\ \bibnamefont
  {Kucharski}}\ and\ \bibinfo {author} {\bibfnamefont {R.~J.}\ \bibnamefont
  {Bartlett}},\ }\href@noop {} {\bibfield  {journal} {\bibinfo  {journal} {J.
  Chem. Phys.}\ }\textbf {\bibinfo {volume} {97}},\ \bibinfo {pages} {4282}
  (\bibinfo {year} {1992})}\BibitemShut {NoStop}%
\bibitem [{\citenamefont {Piecuch}\ and\ \citenamefont
  {Adamowicz}(1994)}]{ccsdtq3}%
  \BibitemOpen
  \bibfield  {author} {\bibinfo {author} {\bibfnamefont {P.}~\bibnamefont
  {Piecuch}}\ and\ \bibinfo {author} {\bibfnamefont {L.}~\bibnamefont
  {Adamowicz}},\ }\href@noop {} {\bibfield  {journal} {\bibinfo  {journal} {J.
  Chem. Phys.}\ }\textbf {\bibinfo {volume} {100}},\ \bibinfo {pages} {5792}
  (\bibinfo {year} {1994})}\BibitemShut {NoStop}%
\bibitem [{\citenamefont {Emrich}(1981)}]{emrich}%
  \BibitemOpen
  \bibfield  {author} {\bibinfo {author} {\bibfnamefont {K.}~\bibnamefont
  {Emrich}},\ }\href@noop {} {\bibfield  {journal} {\bibinfo  {journal} {Nucl.
  Phys. A}\ }\textbf {\bibinfo {volume} {351}},\ \bibinfo {pages} {379}
  (\bibinfo {year} {1981})}\BibitemShut {NoStop}%
\bibitem [{\citenamefont {Geertsen}, \citenamefont {Rittby},\ and\
  \citenamefont {Bartlett}(1989)}]{eomcc1}%
  \BibitemOpen
  \bibfield  {author} {\bibinfo {author} {\bibfnamefont {J.}~\bibnamefont
  {Geertsen}}, \bibinfo {author} {\bibfnamefont {M.}~\bibnamefont {Rittby}}, \
  and\ \bibinfo {author} {\bibfnamefont {R.~J.}\ \bibnamefont {Bartlett}},\
  }\href@noop {} {\bibfield  {journal} {\bibinfo  {journal} {Chem. Phys.
  Lett.}\ }\textbf {\bibinfo {volume} {164}},\ \bibinfo {pages} {57} (\bibinfo
  {year} {1989})}\BibitemShut {NoStop}%
\bibitem [{\citenamefont {Stanton}\ and\ \citenamefont
  {Bartlett}(1993)}]{eomcc3}%
  \BibitemOpen
  \bibfield  {author} {\bibinfo {author} {\bibfnamefont {J.~F.}\ \bibnamefont
  {Stanton}}\ and\ \bibinfo {author} {\bibfnamefont {R.~J.}\ \bibnamefont
  {Bartlett}},\ }\href@noop {} {\bibfield  {journal} {\bibinfo  {journal} {J.
  Chem. Phys.}\ }\textbf {\bibinfo {volume} {98}},\ \bibinfo {pages} {7029}
  (\bibinfo {year} {1993})}\BibitemShut {NoStop}%
\bibitem [{\citenamefont {Kowalski}\ and\ \citenamefont
  {Piecuch}(2001{\natexlab{a}})}]{eomccsdt1}%
  \BibitemOpen
  \bibfield  {author} {\bibinfo {author} {\bibfnamefont {K.}~\bibnamefont
  {Kowalski}}\ and\ \bibinfo {author} {\bibfnamefont {P.}~\bibnamefont
  {Piecuch}},\ }\href@noop {} {\bibfield  {journal} {\bibinfo  {journal} {J.
  Chem. Phys.}\ }\textbf {\bibinfo {volume} {115}},\ \bibinfo {pages} {643}
  (\bibinfo {year} {2001}{\natexlab{a}})}\BibitemShut {NoStop}%
\bibitem [{\citenamefont {Kowalski}\ and\ \citenamefont
  {Piecuch}(2001{\natexlab{b}})}]{eomccsdt2}%
  \BibitemOpen
  \bibfield  {author} {\bibinfo {author} {\bibfnamefont {K.}~\bibnamefont
  {Kowalski}}\ and\ \bibinfo {author} {\bibfnamefont {P.}~\bibnamefont
  {Piecuch}},\ }\href@noop {} {\bibfield  {journal} {\bibinfo  {journal} {Chem.
  Phys. Lett.}\ }\textbf {\bibinfo {volume} {347}},\ \bibinfo {pages} {237}
  (\bibinfo {year} {2001}{\natexlab{b}})}\BibitemShut {NoStop}%
\bibitem [{\citenamefont {Kucharski}\ \emph {et~al.}(2001)\citenamefont
  {Kucharski}, \citenamefont {W{\l}och}, \citenamefont {Musia{\l}},\ and\
  \citenamefont {Bartlett}}]{eomccsdt3}%
  \BibitemOpen
  \bibfield  {author} {\bibinfo {author} {\bibfnamefont {S.~A.}\ \bibnamefont
  {Kucharski}}, \bibinfo {author} {\bibfnamefont {M.}~\bibnamefont {W{\l}och}},
  \bibinfo {author} {\bibfnamefont {M.}~\bibnamefont {Musia{\l}}}, \ and\
  \bibinfo {author} {\bibfnamefont {R.~J.}\ \bibnamefont {Bartlett}},\
  }\href@noop {} {\bibfield  {journal} {\bibinfo  {journal} {J. Chem. Phys.}\
  }\textbf {\bibinfo {volume} {115}},\ \bibinfo {pages} {8263} (\bibinfo {year}
  {2001})}\BibitemShut {NoStop}%
\bibitem [{\citenamefont {K{\' a}llay}\ and\ \citenamefont
  {Gauss}(2004)}]{kallaygauss}%
  \BibitemOpen
  \bibfield  {author} {\bibinfo {author} {\bibfnamefont {M.}~\bibnamefont {K{\'
  a}llay}}\ and\ \bibinfo {author} {\bibfnamefont {J.}~\bibnamefont {Gauss}},\
  }\href@noop {} {\bibfield  {journal} {\bibinfo  {journal} {J. Chem. Phys.}\
  }\textbf {\bibinfo {volume} {121}},\ \bibinfo {pages} {9257} (\bibinfo {year}
  {2004})}\BibitemShut {NoStop}%
\bibitem [{\citenamefont {Hirata}(2004)}]{hirata1}%
  \BibitemOpen
  \bibfield  {author} {\bibinfo {author} {\bibfnamefont {S.}~\bibnamefont
  {Hirata}},\ }\href@noop {} {\bibfield  {journal} {\bibinfo  {journal} {J.
  Chem. Phys.}\ }\textbf {\bibinfo {volume} {121}},\ \bibinfo {pages} {51}
  (\bibinfo {year} {2004})}\BibitemShut {NoStop}%
\bibitem [{\citenamefont {Monkhorst}(1977)}]{monk}%
  \BibitemOpen
  \bibfield  {author} {\bibinfo {author} {\bibfnamefont {H.~J.}\ \bibnamefont
  {Monkhorst}},\ }\href@noop {} {\bibfield  {journal} {\bibinfo  {journal}
  {Int. J. Quantum Chem. Symp.}\ }\textbf {\bibinfo {volume} {11}},\ \bibinfo
  {pages} {421} (\bibinfo {year} {1977})}\BibitemShut {NoStop}%
\bibitem [{\citenamefont {Dalgaard}\ and\ \citenamefont
  {Monkhorst}(1983)}]{monk2}%
  \BibitemOpen
  \bibfield  {author} {\bibinfo {author} {\bibfnamefont {E.}~\bibnamefont
  {Dalgaard}}\ and\ \bibinfo {author} {\bibfnamefont {H.~J.}\ \bibnamefont
  {Monkhorst}},\ }\href@noop {} {\bibfield  {journal} {\bibinfo  {journal}
  {Phys. Rev. A}\ }\textbf {\bibinfo {volume} {28}},\ \bibinfo {pages} {1217}
  (\bibinfo {year} {1983})}\BibitemShut {NoStop}%
\bibitem [{\citenamefont {Mukherjee}\ and\ \citenamefont
  {Mukherjee}(1979)}]{mukherjee_lrcc}%
  \BibitemOpen
  \bibfield  {author} {\bibinfo {author} {\bibfnamefont {D.}~\bibnamefont
  {Mukherjee}}\ and\ \bibinfo {author} {\bibfnamefont {P.~K.}\ \bibnamefont
  {Mukherjee}},\ }\href@noop {} {\bibfield  {journal} {\bibinfo  {journal}
  {Chem. Phys.}\ }\textbf {\bibinfo {volume} {39}},\ \bibinfo {pages} {325}
  (\bibinfo {year} {1979})}\BibitemShut {NoStop}%
\bibitem [{\citenamefont {Sekino}\ and\ \citenamefont
  {Bartlett}(1984)}]{sekino-rjb-1984}%
  \BibitemOpen
  \bibfield  {author} {\bibinfo {author} {\bibfnamefont {H.}~\bibnamefont
  {Sekino}}\ and\ \bibinfo {author} {\bibfnamefont {R.~J.}\ \bibnamefont
  {Bartlett}},\ }\href@noop {} {\bibfield  {journal} {\bibinfo  {journal} {Int.
  J. Quantum Chem. Symp.}\ }\textbf {\bibinfo {volume} {18}},\ \bibinfo {pages}
  {255} (\bibinfo {year} {1984})}\BibitemShut {NoStop}%
\bibitem [{\citenamefont {Takahashi}\ and\ \citenamefont
  {Paldus}(1986)}]{lrcc3}%
  \BibitemOpen
  \bibfield  {author} {\bibinfo {author} {\bibfnamefont {M.}~\bibnamefont
  {Takahashi}}\ and\ \bibinfo {author} {\bibfnamefont {J.}~\bibnamefont
  {Paldus}},\ }\href@noop {} {\bibfield  {journal} {\bibinfo  {journal} {J.
  Chem. Phys.}\ }\textbf {\bibinfo {volume} {85}},\ \bibinfo {pages} {1486}
  (\bibinfo {year} {1986})}\BibitemShut {NoStop}%
\bibitem [{\citenamefont {Koch}\ and\ \citenamefont
  {J{\o}rgensen}(1990)}]{lrcc4}%
  \BibitemOpen
  \bibfield  {author} {\bibinfo {author} {\bibfnamefont {H.}~\bibnamefont
  {Koch}}\ and\ \bibinfo {author} {\bibfnamefont {P.}~\bibnamefont
  {J{\o}rgensen}},\ }\href@noop {} {\bibfield  {journal} {\bibinfo  {journal}
  {J. Chem. Phys.}\ }\textbf {\bibinfo {volume} {93}},\ \bibinfo {pages} {3333}
  (\bibinfo {year} {1990})}\BibitemShut {NoStop}%
\bibitem [{\citenamefont {Koch}\ \emph {et~al.}(1990)\citenamefont {Koch},
  \citenamefont {Jensen}, \citenamefont {J{\o}rgensen},\ and\ \citenamefont
  {Helgaker}}]{jorgensen}%
  \BibitemOpen
  \bibfield  {author} {\bibinfo {author} {\bibfnamefont {H.}~\bibnamefont
  {Koch}}, \bibinfo {author} {\bibfnamefont {H.~J.~A.}\ \bibnamefont {Jensen}},
  \bibinfo {author} {\bibfnamefont {P.}~\bibnamefont {J{\o}rgensen}}, \ and\
  \bibinfo {author} {\bibfnamefont {T.}~\bibnamefont {Helgaker}},\ }\href@noop
  {} {\bibfield  {journal} {\bibinfo  {journal} {J. Chem. Phys.}\ }\textbf
  {\bibinfo {volume} {93}},\ \bibinfo {pages} {3345} (\bibinfo {year}
  {1990})}\BibitemShut {NoStop}%
\bibitem [{\citenamefont {Kondo}, \citenamefont {Piecuch},\ and\ \citenamefont
  {Paldus}(1995)}]{kondo-1995}%
  \BibitemOpen
  \bibfield  {author} {\bibinfo {author} {\bibfnamefont {A.~E.}\ \bibnamefont
  {Kondo}}, \bibinfo {author} {\bibfnamefont {P.}~\bibnamefont {Piecuch}}, \
  and\ \bibinfo {author} {\bibfnamefont {J.}~\bibnamefont {Paldus}},\
  }\href@noop {} {\bibfield  {journal} {\bibinfo  {journal} {J. Chem. Phys.}\
  }\textbf {\bibinfo {volume} {102}},\ \bibinfo {pages} {6511} (\bibinfo {year}
  {1995})}\BibitemShut {NoStop}%
\bibitem [{\citenamefont {Kondo}, \citenamefont {Piecuch},\ and\ \citenamefont
  {Paldus}(1996)}]{kondo-1996}%
  \BibitemOpen
  \bibfield  {author} {\bibinfo {author} {\bibfnamefont {A.~E.}\ \bibnamefont
  {Kondo}}, \bibinfo {author} {\bibfnamefont {P.}~\bibnamefont {Piecuch}}, \
  and\ \bibinfo {author} {\bibfnamefont {J.}~\bibnamefont {Paldus}},\
  }\href@noop {} {\bibfield  {journal} {\bibinfo  {journal} {J. Chem. Phys.}\
  }\textbf {\bibinfo {volume} {104}},\ \bibinfo {pages} {8566} (\bibinfo {year}
  {1996})}\BibitemShut {NoStop}%
\bibitem [{\citenamefont {Raghavachari}\ \emph {et~al.}(1989)\citenamefont
  {Raghavachari}, \citenamefont {Trucks}, \citenamefont {Pople},\ and\
  \citenamefont {Head-Gordon}}]{ccsdpt}%
  \BibitemOpen
  \bibfield  {author} {\bibinfo {author} {\bibfnamefont {K.}~\bibnamefont
  {Raghavachari}}, \bibinfo {author} {\bibfnamefont {G.~W.}\ \bibnamefont
  {Trucks}}, \bibinfo {author} {\bibfnamefont {J.~A.}\ \bibnamefont {Pople}}, \
  and\ \bibinfo {author} {\bibfnamefont {M.}~\bibnamefont {Head-Gordon}},\
  }\href@noop {} {\bibfield  {journal} {\bibinfo  {journal} {Chem. Phys.
  Lett.}\ }\textbf {\bibinfo {volume} {157}},\ \bibinfo {pages} {479} (\bibinfo
  {year} {1989})}\BibitemShut {NoStop}%
\bibitem [{\citenamefont {Watts}, \citenamefont {Gauss},\ and\ \citenamefont
  {Bartlett}(1993)}]{watts-gauss-bartlett-1993}%
  \BibitemOpen
  \bibfield  {author} {\bibinfo {author} {\bibfnamefont {J.~D.}\ \bibnamefont
  {Watts}}, \bibinfo {author} {\bibfnamefont {J.}~\bibnamefont {Gauss}}, \ and\
  \bibinfo {author} {\bibfnamefont {R.~J.}\ \bibnamefont {Bartlett}},\
  }\href@noop {} {\bibfield  {journal} {\bibinfo  {journal} {J. Chem. Phys.}\
  }\textbf {\bibinfo {volume} {98}},\ \bibinfo {pages} {8718} (\bibinfo {year}
  {1993})}\BibitemShut {NoStop}%
\bibitem [{\citenamefont {Piecuch}\ \emph {et~al.}(2002)\citenamefont
  {Piecuch}, \citenamefont {Kowalski}, \citenamefont {Pimienta},\ and\
  \citenamefont {McGuire}}]{irpc}%
  \BibitemOpen
  \bibfield  {author} {\bibinfo {author} {\bibfnamefont {P.}~\bibnamefont
  {Piecuch}}, \bibinfo {author} {\bibfnamefont {K.}~\bibnamefont {Kowalski}},
  \bibinfo {author} {\bibfnamefont {I.~S.~O.}\ \bibnamefont {Pimienta}}, \ and\
  \bibinfo {author} {\bibfnamefont {M.~J.}\ \bibnamefont {McGuire}},\
  }\href@noop {} {\bibfield  {journal} {\bibinfo  {journal} {Int. Rev. Phys.
  Chem.}\ }\textbf {\bibinfo {volume} {21}},\ \bibinfo {pages} {527} (\bibinfo
  {year} {2002})}\BibitemShut {NoStop}%
\bibitem [{\citenamefont {Piecuch}\ \emph {et~al.}(2004)\citenamefont
  {Piecuch}, \citenamefont {Kowalski}, \citenamefont {Pimienta}, \citenamefont
  {Fan}, \citenamefont {Lodriguito}, \citenamefont {McGuire}, \citenamefont
  {Kucharski}, \citenamefont {Ku{\' s}},\ and\ \citenamefont
  {Musia{\l}}}]{PP:TCA}%
  \BibitemOpen
  \bibfield  {author} {\bibinfo {author} {\bibfnamefont {P.}~\bibnamefont
  {Piecuch}}, \bibinfo {author} {\bibfnamefont {K.}~\bibnamefont {Kowalski}},
  \bibinfo {author} {\bibfnamefont {I.~S.~O.}\ \bibnamefont {Pimienta}},
  \bibinfo {author} {\bibfnamefont {P.-D.}\ \bibnamefont {Fan}}, \bibinfo
  {author} {\bibfnamefont {M.}~\bibnamefont {Lodriguito}}, \bibinfo {author}
  {\bibfnamefont {M.~J.}\ \bibnamefont {McGuire}}, \bibinfo {author}
  {\bibfnamefont {S.~A.}\ \bibnamefont {Kucharski}}, \bibinfo {author}
  {\bibfnamefont {T.}~\bibnamefont {Ku{\' s}}}, \ and\ \bibinfo {author}
  {\bibfnamefont {M.}~\bibnamefont {Musia{\l}}},\ }\href@noop {} {\bibfield
  {journal} {\bibinfo  {journal} {Theor. Chem. Acc.}\ }\textbf {\bibinfo
  {volume} {112}},\ \bibinfo {pages} {349} (\bibinfo {year}
  {2004})}\BibitemShut {NoStop}%
\bibitem [{\citenamefont {Piecuch}(2010)}]{piecuch-qtp}%
  \BibitemOpen
  \bibfield  {author} {\bibinfo {author} {\bibfnamefont {P.}~\bibnamefont
  {Piecuch}},\ }\href@noop {} {\bibfield  {journal} {\bibinfo  {journal} {Mol.
  Phys.}\ }\textbf {\bibinfo {volume} {108}},\ \bibinfo {pages} {2987}
  (\bibinfo {year} {2010})}\BibitemShut {NoStop}%
\bibitem [{\citenamefont {Shen}\ and\ \citenamefont
  {Piecuch}(2012{\natexlab{a}})}]{jspp-chemphys2012}%
  \BibitemOpen
  \bibfield  {author} {\bibinfo {author} {\bibfnamefont {J.}~\bibnamefont
  {Shen}}\ and\ \bibinfo {author} {\bibfnamefont {P.}~\bibnamefont {Piecuch}},\
  }\href@noop {} {\bibfield  {journal} {\bibinfo  {journal} {Chem. Phys.}\
  }\textbf {\bibinfo {volume} {401}},\ \bibinfo {pages} {180} (\bibinfo {year}
  {2012}{\natexlab{a}})}\BibitemShut {NoStop}%
\bibitem [{\citenamefont {Shen}\ and\ \citenamefont
  {Piecuch}(2012{\natexlab{b}})}]{jspp-jcp2012}%
  \BibitemOpen
  \bibfield  {author} {\bibinfo {author} {\bibfnamefont {J.}~\bibnamefont
  {Shen}}\ and\ \bibinfo {author} {\bibfnamefont {P.}~\bibnamefont {Piecuch}},\
  }\href@noop {} {\bibfield  {journal} {\bibinfo  {journal} {J. Chem. Phys.}\
  }\textbf {\bibinfo {volume} {136}},\ \bibinfo {pages} {144104} (\bibinfo
  {year} {2012}{\natexlab{b}})}\BibitemShut {NoStop}%
\bibitem [{\citenamefont {Shen}\ and\ \citenamefont
  {Piecuch}(2012{\natexlab{c}})}]{jspp-jctc2012}%
  \BibitemOpen
  \bibfield  {author} {\bibinfo {author} {\bibfnamefont {J.}~\bibnamefont
  {Shen}}\ and\ \bibinfo {author} {\bibfnamefont {P.}~\bibnamefont {Piecuch}},\
  }\href@noop {} {\bibfield  {journal} {\bibinfo  {journal} {J. Chem. Theory
  Comput.}\ }\textbf {\bibinfo {volume} {8}},\ \bibinfo {pages} {4968}
  (\bibinfo {year} {2012}{\natexlab{c}})}\BibitemShut {NoStop}%
\bibitem [{\citenamefont {Bauman}, \citenamefont {Shen},\ and\ \citenamefont
  {Piecuch}(2017)}]{nbjspp-molphys2017}%
  \BibitemOpen
  \bibfield  {author} {\bibinfo {author} {\bibfnamefont {N.~P.}\ \bibnamefont
  {Bauman}}, \bibinfo {author} {\bibfnamefont {J.}~\bibnamefont {Shen}}, \ and\
  \bibinfo {author} {\bibfnamefont {P.}~\bibnamefont {Piecuch}},\ }\href@noop
  {} {\bibfield  {journal} {\bibinfo  {journal} {Mol. Phys.}\ }\textbf
  {\bibinfo {volume} {115}},\ \bibinfo {pages} {2860} (\bibinfo {year}
  {2017})}\BibitemShut {NoStop}%
\bibitem [{\citenamefont {Piecuch}\ and\ \citenamefont
  {W{\l}och}(2005)}]{crccl_jcp}%
  \BibitemOpen
  \bibfield  {author} {\bibinfo {author} {\bibfnamefont {P.}~\bibnamefont
  {Piecuch}}\ and\ \bibinfo {author} {\bibfnamefont {M.}~\bibnamefont
  {W{\l}och}},\ }\href@noop {} {\bibfield  {journal} {\bibinfo  {journal} {J.
  Chem. Phys.}\ }\textbf {\bibinfo {volume} {123}},\ \bibinfo {pages} {224105}
  (\bibinfo {year} {2005})}\BibitemShut {NoStop}%
\bibitem [{\citenamefont {Piecuch}\ \emph {et~al.}(2006)\citenamefont
  {Piecuch}, \citenamefont {W{\l}och}, \citenamefont {Gour},\ and\
  \citenamefont {Kinal}}]{crccl_cpl}%
  \BibitemOpen
  \bibfield  {author} {\bibinfo {author} {\bibfnamefont {P.}~\bibnamefont
  {Piecuch}}, \bibinfo {author} {\bibfnamefont {M.}~\bibnamefont {W{\l}och}},
  \bibinfo {author} {\bibfnamefont {J.~R.}\ \bibnamefont {Gour}}, \ and\
  \bibinfo {author} {\bibfnamefont {A.}~\bibnamefont {Kinal}},\ }\href@noop {}
  {\bibfield  {journal} {\bibinfo  {journal} {Chem. Phys. Lett.}\ }\textbf
  {\bibinfo {volume} {418}},\ \bibinfo {pages} {467} (\bibinfo {year}
  {2006})}\BibitemShut {NoStop}%
\bibitem [{\citenamefont {W{\l}och}\ \emph {et~al.}(2006)\citenamefont
  {W{\l}och}, \citenamefont {Lodriguito}, \citenamefont {Piecuch},\ and\
  \citenamefont {Gour}}]{crccl_molphys}%
  \BibitemOpen
  \bibfield  {author} {\bibinfo {author} {\bibfnamefont {M.}~\bibnamefont
  {W{\l}och}}, \bibinfo {author} {\bibfnamefont {M.~D.}\ \bibnamefont
  {Lodriguito}}, \bibinfo {author} {\bibfnamefont {P.}~\bibnamefont {Piecuch}},
  \ and\ \bibinfo {author} {\bibfnamefont {J.~R.}\ \bibnamefont {Gour}},\
  }\href@noop {} {\bibfield  {journal} {\bibinfo  {journal} {Mol. Phys.}\
  }\textbf {\bibinfo {volume} {104}},\ \bibinfo {pages} {2149} (\bibinfo {year}
  {2006})},\ \bibinfo {note} {{\bf 104}, 2991 (2006) [Erratum]}\BibitemShut
  {NoStop}%
\bibitem [{\citenamefont {W{\l}och}, \citenamefont {Gour},\ and\ \citenamefont
  {Piecuch}(2007)}]{crccl_jpc}%
  \BibitemOpen
  \bibfield  {author} {\bibinfo {author} {\bibfnamefont {M.}~\bibnamefont
  {W{\l}och}}, \bibinfo {author} {\bibfnamefont {J.~R.}\ \bibnamefont {Gour}},
  \ and\ \bibinfo {author} {\bibfnamefont {P.}~\bibnamefont {Piecuch}},\
  }\href@noop {} {\bibfield  {journal} {\bibinfo  {journal} {J. Phys. Chem. A}\
  }\textbf {\bibinfo {volume} {111}},\ \bibinfo {pages} {11359} (\bibinfo
  {year} {2007})}\BibitemShut {NoStop}%
\bibitem [{\citenamefont {Piecuch}, \citenamefont {Gour},\ and\ \citenamefont
  {W{\l}och}(2008)}]{crccl_ijqc}%
  \BibitemOpen
  \bibfield  {author} {\bibinfo {author} {\bibfnamefont {P.}~\bibnamefont
  {Piecuch}}, \bibinfo {author} {\bibfnamefont {J.~R.}\ \bibnamefont {Gour}}, \
  and\ \bibinfo {author} {\bibfnamefont {M.}~\bibnamefont {W{\l}och}},\
  }\href@noop {} {\bibfield  {journal} {\bibinfo  {journal} {Int. J. Quantum
  Chem.}\ }\textbf {\bibinfo {volume} {108}},\ \bibinfo {pages} {2128}
  (\bibinfo {year} {2008})}\BibitemShut {NoStop}%
\bibitem [{\citenamefont {Magoulas}\ \emph {et~al.}(2018)\citenamefont
  {Magoulas}, \citenamefont {Bauman}, \citenamefont {Shen},\ and\ \citenamefont
  {Piecuch}}]{ccpq-be2-jpca-2018}%
  \BibitemOpen
  \bibfield  {author} {\bibinfo {author} {\bibfnamefont {I.}~\bibnamefont
  {Magoulas}}, \bibinfo {author} {\bibfnamefont {N.~P.}\ \bibnamefont
  {Bauman}}, \bibinfo {author} {\bibfnamefont {J.}~\bibnamefont {Shen}}, \ and\
  \bibinfo {author} {\bibfnamefont {P.}~\bibnamefont {Piecuch}},\ }\href@noop
  {} {\bibfield  {journal} {\bibinfo  {journal} {J. Phys. Chem. A}\ }\textbf
  {\bibinfo {volume} {122}},\ \bibinfo {pages} {1350} (\bibinfo {year}
  {2018})}\BibitemShut {NoStop}%
\bibitem [{\citenamefont {Piecuch}, \citenamefont {W{\l}och},\ and\
  \citenamefont {Varandas}(2007)}]{ptcp2007}%
  \BibitemOpen
  \bibfield  {author} {\bibinfo {author} {\bibfnamefont {P.}~\bibnamefont
  {Piecuch}}, \bibinfo {author} {\bibfnamefont {M.}~\bibnamefont {W{\l}och}}, \
  and\ \bibinfo {author} {\bibfnamefont {A.~J.~C.}\ \bibnamefont {Varandas}},\
  }in\ \href@noop {} {\emph {\bibinfo {booktitle} {Topics in the Theory of
  Chemical and Physical Systems}}},\ \bibinfo {series} {Progress in Theoretical
  Chemistry and Physics}, Vol.~\bibinfo {volume} {16},\ \bibinfo {editor}
  {edited by\ \bibinfo {editor} {\bibfnamefont {S.}~\bibnamefont {Lahmar}},
  \bibinfo {editor} {\bibfnamefont {J.}~\bibnamefont {Maruani}}, \bibinfo
  {editor} {\bibfnamefont {S.}~\bibnamefont {Wilson}}, \ and\ \bibinfo {editor}
  {\bibfnamefont {G.}~\bibnamefont {Delgado-Barrio}}}\ (\bibinfo  {publisher}
  {Springer},\ \bibinfo {address} {Dordrecht},\ \bibinfo {year} {2007})\ pp.\
  \bibinfo {pages} {63--121}\BibitemShut {NoStop}%
\bibitem [{\citenamefont {Piecuch}, \citenamefont {W{\l}och},\ and\
  \citenamefont {Varandas}(2008)}]{msg65}%
  \BibitemOpen
  \bibfield  {author} {\bibinfo {author} {\bibfnamefont {P.}~\bibnamefont
  {Piecuch}}, \bibinfo {author} {\bibfnamefont {M.}~\bibnamefont {W{\l}och}}, \
  and\ \bibinfo {author} {\bibfnamefont {A.~J.~C.}\ \bibnamefont {Varandas}},\
  }\href@noop {} {\bibfield  {journal} {\bibinfo  {journal} {Theor. Chem.
  Acc.}\ }\textbf {\bibinfo {volume} {120}},\ \bibinfo {pages} {59} (\bibinfo
  {year} {2008})}\BibitemShut {NoStop}%
\bibitem [{\citenamefont {Horoi}\ \emph {et~al.}(2007)\citenamefont {Horoi},
  \citenamefont {Gour}, \citenamefont {W{\l}och}, \citenamefont {Lodriguito},
  \citenamefont {Brown},\ and\ \citenamefont {Piecuch}}]{nuclei8}%
  \BibitemOpen
  \bibfield  {author} {\bibinfo {author} {\bibfnamefont {M.}~\bibnamefont
  {Horoi}}, \bibinfo {author} {\bibfnamefont {J.~R.}\ \bibnamefont {Gour}},
  \bibinfo {author} {\bibfnamefont {M.}~\bibnamefont {W{\l}och}}, \bibinfo
  {author} {\bibfnamefont {M.~D.}\ \bibnamefont {Lodriguito}}, \bibinfo
  {author} {\bibfnamefont {B.~A.}\ \bibnamefont {Brown}}, \ and\ \bibinfo
  {author} {\bibfnamefont {P.}~\bibnamefont {Piecuch}},\ }\href@noop {}
  {\bibfield  {journal} {\bibinfo  {journal} {Phys. Rev. Lett.}\ }\textbf
  {\bibinfo {volume} {98}},\ \bibinfo {pages} {112501} (\bibinfo {year}
  {2007})}\BibitemShut {NoStop}%
\bibitem [{\citenamefont {Ge}, \citenamefont {Gordon},\ and\ \citenamefont
  {Piecuch}(2007)}]{ge1}%
  \BibitemOpen
  \bibfield  {author} {\bibinfo {author} {\bibfnamefont {Y.}~\bibnamefont
  {Ge}}, \bibinfo {author} {\bibfnamefont {M.~S.}\ \bibnamefont {Gordon}}, \
  and\ \bibinfo {author} {\bibfnamefont {P.}~\bibnamefont {Piecuch}},\
  }\href@noop {} {\bibfield  {journal} {\bibinfo  {journal} {J. Chem. Phys.}\
  }\textbf {\bibinfo {volume} {127}},\ \bibinfo {pages} {174106} (\bibinfo
  {year} {2007})}\BibitemShut {NoStop}%
\bibitem [{\citenamefont {Ge}\ \emph {et~al.}(2008)\citenamefont {Ge},
  \citenamefont {Gordon}, \citenamefont {Piecuch}, \citenamefont {W{\l}och},\
  and\ \citenamefont {Gour}}]{ge2}%
  \BibitemOpen
  \bibfield  {author} {\bibinfo {author} {\bibfnamefont {Y.}~\bibnamefont
  {Ge}}, \bibinfo {author} {\bibfnamefont {M.~S.}\ \bibnamefont {Gordon}},
  \bibinfo {author} {\bibfnamefont {P.}~\bibnamefont {Piecuch}}, \bibinfo
  {author} {\bibfnamefont {M.}~\bibnamefont {W{\l}och}}, \ and\ \bibinfo
  {author} {\bibfnamefont {J.~R.}\ \bibnamefont {Gour}},\ }\href@noop {}
  {\bibfield  {journal} {\bibinfo  {journal} {J. Phys. Chem. A}\ }\textbf
  {\bibinfo {volume} {112}},\ \bibinfo {pages} {11873} (\bibinfo {year}
  {2008})}\BibitemShut {NoStop}%
\bibitem [{\citenamefont {Yuwono}\ \emph {et~al.}(2019)\citenamefont {Yuwono},
  \citenamefont {Magoulas}, \citenamefont {Shen},\ and\ \citenamefont
  {Piecuch}}]{ccpq-mg2-mp-2019}%
  \BibitemOpen
  \bibfield  {author} {\bibinfo {author} {\bibfnamefont {S.~H.}\ \bibnamefont
  {Yuwono}}, \bibinfo {author} {\bibfnamefont {I.}~\bibnamefont {Magoulas}},
  \bibinfo {author} {\bibfnamefont {J.}~\bibnamefont {Shen}}, \ and\ \bibinfo
  {author} {\bibfnamefont {P.}~\bibnamefont {Piecuch}},\ }\href@noop {}
  {\bibfield  {journal} {\bibinfo  {journal} {Mol. Phys.}\ }\textbf {\bibinfo
  {volume} {117}},\ \bibinfo {pages} {1486} (\bibinfo {year}
  {2019})}\BibitemShut {NoStop}%
\bibitem [{\citenamefont {Piecuch}\ and\ \citenamefont
  {Kowalski}(2000)}]{leszcz}%
  \BibitemOpen
  \bibfield  {author} {\bibinfo {author} {\bibfnamefont {P.}~\bibnamefont
  {Piecuch}}\ and\ \bibinfo {author} {\bibfnamefont {K.}~\bibnamefont
  {Kowalski}},\ }in\ \href@noop {} {\emph {\bibinfo {booktitle} {Computational
  Chemistry: Reviews of Current Trends}}},\ Vol.~\bibinfo {volume} {5},\
  \bibinfo {editor} {edited by\ \bibinfo {editor} {\bibfnamefont
  {J.}~\bibnamefont {Leszczy{\' n}ski}}}\ (\bibinfo  {publisher} {World
  Scientific},\ \bibinfo {address} {Singapore},\ \bibinfo {year} {2000})\ pp.\
  \bibinfo {pages} {1--104}\BibitemShut {NoStop}%
\bibitem [{\citenamefont {Kowalski}\ and\ \citenamefont
  {Piecuch}(2000)}]{ren1}%
  \BibitemOpen
  \bibfield  {author} {\bibinfo {author} {\bibfnamefont {K.}~\bibnamefont
  {Kowalski}}\ and\ \bibinfo {author} {\bibfnamefont {P.}~\bibnamefont
  {Piecuch}},\ }\href@noop {} {\bibfield  {journal} {\bibinfo  {journal} {J.
  Chem. Phys.}\ }\textbf {\bibinfo {volume} {113}},\ \bibinfo {pages} {18}
  (\bibinfo {year} {2000})}\BibitemShut {NoStop}%
\bibitem [{\citenamefont {Kowalski}\ and\ \citenamefont
  {Piecuch}(2005)}]{ndcmmcc}%
  \BibitemOpen
  \bibfield  {author} {\bibinfo {author} {\bibfnamefont {K.}~\bibnamefont
  {Kowalski}}\ and\ \bibinfo {author} {\bibfnamefont {P.}~\bibnamefont
  {Piecuch}},\ }\href@noop {} {\bibfield  {journal} {\bibinfo  {journal} {J.
  Chem. Phys.}\ }\textbf {\bibinfo {volume} {122}},\ \bibinfo {pages} {074107}
  (\bibinfo {year} {2005})}\BibitemShut {NoStop}%
\bibitem [{\citenamefont {Hirata}\ \emph {et~al.}(2001)\citenamefont {Hirata},
  \citenamefont {Nooijen}, \citenamefont {Grabowski},\ and\ \citenamefont
  {Bartlett}}]{eomccpt}%
  \BibitemOpen
  \bibfield  {author} {\bibinfo {author} {\bibfnamefont {S.}~\bibnamefont
  {Hirata}}, \bibinfo {author} {\bibfnamefont {M.}~\bibnamefont {Nooijen}},
  \bibinfo {author} {\bibfnamefont {I.}~\bibnamefont {Grabowski}}, \ and\
  \bibinfo {author} {\bibfnamefont {R.~J.}\ \bibnamefont {Bartlett}},\
  }\href@noop {} {\bibfield  {journal} {\bibinfo  {journal} {J. Chem. Phys.}\
  }\textbf {\bibinfo {volume} {114}},\ \bibinfo {pages} {3919} (\bibinfo {year}
  {2001})},\ \bibinfo {note} {{\bf 115}, 3967 (2001) [Erratum]}\BibitemShut
  {NoStop}%
\bibitem [{\citenamefont {Hirata}\ \emph {et~al.}(2004)\citenamefont {Hirata},
  \citenamefont {Fan}, \citenamefont {Auer}, \citenamefont {Nooijen},\ and\
  \citenamefont {Piecuch}}]{ccsdpt2}%
  \BibitemOpen
  \bibfield  {author} {\bibinfo {author} {\bibfnamefont {S.}~\bibnamefont
  {Hirata}}, \bibinfo {author} {\bibfnamefont {P.-D.}\ \bibnamefont {Fan}},
  \bibinfo {author} {\bibfnamefont {A.~A.}\ \bibnamefont {Auer}}, \bibinfo
  {author} {\bibfnamefont {M.}~\bibnamefont {Nooijen}}, \ and\ \bibinfo
  {author} {\bibfnamefont {P.}~\bibnamefont {Piecuch}},\ }\href@noop {}
  {\bibfield  {journal} {\bibinfo  {journal} {J. Chem. Phys.}\ }\textbf
  {\bibinfo {volume} {121}},\ \bibinfo {pages} {12197} (\bibinfo {year}
  {2004})}\BibitemShut {NoStop}%
\bibitem [{\citenamefont {Gwaltney}\ and\ \citenamefont
  {Head-Gordon}(2000)}]{gwaltney1}%
  \BibitemOpen
  \bibfield  {author} {\bibinfo {author} {\bibfnamefont {S.~R.}\ \bibnamefont
  {Gwaltney}}\ and\ \bibinfo {author} {\bibfnamefont {M.}~\bibnamefont
  {Head-Gordon}},\ }\href@noop {} {\bibfield  {journal} {\bibinfo  {journal}
  {Chem. Phys. Lett.}\ }\textbf {\bibinfo {volume} {323}},\ \bibinfo {pages}
  {21} (\bibinfo {year} {2000})}\BibitemShut {NoStop}%
\bibitem [{\citenamefont {Gwaltney}\ and\ \citenamefont
  {Head-Gordon}(2001)}]{gwaltney3}%
  \BibitemOpen
  \bibfield  {author} {\bibinfo {author} {\bibfnamefont {S.~R.}\ \bibnamefont
  {Gwaltney}}\ and\ \bibinfo {author} {\bibfnamefont {M.}~\bibnamefont
  {Head-Gordon}},\ }\href@noop {} {\bibfield  {journal} {\bibinfo  {journal}
  {J. Chem. Phys.}\ }\textbf {\bibinfo {volume} {115}},\ \bibinfo {pages}
  {2014} (\bibinfo {year} {2001})}\BibitemShut {NoStop}%
\bibitem [{\citenamefont {Stanton}(1997)}]{stanton1997}%
  \BibitemOpen
  \bibfield  {author} {\bibinfo {author} {\bibfnamefont {J.~F.}\ \bibnamefont
  {Stanton}},\ }\href@noop {} {\bibfield  {journal} {\bibinfo  {journal} {Chem.
  Phys. Lett.}\ }\textbf {\bibinfo {volume} {281}},\ \bibinfo {pages} {130}
  (\bibinfo {year} {1997})}\BibitemShut {NoStop}%
\bibitem [{\citenamefont {Crawford}\ and\ \citenamefont
  {Stanton}(1998)}]{crawford1998}%
  \BibitemOpen
  \bibfield  {author} {\bibinfo {author} {\bibfnamefont {T.~D.}\ \bibnamefont
  {Crawford}}\ and\ \bibinfo {author} {\bibfnamefont {J.~F.}\ \bibnamefont
  {Stanton}},\ }\href@noop {} {\bibfield  {journal} {\bibinfo  {journal} {Int.
  J. Quantum Chem.}\ }\textbf {\bibinfo {volume} {70}},\ \bibinfo {pages} {601}
  (\bibinfo {year} {1998})}\BibitemShut {NoStop}%
\bibitem [{\citenamefont {Kucharski}\ and\ \citenamefont
  {Bartlett}(1998)}]{ref:26}%
  \BibitemOpen
  \bibfield  {author} {\bibinfo {author} {\bibfnamefont {S.~A.}\ \bibnamefont
  {Kucharski}}\ and\ \bibinfo {author} {\bibfnamefont {R.~J.}\ \bibnamefont
  {Bartlett}},\ }\href@noop {} {\bibfield  {journal} {\bibinfo  {journal} {J.
  Chem. Phys.}\ }\textbf {\bibinfo {volume} {108}},\ \bibinfo {pages} {5243}
  (\bibinfo {year} {1998})}\BibitemShut {NoStop}%
\bibitem [{\citenamefont {Taube}\ and\ \citenamefont
  {Bartlett}(2008{\natexlab{a}})}]{bartlett2008a}%
  \BibitemOpen
  \bibfield  {author} {\bibinfo {author} {\bibfnamefont {A.~G.}\ \bibnamefont
  {Taube}}\ and\ \bibinfo {author} {\bibfnamefont {R.~J.}\ \bibnamefont
  {Bartlett}},\ }\href@noop {} {\bibfield  {journal} {\bibinfo  {journal} {J.
  Chem. Phys.}\ }\textbf {\bibinfo {volume} {128}},\ \bibinfo {pages} {044110}
  (\bibinfo {year} {2008}{\natexlab{a}})}\BibitemShut {NoStop}%
\bibitem [{\citenamefont {Taube}\ and\ \citenamefont
  {Bartlett}(2008{\natexlab{b}})}]{bartlett2008b}%
  \BibitemOpen
  \bibfield  {author} {\bibinfo {author} {\bibfnamefont {A.~G.}\ \bibnamefont
  {Taube}}\ and\ \bibinfo {author} {\bibfnamefont {R.~J.}\ \bibnamefont
  {Bartlett}},\ }\href@noop {} {\bibfield  {journal} {\bibinfo  {journal} {J.
  Chem. Phys.}\ }\textbf {\bibinfo {volume} {128}},\ \bibinfo {pages} {044111}
  (\bibinfo {year} {2008}{\natexlab{b}})}\BibitemShut {NoStop}%
\bibitem [{\citenamefont {Eriksen}\ \emph
  {et~al.}(2014{\natexlab{a}})\citenamefont {Eriksen}, \citenamefont
  {Kristensen}, \citenamefont {Kj{\ae}rgaard}, \citenamefont {J{\o}rgensen},\
  and\ \citenamefont {Gauss}}]{eriksen1}%
  \BibitemOpen
  \bibfield  {author} {\bibinfo {author} {\bibfnamefont {J.~J.}\ \bibnamefont
  {Eriksen}}, \bibinfo {author} {\bibfnamefont {K.}~\bibnamefont {Kristensen}},
  \bibinfo {author} {\bibfnamefont {T.}~\bibnamefont {Kj{\ae}rgaard}}, \bibinfo
  {author} {\bibfnamefont {P.}~\bibnamefont {J{\o}rgensen}}, \ and\ \bibinfo
  {author} {\bibfnamefont {J.}~\bibnamefont {Gauss}},\ }\href@noop {}
  {\bibfield  {journal} {\bibinfo  {journal} {J. Chem. Phys.}\ }\textbf
  {\bibinfo {volume} {140}},\ \bibinfo {pages} {064108} (\bibinfo {year}
  {2014}{\natexlab{a}})}\BibitemShut {NoStop}%
\bibitem [{\citenamefont {Eriksen}\ \emph
  {et~al.}(2014{\natexlab{b}})\citenamefont {Eriksen}, \citenamefont
  {J{\o}rgensen}, \citenamefont {Olsen},\ and\ \citenamefont
  {Gauss}}]{eriksen2}%
  \BibitemOpen
  \bibfield  {author} {\bibinfo {author} {\bibfnamefont {J.~J.}\ \bibnamefont
  {Eriksen}}, \bibinfo {author} {\bibfnamefont {P.}~\bibnamefont
  {J{\o}rgensen}}, \bibinfo {author} {\bibfnamefont {J.}~\bibnamefont {Olsen}},
  \ and\ \bibinfo {author} {\bibfnamefont {J.}~\bibnamefont {Gauss}},\
  }\href@noop {} {\bibfield  {journal} {\bibinfo  {journal} {J. Chem. Phys.}\
  }\textbf {\bibinfo {volume} {140}},\ \bibinfo {pages} {174114} (\bibinfo
  {year} {2014}{\natexlab{b}})}\BibitemShut {NoStop}%
\bibitem [{\citenamefont {Oliphant}\ and\ \citenamefont
  {Adamowicz}(1991{\natexlab{b}})}]{semi0a}%
  \BibitemOpen
  \bibfield  {author} {\bibinfo {author} {\bibfnamefont {N.}~\bibnamefont
  {Oliphant}}\ and\ \bibinfo {author} {\bibfnamefont {L.}~\bibnamefont
  {Adamowicz}},\ }\href@noop {} {\bibfield  {journal} {\bibinfo  {journal} {J.
  Chem. Phys.}\ }\textbf {\bibinfo {volume} {94}},\ \bibinfo {pages} {1229}
  (\bibinfo {year} {1991}{\natexlab{b}})}\BibitemShut {NoStop}%
\bibitem [{\citenamefont {Oliphant}\ and\ \citenamefont
  {Adamowicz}(1992)}]{semi0b}%
  \BibitemOpen
  \bibfield  {author} {\bibinfo {author} {\bibfnamefont {N.}~\bibnamefont
  {Oliphant}}\ and\ \bibinfo {author} {\bibfnamefont {L.}~\bibnamefont
  {Adamowicz}},\ }\href@noop {} {\bibfield  {journal} {\bibinfo  {journal} {J.
  Chem. Phys.}\ }\textbf {\bibinfo {volume} {96}},\ \bibinfo {pages} {3739}
  (\bibinfo {year} {1992})}\BibitemShut {NoStop}%
\bibitem [{\citenamefont {Piecuch}, \citenamefont {Oliphant},\ and\
  \citenamefont {Adamowicz}(1993)}]{semi2}%
  \BibitemOpen
  \bibfield  {author} {\bibinfo {author} {\bibfnamefont {P.}~\bibnamefont
  {Piecuch}}, \bibinfo {author} {\bibfnamefont {N.}~\bibnamefont {Oliphant}}, \
  and\ \bibinfo {author} {\bibfnamefont {L.}~\bibnamefont {Adamowicz}},\
  }\href@noop {} {\bibfield  {journal} {\bibinfo  {journal} {J. Chem. Phys.}\
  }\textbf {\bibinfo {volume} {99}},\ \bibinfo {pages} {1875} (\bibinfo {year}
  {1993})}\BibitemShut {NoStop}%
\bibitem [{\citenamefont {Piecuch}\ and\ \citenamefont
  {Adamowicz}(1995)}]{semih2o}%
  \BibitemOpen
  \bibfield  {author} {\bibinfo {author} {\bibfnamefont {P.}~\bibnamefont
  {Piecuch}}\ and\ \bibinfo {author} {\bibfnamefont {L.}~\bibnamefont
  {Adamowicz}},\ }\href@noop {} {\bibfield  {journal} {\bibinfo  {journal} {J.
  Chem. Phys.}\ }\textbf {\bibinfo {volume} {102}},\ \bibinfo {pages} {898}
  (\bibinfo {year} {1995})}\BibitemShut {NoStop}%
\bibitem [{\citenamefont {Adamowicz}, \citenamefont {Piecuch},\ and\
  \citenamefont {Ghose}(1998)}]{semi3c}%
  \BibitemOpen
  \bibfield  {author} {\bibinfo {author} {\bibfnamefont {L.}~\bibnamefont
  {Adamowicz}}, \bibinfo {author} {\bibfnamefont {P.}~\bibnamefont {Piecuch}},
  \ and\ \bibinfo {author} {\bibfnamefont {K.~B.}\ \bibnamefont {Ghose}},\
  }\href@noop {} {\bibfield  {journal} {\bibinfo  {journal} {Mol. Phys.}\
  }\textbf {\bibinfo {volume} {94}},\ \bibinfo {pages} {225} (\bibinfo {year}
  {1998})}\BibitemShut {NoStop}%
\bibitem [{\citenamefont {Piecuch}, \citenamefont {Kucharski},\ and\
  \citenamefont {Bartlett}(1999)}]{semi4}%
  \BibitemOpen
  \bibfield  {author} {\bibinfo {author} {\bibfnamefont {P.}~\bibnamefont
  {Piecuch}}, \bibinfo {author} {\bibfnamefont {S.~A.}\ \bibnamefont
  {Kucharski}}, \ and\ \bibinfo {author} {\bibfnamefont {R.~J.}\ \bibnamefont
  {Bartlett}},\ }\href@noop {} {\bibfield  {journal} {\bibinfo  {journal} {J.
  Chem. Phys.}\ }\textbf {\bibinfo {volume} {110}},\ \bibinfo {pages} {6103}
  (\bibinfo {year} {1999})}\BibitemShut {NoStop}%
\bibitem [{\citenamefont {Booth}, \citenamefont {Thom},\ and\ \citenamefont
  {Alavi}(2009)}]{Booth2009}%
  \BibitemOpen
  \bibfield  {author} {\bibinfo {author} {\bibfnamefont {G.~H.}\ \bibnamefont
  {Booth}}, \bibinfo {author} {\bibfnamefont {A.~J.~W.}\ \bibnamefont {Thom}},
  \ and\ \bibinfo {author} {\bibfnamefont {A.}~\bibnamefont {Alavi}},\
  }\href@noop {} {\bibfield  {journal} {\bibinfo  {journal} {J. Chem. Phys.}\
  }\textbf {\bibinfo {volume} {131}},\ \bibinfo {pages} {054106} (\bibinfo
  {year} {2009})}\BibitemShut {NoStop}%
\bibitem [{\citenamefont {Cleland}, \citenamefont {Booth},\ and\ \citenamefont
  {Alavi}(2010)}]{Cleland2010}%
  \BibitemOpen
  \bibfield  {author} {\bibinfo {author} {\bibfnamefont {D.}~\bibnamefont
  {Cleland}}, \bibinfo {author} {\bibfnamefont {G.~H.}\ \bibnamefont {Booth}},
  \ and\ \bibinfo {author} {\bibfnamefont {A.}~\bibnamefont {Alavi}},\
  }\href@noop {} {\bibfield  {journal} {\bibinfo  {journal} {J. Chem. Phys.}\
  }\textbf {\bibinfo {volume} {132}},\ \bibinfo {pages} {041103} (\bibinfo
  {year} {2010})}\BibitemShut {NoStop}%
\bibitem [{\citenamefont {Dobrautz}, \citenamefont {Smart},\ and\ \citenamefont
  {Alavi}(2019)}]{fciqmc-uga-2019}%
  \BibitemOpen
  \bibfield  {author} {\bibinfo {author} {\bibfnamefont {W.}~\bibnamefont
  {Dobrautz}}, \bibinfo {author} {\bibfnamefont {S.~D.}\ \bibnamefont {Smart}},
  \ and\ \bibinfo {author} {\bibfnamefont {A.}~\bibnamefont {Alavi}},\
  }\href@noop {} {\bibfield  {journal} {\bibinfo  {journal} {J. Chem. Phys.}\
  }\textbf {\bibinfo {volume} {151}},\ \bibinfo {pages} {094104} (\bibinfo
  {year} {2019})}\BibitemShut {NoStop}%
\bibitem [{\citenamefont {Ghanem}, \citenamefont {Lozovoi},\ and\ \citenamefont
  {Alavi}(2019)}]{ghanem_alavi_fciqmc_jcp_2019}%
  \BibitemOpen
  \bibfield  {author} {\bibinfo {author} {\bibfnamefont {K.}~\bibnamefont
  {Ghanem}}, \bibinfo {author} {\bibfnamefont {A.~Y.}\ \bibnamefont {Lozovoi}},
  \ and\ \bibinfo {author} {\bibfnamefont {A.}~\bibnamefont {Alavi}},\
  }\href@noop {} {\bibfield  {journal} {\bibinfo  {journal} {J. Chem. Phys.}\
  }\textbf {\bibinfo {volume} {151}},\ \bibinfo {pages} {224108} (\bibinfo
  {year} {2019})}\BibitemShut {NoStop}%
\bibitem [{\citenamefont {Ghanem}, \citenamefont {Guther},\ and\ \citenamefont
  {Alavi}(2020)}]{ghanem_alavi_fciqmc_2020}%
  \BibitemOpen
  \bibfield  {author} {\bibinfo {author} {\bibfnamefont {K.}~\bibnamefont
  {Ghanem}}, \bibinfo {author} {\bibfnamefont {K.}~\bibnamefont {Guther}}, \
  and\ \bibinfo {author} {\bibfnamefont {A.}~\bibnamefont {Alavi}},\
  }\href@noop {} {\bibfield  {journal} {\bibinfo  {journal} {J. Chem. Phys.}\
  }\textbf {\bibinfo {volume} {153}},\ \bibinfo {pages} {224115} (\bibinfo
  {year} {2020})}\BibitemShut {NoStop}%
\bibitem [{\citenamefont {Thom}(2010)}]{Thom2010}%
  \BibitemOpen
  \bibfield  {author} {\bibinfo {author} {\bibfnamefont {A.~J.~W.}\
  \bibnamefont {Thom}},\ }\href@noop {} {\bibfield  {journal} {\bibinfo
  {journal} {Phys. Rev. Lett.}\ }\textbf {\bibinfo {volume} {105}},\ \bibinfo
  {pages} {263004} (\bibinfo {year} {2010})}\BibitemShut {NoStop}%
\bibitem [{\citenamefont {Franklin}\ \emph {et~al.}(2016)\citenamefont
  {Franklin}, \citenamefont {Spencer}, \citenamefont {Zoccante},\ and\
  \citenamefont {Thom}}]{Franklin2016}%
  \BibitemOpen
  \bibfield  {author} {\bibinfo {author} {\bibfnamefont {R.~S.~T.}\
  \bibnamefont {Franklin}}, \bibinfo {author} {\bibfnamefont {J.~S.}\
  \bibnamefont {Spencer}}, \bibinfo {author} {\bibfnamefont {A.}~\bibnamefont
  {Zoccante}}, \ and\ \bibinfo {author} {\bibfnamefont {A.~J.~W.}\ \bibnamefont
  {Thom}},\ }\href@noop {} {\bibfield  {journal} {\bibinfo  {journal} {J. Chem.
  Phys.}\ }\textbf {\bibinfo {volume} {144}},\ \bibinfo {pages} {044111}
  (\bibinfo {year} {2016})}\BibitemShut {NoStop}%
\bibitem [{\citenamefont {Spencer}\ and\ \citenamefont
  {Thom}(2016)}]{Spencer2016}%
  \BibitemOpen
  \bibfield  {author} {\bibinfo {author} {\bibfnamefont {J.~S.}\ \bibnamefont
  {Spencer}}\ and\ \bibinfo {author} {\bibfnamefont {A.~J.~W.}\ \bibnamefont
  {Thom}},\ }\href@noop {} {\bibfield  {journal} {\bibinfo  {journal} {J. Chem.
  Phys.}\ }\textbf {\bibinfo {volume} {144}},\ \bibinfo {pages} {084108}
  (\bibinfo {year} {2016})}\BibitemShut {NoStop}%
\bibitem [{\citenamefont {Scott}\ and\ \citenamefont {Thom}(2017)}]{Scott2017}%
  \BibitemOpen
  \bibfield  {author} {\bibinfo {author} {\bibfnamefont {C.~J.~C.}\
  \bibnamefont {Scott}}\ and\ \bibinfo {author} {\bibfnamefont {A.~J.~W.}\
  \bibnamefont {Thom}},\ }\href@noop {} {\bibfield  {journal} {\bibinfo
  {journal} {J. Chem. Phys.}\ }\textbf {\bibinfo {volume} {147}},\ \bibinfo
  {pages} {124105} (\bibinfo {year} {2017})}\BibitemShut {NoStop}%
\bibitem [{\citenamefont {Deustua}, \citenamefont {Shen},\ and\ \citenamefont
  {Piecuch}(2017)}]{stochastic-ccpq-prl-2017}%
  \BibitemOpen
  \bibfield  {author} {\bibinfo {author} {\bibfnamefont {J.~E.}\ \bibnamefont
  {Deustua}}, \bibinfo {author} {\bibfnamefont {J.}~\bibnamefont {Shen}}, \
  and\ \bibinfo {author} {\bibfnamefont {P.}~\bibnamefont {Piecuch}},\
  }\href@noop {} {\bibfield  {journal} {\bibinfo  {journal} {Phys. Rev. Lett.}\
  }\textbf {\bibinfo {volume} {119}},\ \bibinfo {pages} {223003} (\bibinfo
  {year} {2017})}\BibitemShut {NoStop}%
\bibitem [{\citenamefont {Yuwono}\ \emph {et~al.}(2020)\citenamefont {Yuwono},
  \citenamefont {Chakraborty}, \citenamefont {Deustua}, \citenamefont {Shen},\
  and\ \citenamefont {Piecuch}}]{stochastic-ccpq-molphys-2020}%
  \BibitemOpen
  \bibfield  {author} {\bibinfo {author} {\bibfnamefont {S.~H.}\ \bibnamefont
  {Yuwono}}, \bibinfo {author} {\bibfnamefont {A.}~\bibnamefont {Chakraborty}},
  \bibinfo {author} {\bibfnamefont {J.~E.}\ \bibnamefont {Deustua}}, \bibinfo
  {author} {\bibfnamefont {J.}~\bibnamefont {Shen}}, \ and\ \bibinfo {author}
  {\bibfnamefont {P.}~\bibnamefont {Piecuch}},\ }\href {\doibase
  10.1080/00268976.2020.1817592} {\bibfield  {journal} {\bibinfo  {journal}
  {Mol. Phys.}\ }\textbf {\bibinfo {volume} {118}},\ \bibinfo {pages}
  {e1817592} (\bibinfo {year} {2020})}\BibitemShut {NoStop}%
\bibitem [{\citenamefont {Deustua}, \citenamefont {Shen},\ and\ \citenamefont
  {Piecuch}(2021)}]{jed-js-pp-jcp-2021}%
  \BibitemOpen
  \bibfield  {author} {\bibinfo {author} {\bibfnamefont {J.~E.}\ \bibnamefont
  {Deustua}}, \bibinfo {author} {\bibfnamefont {J.}~\bibnamefont {Shen}}, \
  and\ \bibinfo {author} {\bibfnamefont {P.}~\bibnamefont {Piecuch}},\
  }\href@noop {} {\bibfield  {journal} {\bibinfo  {journal} {J. Chem. Phys.}\
  }\textbf {\bibinfo {volume} {154}},\ \bibinfo {pages} {124103} (\bibinfo
  {year} {2021})}\BibitemShut {NoStop}%
\bibitem [{\citenamefont {Deustua}\ \emph {et~al.}(2019)\citenamefont
  {Deustua}, \citenamefont {Yuwono}, \citenamefont {Shen},\ and\ \citenamefont
  {Piecuch}}]{eomccp-jcp-2019}%
  \BibitemOpen
  \bibfield  {author} {\bibinfo {author} {\bibfnamefont {J.~E.}\ \bibnamefont
  {Deustua}}, \bibinfo {author} {\bibfnamefont {S.~H.}\ \bibnamefont {Yuwono}},
  \bibinfo {author} {\bibfnamefont {J.}~\bibnamefont {Shen}}, \ and\ \bibinfo
  {author} {\bibfnamefont {P.}~\bibnamefont {Piecuch}},\ }\href@noop {}
  {\bibfield  {journal} {\bibinfo  {journal} {J. Chem. Phys.}\ }\textbf
  {\bibinfo {volume} {150}},\ \bibinfo {pages} {111101} (\bibinfo {year}
  {2019})}\BibitemShut {NoStop}%
\bibitem [{\citenamefont {Huron}, \citenamefont {Malrieu},\ and\ \citenamefont
  {Rancurel}(1973)}]{sci_3}%
  \BibitemOpen
  \bibfield  {author} {\bibinfo {author} {\bibfnamefont {B.}~\bibnamefont
  {Huron}}, \bibinfo {author} {\bibfnamefont {J.~P.}\ \bibnamefont {Malrieu}},
  \ and\ \bibinfo {author} {\bibfnamefont {P.}~\bibnamefont {Rancurel}},\
  }\href@noop {} {\bibfield  {journal} {\bibinfo  {journal} {J. Chem. Phys.}\
  }\textbf {\bibinfo {volume} {58}},\ \bibinfo {pages} {5745} (\bibinfo {year}
  {1973})}\BibitemShut {NoStop}%
\bibitem [{\citenamefont {Garniron}\ \emph {et~al.}(2017)\citenamefont
  {Garniron}, \citenamefont {Scemama}, \citenamefont {Loos},\ and\
  \citenamefont {Caffarel}}]{cipsi_1}%
  \BibitemOpen
  \bibfield  {author} {\bibinfo {author} {\bibfnamefont {Y.}~\bibnamefont
  {Garniron}}, \bibinfo {author} {\bibfnamefont {A.}~\bibnamefont {Scemama}},
  \bibinfo {author} {\bibfnamefont {P.-F.}\ \bibnamefont {Loos}}, \ and\
  \bibinfo {author} {\bibfnamefont {M.}~\bibnamefont {Caffarel}},\ }\href@noop
  {} {\bibfield  {journal} {\bibinfo  {journal} {J. Chem. Phys.}\ }\textbf
  {\bibinfo {volume} {147}},\ \bibinfo {pages} {034101} (\bibinfo {year}
  {2017})}\BibitemShut {NoStop}%
\bibitem [{\citenamefont {Garniron}\ \emph {et~al.}(2019)\citenamefont
  {Garniron}, \citenamefont {Applencourt}, \citenamefont {Gasperich},
  \citenamefont {Benali}, \citenamefont {Fert{\' e}}, \citenamefont {Paquier},
  \citenamefont {Pradines}, \citenamefont {Assaraf}, \citenamefont {Reinhardt},
  \citenamefont {Toulouse}, \citenamefont {Barbaresco}, \citenamefont {Renon},
  \citenamefont {David}, \citenamefont {Malrieu}, \citenamefont {V{\' e}ril},
  \citenamefont {Caffarel}, \citenamefont {Loos}, \citenamefont {Giner},\ and\
  \citenamefont {Scemama}}]{cipsi_2}%
  \BibitemOpen
  \bibfield  {author} {\bibinfo {author} {\bibfnamefont {Y.}~\bibnamefont
  {Garniron}}, \bibinfo {author} {\bibfnamefont {T.}~\bibnamefont
  {Applencourt}}, \bibinfo {author} {\bibfnamefont {K.}~\bibnamefont
  {Gasperich}}, \bibinfo {author} {\bibfnamefont {A.}~\bibnamefont {Benali}},
  \bibinfo {author} {\bibfnamefont {A.}~\bibnamefont {Fert{\' e}}}, \bibinfo
  {author} {\bibfnamefont {J.}~\bibnamefont {Paquier}}, \bibinfo {author}
  {\bibfnamefont {B.}~\bibnamefont {Pradines}}, \bibinfo {author}
  {\bibfnamefont {R.}~\bibnamefont {Assaraf}}, \bibinfo {author} {\bibfnamefont
  {P.}~\bibnamefont {Reinhardt}}, \bibinfo {author} {\bibfnamefont
  {J.}~\bibnamefont {Toulouse}}, \bibinfo {author} {\bibfnamefont
  {P.}~\bibnamefont {Barbaresco}}, \bibinfo {author} {\bibfnamefont
  {N.}~\bibnamefont {Renon}}, \bibinfo {author} {\bibfnamefont
  {G.}~\bibnamefont {David}}, \bibinfo {author} {\bibfnamefont {J.-P.}\
  \bibnamefont {Malrieu}}, \bibinfo {author} {\bibfnamefont {M.}~\bibnamefont
  {V{\' e}ril}}, \bibinfo {author} {\bibfnamefont {M.}~\bibnamefont
  {Caffarel}}, \bibinfo {author} {\bibfnamefont {P.-F.}\ \bibnamefont {Loos}},
  \bibinfo {author} {\bibfnamefont {E.}~\bibnamefont {Giner}}, \ and\ \bibinfo
  {author} {\bibfnamefont {A.}~\bibnamefont {Scemama}},\ }\href@noop {}
  {\bibfield  {journal} {\bibinfo  {journal} {J. Chem. Theory Comput.}\
  }\textbf {\bibinfo {volume} {15}},\ \bibinfo {pages} {3591} (\bibinfo {year}
  {2019})}\BibitemShut {NoStop}%
\bibitem [{\citenamefont {Gururangan}\ \emph {et~al.}(2021)\citenamefont
  {Gururangan}, \citenamefont {Deustua}, \citenamefont {Shen},\ and\
  \citenamefont {Piecuch}}]{cipsi-ccpq-2021}%
  \BibitemOpen
  \bibfield  {author} {\bibinfo {author} {\bibfnamefont {K.}~\bibnamefont
  {Gururangan}}, \bibinfo {author} {\bibfnamefont {J.~E.}\ \bibnamefont
  {Deustua}}, \bibinfo {author} {\bibfnamefont {J.}~\bibnamefont {Shen}}, \
  and\ \bibinfo {author} {\bibfnamefont {P.}~\bibnamefont {Piecuch}},\
  }\href@noop {} {\bibfield  {journal} {\bibinfo  {journal} {J. Chem. Phys.}\
  }\textbf {\bibinfo {volume} {155}},\ \bibinfo {pages} {174114} (\bibinfo
  {year} {2021})}\BibitemShut {NoStop}%
\bibitem [{\citenamefont {Pedersen}, \citenamefont {Herek},\ and\ \citenamefont
  {Zewail}(1994)}]{sjs-ref1}%
  \BibitemOpen
  \bibfield  {author} {\bibinfo {author} {\bibfnamefont {S.}~\bibnamefont
  {Pedersen}}, \bibinfo {author} {\bibfnamefont {J.~L.}\ \bibnamefont {Herek}},
  \ and\ \bibinfo {author} {\bibfnamefont {A.~H.}\ \bibnamefont {Zewail}},\
  }\href@noop {} {\bibfield  {journal} {\bibinfo  {journal} {Science}\ }\textbf
  {\bibinfo {volume} {266}},\ \bibinfo {pages} {1359} (\bibinfo {year}
  {1994})}\BibitemShut {NoStop}%
\bibitem [{\citenamefont {Cho}, \citenamefont {Ko},\ and\ \citenamefont
  {Lee}(2016)}]{sjs-ref3}%
  \BibitemOpen
  \bibfield  {author} {\bibinfo {author} {\bibfnamefont {D.}~\bibnamefont
  {Cho}}, \bibinfo {author} {\bibfnamefont {K.~C.}\ \bibnamefont {Ko}}, \ and\
  \bibinfo {author} {\bibfnamefont {J.~Y.}\ \bibnamefont {Lee}},\ }\href@noop
  {} {\bibfield  {journal} {\bibinfo  {journal} {Int. J. Quantum Chem.}\
  }\textbf {\bibinfo {volume} {116}},\ \bibinfo {pages} {578} (\bibinfo {year}
  {2016})}\BibitemShut {NoStop}%
\bibitem [{\citenamefont {Sugawara}, \citenamefont {Komatsu},\ and\
  \citenamefont {Suzuki}(2011)}]{sjs-ref5}%
  \BibitemOpen
  \bibfield  {author} {\bibinfo {author} {\bibfnamefont {T.}~\bibnamefont
  {Sugawara}}, \bibinfo {author} {\bibfnamefont {H.}~\bibnamefont {Komatsu}}, \
  and\ \bibinfo {author} {\bibfnamefont {K.}~\bibnamefont {Suzuki}},\
  }\href@noop {} {\bibfield  {journal} {\bibinfo  {journal} {Chem. Soc. Rev.}\
  }\textbf {\bibinfo {volume} {40}},\ \bibinfo {pages} {3105} (\bibinfo {year}
  {2011})}\BibitemShut {NoStop}%
\bibitem [{\citenamefont {Sanvito}(2011)}]{sjs-ref6}%
  \BibitemOpen
  \bibfield  {author} {\bibinfo {author} {\bibfnamefont {S.}~\bibnamefont
  {Sanvito}},\ }\href@noop {} {\bibfield  {journal} {\bibinfo  {journal} {Chem.
  Soc. Rev.}\ }\textbf {\bibinfo {volume} {40}},\ \bibinfo {pages} {3336}
  (\bibinfo {year} {2011})}\BibitemShut {NoStop}%
\bibitem [{\citenamefont {Nakano}\ and\ \citenamefont
  {Champagne}(2016)}]{sjs-ref7}%
  \BibitemOpen
  \bibfield  {author} {\bibinfo {author} {\bibfnamefont {M.}~\bibnamefont
  {Nakano}}\ and\ \bibinfo {author} {\bibfnamefont {B.}~\bibnamefont
  {Champagne}},\ }\href@noop {} {\bibfield  {journal} {\bibinfo  {journal}
  {WIREs Comput. Mol. Sci.}\ }\textbf {\bibinfo {volume} {6}},\ \bibinfo
  {pages} {198} (\bibinfo {year} {2016})}\BibitemShut {NoStop}%
\bibitem [{\citenamefont {Zgierski}, \citenamefont {Patchkovskii},\ and\
  \citenamefont {Lim}(2005)}]{zgierski1}%
  \BibitemOpen
  \bibfield  {author} {\bibinfo {author} {\bibfnamefont {M.~Z.}\ \bibnamefont
  {Zgierski}}, \bibinfo {author} {\bibfnamefont {S.}~\bibnamefont
  {Patchkovskii}}, \ and\ \bibinfo {author} {\bibfnamefont {E.~C.}\
  \bibnamefont {Lim}},\ }\href@noop {} {\bibfield  {journal} {\bibinfo
  {journal} {J. Chem. Phys.}\ }\textbf {\bibinfo {volume} {123}},\ \bibinfo
  {pages} {081101} (\bibinfo {year} {2005})}\BibitemShut {NoStop}%
\bibitem [{\citenamefont {Zgierski}\ \emph {et~al.}(2005)\citenamefont
  {Zgierski}, \citenamefont {Patchkovskii}, \citenamefont {Fujiwara},\ and\
  \citenamefont {Lim}}]{zgierski2}%
  \BibitemOpen
  \bibfield  {author} {\bibinfo {author} {\bibfnamefont {M.~Z.}\ \bibnamefont
  {Zgierski}}, \bibinfo {author} {\bibfnamefont {S.}~\bibnamefont
  {Patchkovskii}}, \bibinfo {author} {\bibfnamefont {T.}~\bibnamefont
  {Fujiwara}}, \ and\ \bibinfo {author} {\bibfnamefont {E.~C.}\ \bibnamefont
  {Lim}},\ }\href@noop {} {\bibfield  {journal} {\bibinfo  {journal} {J. Phys.
  Chem. A}\ }\textbf {\bibinfo {volume} {109}},\ \bibinfo {pages} {9384}
  (\bibinfo {year} {2005})}\BibitemShut {NoStop}%
\bibitem [{\citenamefont {Park}\ \emph {et~al.}(2021)\citenamefont {Park},
  \citenamefont {Shen}, \citenamefont {Lee}, \citenamefont {Piecuch},
  \citenamefont {Filatov},\ and\ \citenamefont {Choi}}]{jpclett-2021}%
  \BibitemOpen
  \bibfield  {author} {\bibinfo {author} {\bibfnamefont {W.}~\bibnamefont
  {Park}}, \bibinfo {author} {\bibfnamefont {J.}~\bibnamefont {Shen}}, \bibinfo
  {author} {\bibfnamefont {S.}~\bibnamefont {Lee}}, \bibinfo {author}
  {\bibfnamefont {P.}~\bibnamefont {Piecuch}}, \bibinfo {author} {\bibfnamefont
  {M.}~\bibnamefont {Filatov}}, \ and\ \bibinfo {author} {\bibfnamefont
  {C.~H.}\ \bibnamefont {Choi}},\ }\href@noop {} {\bibfield  {journal}
  {\bibinfo  {journal} {J. Phys. Chem. Lett.}\ }\textbf {\bibinfo {volume}
  {12}},\ \bibinfo {pages} {9720} (\bibinfo {year} {2021})}\BibitemShut
  {NoStop}%
\bibitem [{\citenamefont {Minami}\ and\ \citenamefont
  {Nakano}(2012)}]{sjs-ref8}%
  \BibitemOpen
  \bibfield  {author} {\bibinfo {author} {\bibfnamefont {T.}~\bibnamefont
  {Minami}}\ and\ \bibinfo {author} {\bibfnamefont {M.}~\bibnamefont
  {Nakano}},\ }\href@noop {} {\bibfield  {journal} {\bibinfo  {journal} {J.
  Phys. Chem. Lett.}\ }\textbf {\bibinfo {volume} {3}},\ \bibinfo {pages} {145}
  (\bibinfo {year} {2012})}\BibitemShut {NoStop}%
\bibitem [{\citenamefont {Hedley}, \citenamefont {Ruseckas},\ and\
  \citenamefont {Samuel}(2017)}]{sjs-ref11}%
  \BibitemOpen
  \bibfield  {author} {\bibinfo {author} {\bibfnamefont {G.~J.}\ \bibnamefont
  {Hedley}}, \bibinfo {author} {\bibfnamefont {A.}~\bibnamefont {Ruseckas}}, \
  and\ \bibinfo {author} {\bibfnamefont {I.~D.~W.}\ \bibnamefont {Samuel}},\
  }\href@noop {} {\bibfield  {journal} {\bibinfo  {journal} {Chem. Rev.}\
  }\textbf {\bibinfo {volume} {117}},\ \bibinfo {pages} {796} (\bibinfo {year}
  {2017})}\BibitemShut {NoStop}%
\bibitem [{\citenamefont {Niklas}\ and\ \citenamefont
  {Poluektov}(2017)}]{sjs-ref12}%
  \BibitemOpen
  \bibfield  {author} {\bibinfo {author} {\bibfnamefont {J.}~\bibnamefont
  {Niklas}}\ and\ \bibinfo {author} {\bibfnamefont {O.~G.}\ \bibnamefont
  {Poluektov}},\ }\href@noop {} {\bibfield  {journal} {\bibinfo  {journal}
  {Adv. Energy Mat.}\ }\textbf {\bibinfo {volume} {7}},\ \bibinfo {pages}
  {1602226} (\bibinfo {year} {2017})}\BibitemShut {NoStop}%
\bibitem [{\citenamefont {Hart}\ \emph {et~al.}(1992)\citenamefont {Hart},
  \citenamefont {Rappe}, \citenamefont {Gorun},\ and\ \citenamefont
  {Upton}}]{hfh-1992}%
  \BibitemOpen
  \bibfield  {author} {\bibinfo {author} {\bibfnamefont {J.~R.}\ \bibnamefont
  {Hart}}, \bibinfo {author} {\bibfnamefont {A.~K.}\ \bibnamefont {Rappe}},
  \bibinfo {author} {\bibfnamefont {S.~M.}\ \bibnamefont {Gorun}}, \ and\
  \bibinfo {author} {\bibfnamefont {T.~H.}\ \bibnamefont {Upton}},\ }\href@noop
  {} {\bibfield  {journal} {\bibinfo  {journal} {J. Phys. Chem.}\ }\textbf
  {\bibinfo {volume} {96}},\ \bibinfo {pages} {6264} (\bibinfo {year}
  {1992})}\BibitemShut {NoStop}%
\bibitem [{\citenamefont {Slipchenko}\ and\ \citenamefont
  {Krylov}(2002)}]{ch2_krylov}%
  \BibitemOpen
  \bibfield  {author} {\bibinfo {author} {\bibfnamefont {L.~V.}\ \bibnamefont
  {Slipchenko}}\ and\ \bibinfo {author} {\bibfnamefont {A.~I.}\ \bibnamefont
  {Krylov}},\ }\href@noop {} {\bibfield  {journal} {\bibinfo  {journal} {J.
  Chem. Phys.}\ }\textbf {\bibinfo {volume} {117}},\ \bibinfo {pages} {4694}
  (\bibinfo {year} {2002})}\BibitemShut {NoStop}%
\bibitem [{\citenamefont {Li}\ and\ \citenamefont
  {Paldus}(2008)}]{ch2-rmrccsdt}%
  \BibitemOpen
  \bibfield  {author} {\bibinfo {author} {\bibfnamefont {X.}~\bibnamefont
  {Li}}\ and\ \bibinfo {author} {\bibfnamefont {J.}~\bibnamefont {Paldus}},\
  }\href@noop {} {\bibfield  {journal} {\bibinfo  {journal} {J. Chem. Phys.}\
  }\textbf {\bibinfo {volume} {129}},\ \bibinfo {pages} {174101} (\bibinfo
  {year} {2008})}\BibitemShut {NoStop}%
\bibitem [{\citenamefont {Demel}\ \emph {et~al.}(2008)\citenamefont {Demel},
  \citenamefont {Shamasundar}, \citenamefont {Kong},\ and\ \citenamefont
  {Nooijen}}]{icmrcc6}%
  \BibitemOpen
  \bibfield  {author} {\bibinfo {author} {\bibfnamefont {O.}~\bibnamefont
  {Demel}}, \bibinfo {author} {\bibfnamefont {K.~R.}\ \bibnamefont
  {Shamasundar}}, \bibinfo {author} {\bibfnamefont {L.}~\bibnamefont {Kong}}, \
  and\ \bibinfo {author} {\bibfnamefont {M.}~\bibnamefont {Nooijen}},\
  }\href@noop {} {\bibfield  {journal} {\bibinfo  {journal} {J. Phys. Chem. A}\
  }\textbf {\bibinfo {volume} {112}},\ \bibinfo {pages} {11895} (\bibinfo
  {year} {2008})}\BibitemShut {NoStop}%
\bibitem [{\citenamefont {Saito}\ \emph {et~al.}(2011)\citenamefont {Saito},
  \citenamefont {Nishihara}, \citenamefont {Yamanaka}, \citenamefont
  {Kitagawa}, \citenamefont {Kawakami}, \citenamefont {Yamada}, \citenamefont
  {Isobe}, \citenamefont {Okumura},\ and\ \citenamefont
  {Yamaguchi}}]{BiradicalGeom}%
  \BibitemOpen
  \bibfield  {author} {\bibinfo {author} {\bibfnamefont {T.}~\bibnamefont
  {Saito}}, \bibinfo {author} {\bibfnamefont {S.}~\bibnamefont {Nishihara}},
  \bibinfo {author} {\bibfnamefont {S.}~\bibnamefont {Yamanaka}}, \bibinfo
  {author} {\bibfnamefont {Y.}~\bibnamefont {Kitagawa}}, \bibinfo {author}
  {\bibfnamefont {T.}~\bibnamefont {Kawakami}}, \bibinfo {author}
  {\bibfnamefont {S.}~\bibnamefont {Yamada}}, \bibinfo {author} {\bibfnamefont
  {H.}~\bibnamefont {Isobe}}, \bibinfo {author} {\bibfnamefont
  {M.}~\bibnamefont {Okumura}}, \ and\ \bibinfo {author} {\bibfnamefont
  {K.}~\bibnamefont {Yamaguchi}},\ }\href@noop {} {\bibfield  {journal}
  {\bibinfo  {journal} {Theor. Chem. Acc.}\ }\textbf {\bibinfo {volume}
  {130}},\ \bibinfo {pages} {749} (\bibinfo {year} {2011})}\BibitemShut
  {NoStop}%
\bibitem [{\citenamefont {Ess}\ \emph {et~al.}(2011)\citenamefont {Ess},
  \citenamefont {Johnson}, \citenamefont {Hu},\ and\ \citenamefont
  {Yang}}]{sjs-ref15}%
  \BibitemOpen
  \bibfield  {author} {\bibinfo {author} {\bibfnamefont {D.~H.}\ \bibnamefont
  {Ess}}, \bibinfo {author} {\bibfnamefont {E.~R.}\ \bibnamefont {Johnson}},
  \bibinfo {author} {\bibfnamefont {X.}~\bibnamefont {Hu}}, \ and\ \bibinfo
  {author} {\bibfnamefont {W.}~\bibnamefont {Yang}},\ }\href@noop {} {\bibfield
   {journal} {\bibinfo  {journal} {J. Phys. Chem. A}\ }\textbf {\bibinfo
  {volume} {115}},\ \bibinfo {pages} {76} (\bibinfo {year} {2011})}\BibitemShut
  {NoStop}%
\bibitem [{\citenamefont {Abe}(2013)}]{sjs-ref18}%
  \BibitemOpen
  \bibfield  {author} {\bibinfo {author} {\bibfnamefont {M.}~\bibnamefont
  {Abe}},\ }\href@noop {} {\bibfield  {journal} {\bibinfo  {journal} {Chem.
  Rev.}\ }\textbf {\bibinfo {volume} {113}},\ \bibinfo {pages} {7011} (\bibinfo
  {year} {2013})}\BibitemShut {NoStop}%
\bibitem [{\citenamefont {Garza}, \citenamefont {Jim{\'e}nez-Hoyos},\ and\
  \citenamefont {Scuseria}(2014)}]{sjs-ref16}%
  \BibitemOpen
  \bibfield  {author} {\bibinfo {author} {\bibfnamefont {A.~J.}\ \bibnamefont
  {Garza}}, \bibinfo {author} {\bibfnamefont {C.~A.}\ \bibnamefont
  {Jim{\'e}nez-Hoyos}}, \ and\ \bibinfo {author} {\bibfnamefont {G.~E.}\
  \bibnamefont {Scuseria}},\ }\href@noop {} {\bibfield  {journal} {\bibinfo
  {journal} {J. Chem. Phys.}\ }\textbf {\bibinfo {volume} {140}},\ \bibinfo
  {pages} {244102} (\bibinfo {year} {2014})}\BibitemShut {NoStop}%
\bibitem [{\citenamefont {Ibeji}\ and\ \citenamefont
  {Ghosh}(2015)}]{sjs-ref14}%
  \BibitemOpen
  \bibfield  {author} {\bibinfo {author} {\bibfnamefont {C.~U.}\ \bibnamefont
  {Ibeji}}\ and\ \bibinfo {author} {\bibfnamefont {D.}~\bibnamefont {Ghosh}},\
  }\href@noop {} {\bibfield  {journal} {\bibinfo  {journal} {Phys. Chem. Chem.
  Phys.}\ }\textbf {\bibinfo {volume} {17}},\ \bibinfo {pages} {9849} (\bibinfo
  {year} {2015})}\BibitemShut {NoStop}%
\bibitem [{\citenamefont {Ajala}, \citenamefont {Shen},\ and\ \citenamefont
  {Piecuch}(2017)}]{aj-js-pp-jpca2017}%
  \BibitemOpen
  \bibfield  {author} {\bibinfo {author} {\bibfnamefont {A.~O.}\ \bibnamefont
  {Ajala}}, \bibinfo {author} {\bibfnamefont {J.}~\bibnamefont {Shen}}, \ and\
  \bibinfo {author} {\bibfnamefont {P.}~\bibnamefont {Piecuch}},\ }\href@noop
  {} {\bibfield  {journal} {\bibinfo  {journal} {J. Phys. Chem. A}\ }\textbf
  {\bibinfo {volume} {121}},\ \bibinfo {pages} {3469} (\bibinfo {year}
  {2017})}\BibitemShut {NoStop}%
\bibitem [{\citenamefont {Stoneburner}\ \emph {et~al.}(2017)\citenamefont
  {Stoneburner}, \citenamefont {Shen}, \citenamefont {Ajala}, \citenamefont
  {Piecuch}, \citenamefont {Truhlar},\ and\ \citenamefont
  {Gagliardi}}]{stoneburner-js-jcp2017}%
  \BibitemOpen
  \bibfield  {author} {\bibinfo {author} {\bibfnamefont {S.~J.}\ \bibnamefont
  {Stoneburner}}, \bibinfo {author} {\bibfnamefont {J.}~\bibnamefont {Shen}},
  \bibinfo {author} {\bibfnamefont {A.~O.}\ \bibnamefont {Ajala}}, \bibinfo
  {author} {\bibfnamefont {P.}~\bibnamefont {Piecuch}}, \bibinfo {author}
  {\bibfnamefont {D.~G.}\ \bibnamefont {Truhlar}}, \ and\ \bibinfo {author}
  {\bibfnamefont {L.}~\bibnamefont {Gagliardi}},\ }\href@noop {} {\bibfield
  {journal} {\bibinfo  {journal} {J. Chem. Phys.}\ }\textbf {\bibinfo {volume}
  {147}},\ \bibinfo {pages} {164120} (\bibinfo {year} {2017})}\BibitemShut
  {NoStop}%
\bibitem [{\citenamefont {Shen}\ and\ \citenamefont
  {Piecuch}(2021)}]{js-pp-dea2021}%
  \BibitemOpen
  \bibfield  {author} {\bibinfo {author} {\bibfnamefont {J.}~\bibnamefont
  {Shen}}\ and\ \bibinfo {author} {\bibfnamefont {P.}~\bibnamefont {Piecuch}},\
  }\href@noop {} {\bibfield  {journal} {\bibinfo  {journal} {Mol. Phys.}\
  }\textbf {\bibinfo {volume} {119}},\ \bibinfo {pages} {e1966534} (\bibinfo
  {year} {2021})}\BibitemShut {NoStop}%
\bibitem [{\citenamefont {Gulania}\ \emph {et~al.}(2021)\citenamefont
  {Gulania}, \citenamefont {Kj{\o}nstad}, \citenamefont {Stanton},
  \citenamefont {Koch},\ and\ \citenamefont {Krylov}}]{dipea7}%
  \BibitemOpen
  \bibfield  {author} {\bibinfo {author} {\bibfnamefont {S.}~\bibnamefont
  {Gulania}}, \bibinfo {author} {\bibfnamefont {E.~F.}\ \bibnamefont
  {Kj{\o}nstad}}, \bibinfo {author} {\bibfnamefont {J.~F.}\ \bibnamefont
  {Stanton}}, \bibinfo {author} {\bibfnamefont {H.}~\bibnamefont {Koch}}, \
  and\ \bibinfo {author} {\bibfnamefont {A.~I.}\ \bibnamefont {Krylov}},\
  }\href@noop {} {\bibfield  {journal} {\bibinfo  {journal} {J. Chem. Phys.}\
  }\textbf {\bibinfo {volume} {154}},\ \bibinfo {pages} {114115} (\bibinfo
  {year} {2021})}\BibitemShut {NoStop}%
\bibitem [{\citenamefont {Nooijen}\ and\ \citenamefont
  {Bartlett}(1997)}]{dipea1}%
  \BibitemOpen
  \bibfield  {author} {\bibinfo {author} {\bibfnamefont {M.}~\bibnamefont
  {Nooijen}}\ and\ \bibinfo {author} {\bibfnamefont {R.~J.}\ \bibnamefont
  {Bartlett}},\ }\href@noop {} {\bibfield  {journal} {\bibinfo  {journal} {J.
  Chem. Phys.}\ }\textbf {\bibinfo {volume} {106}},\ \bibinfo {pages} {6441}
  (\bibinfo {year} {1997})}\BibitemShut {NoStop}%
\bibitem [{\citenamefont {Wladyslawski}\ and\ \citenamefont
  {Nooijen}(2002)}]{dipea2}%
  \BibitemOpen
  \bibfield  {author} {\bibinfo {author} {\bibfnamefont {M.}~\bibnamefont
  {Wladyslawski}}\ and\ \bibinfo {author} {\bibfnamefont {M.}~\bibnamefont
  {Nooijen}},\ }in\ \href@noop {} {\emph {\bibinfo {booktitle} {Low-Lying
  Potential Energy Surfaces}}},\ \bibinfo {series} {ACS Symposium Series},
  Vol.\ \bibinfo {volume} {828},\ \bibinfo {editor} {edited by\ \bibinfo
  {editor} {\bibfnamefont {M.~R.}\ \bibnamefont {Hoffmann}}\ and\ \bibinfo
  {editor} {\bibfnamefont {K.~G.}\ \bibnamefont {Dyall}}}\ (\bibinfo
  {publisher} {American Chemical Society},\ \bibinfo {address} {{Washington,
  D.C.}},\ \bibinfo {year} {2002})\ pp.\ \bibinfo {pages} {65--92}\BibitemShut
  {NoStop}%
\bibitem [{\citenamefont {Nooijen}(2002)}]{dipea3}%
  \BibitemOpen
  \bibfield  {author} {\bibinfo {author} {\bibfnamefont {M.}~\bibnamefont
  {Nooijen}},\ }\href@noop {} {\bibfield  {journal} {\bibinfo  {journal} {Int.
  J. Mol. Sci.}\ }\textbf {\bibinfo {volume} {3}},\ \bibinfo {pages} {656}
  (\bibinfo {year} {2002})}\BibitemShut {NoStop}%
\bibitem [{\citenamefont {Sattelmeyer}, \citenamefont {Schaefer},\ and\
  \citenamefont {Stanton}(2003)}]{dip-stanton}%
  \BibitemOpen
  \bibfield  {author} {\bibinfo {author} {\bibfnamefont {K.~W.}\ \bibnamefont
  {Sattelmeyer}}, \bibinfo {author} {\bibfnamefont {H.~F.}\ \bibnamefont
  {Schaefer}, \bibfnamefont {III}}, \ and\ \bibinfo {author} {\bibfnamefont
  {J.~F.}\ \bibnamefont {Stanton}},\ }\href@noop {} {\bibfield  {journal}
  {\bibinfo  {journal} {Chem. Phys. Lett.}\ }\textbf {\bibinfo {volume}
  {378}},\ \bibinfo {pages} {42} (\bibinfo {year} {2003})}\BibitemShut
  {NoStop}%
\bibitem [{\citenamefont {Musia{\l}}, \citenamefont {Perera},\ and\
  \citenamefont {Bartlett}(2011)}]{dipea5}%
  \BibitemOpen
  \bibfield  {author} {\bibinfo {author} {\bibfnamefont {M.}~\bibnamefont
  {Musia{\l}}}, \bibinfo {author} {\bibfnamefont {A.}~\bibnamefont {Perera}}, \
  and\ \bibinfo {author} {\bibfnamefont {R.~J.}\ \bibnamefont {Bartlett}},\
  }\href@noop {} {\bibfield  {journal} {\bibinfo  {journal} {J. Chem. Phys.}\
  }\textbf {\bibinfo {volume} {134}},\ \bibinfo {pages} {114108} (\bibinfo
  {year} {2011})}\BibitemShut {NoStop}%
\bibitem [{\citenamefont {Musia{\l}}, \citenamefont {Kucharski},\ and\
  \citenamefont {Bartlett}(2011)}]{dipea6}%
  \BibitemOpen
  \bibfield  {author} {\bibinfo {author} {\bibfnamefont {M.}~\bibnamefont
  {Musia{\l}}}, \bibinfo {author} {\bibfnamefont {S.~A.}\ \bibnamefont
  {Kucharski}}, \ and\ \bibinfo {author} {\bibfnamefont {R.~J.}\ \bibnamefont
  {Bartlett}},\ }\href@noop {} {\bibfield  {journal} {\bibinfo  {journal} {J.
  Chem. Theory Comput.}\ }\textbf {\bibinfo {volume} {7}},\ \bibinfo {pages}
  {3088} (\bibinfo {year} {2011})}\BibitemShut {NoStop}%
\bibitem [{\citenamefont {Ku{\' s}}\ and\ \citenamefont
  {Krylov}(2011)}]{kus-krylov-2011}%
  \BibitemOpen
  \bibfield  {author} {\bibinfo {author} {\bibfnamefont {T.}~\bibnamefont
  {Ku{\' s}}}\ and\ \bibinfo {author} {\bibfnamefont {A.~I.}\ \bibnamefont
  {Krylov}},\ }\href@noop {} {\bibfield  {journal} {\bibinfo  {journal} {J.
  Chem. Phys.}\ }\textbf {\bibinfo {volume} {135}},\ \bibinfo {pages} {084109}
  (\bibinfo {year} {2011})}\BibitemShut {NoStop}%
\bibitem [{\citenamefont {Ku{\' s}}\ and\ \citenamefont
  {Krylov}(2012)}]{kus-krylov-2012}%
  \BibitemOpen
  \bibfield  {author} {\bibinfo {author} {\bibfnamefont {T.}~\bibnamefont
  {Ku{\' s}}}\ and\ \bibinfo {author} {\bibfnamefont {A.~I.}\ \bibnamefont
  {Krylov}},\ }\href@noop {} {\bibfield  {journal} {\bibinfo  {journal} {J.
  Chem. Phys.}\ }\textbf {\bibinfo {volume} {136}},\ \bibinfo {pages} {244109}
  (\bibinfo {year} {2012})}\BibitemShut {NoStop}%
\bibitem [{\citenamefont {Shen}\ and\ \citenamefont
  {Piecuch}(2013)}]{jspp-dea-dip-2013}%
  \BibitemOpen
  \bibfield  {author} {\bibinfo {author} {\bibfnamefont {J.}~\bibnamefont
  {Shen}}\ and\ \bibinfo {author} {\bibfnamefont {P.}~\bibnamefont {Piecuch}},\
  }\href@noop {} {\bibfield  {journal} {\bibinfo  {journal} {J. Chem. Phys.}\
  }\textbf {\bibinfo {volume} {138}},\ \bibinfo {pages} {194102} (\bibinfo
  {year} {2013})}\BibitemShut {NoStop}%
\bibitem [{\citenamefont {Shen}\ and\ \citenamefont
  {Piecuch}(2014)}]{jspp-dea-dip-2014}%
  \BibitemOpen
  \bibfield  {author} {\bibinfo {author} {\bibfnamefont {J.}~\bibnamefont
  {Shen}}\ and\ \bibinfo {author} {\bibfnamefont {P.}~\bibnamefont {Piecuch}},\
  }\href@noop {} {\bibfield  {journal} {\bibinfo  {journal} {Mol. Phys.}\
  }\textbf {\bibinfo {volume} {112}},\ \bibinfo {pages} {868} (\bibinfo {year}
  {2014})}\BibitemShut {NoStop}%
\bibitem [{\citenamefont {Piecuch}, \citenamefont {Gour},\ and\ \citenamefont
  {W{\l}och}(2009)}]{crccl_ijqc2}%
  \BibitemOpen
  \bibfield  {author} {\bibinfo {author} {\bibfnamefont {P.}~\bibnamefont
  {Piecuch}}, \bibinfo {author} {\bibfnamefont {J.~R.}\ \bibnamefont {Gour}}, \
  and\ \bibinfo {author} {\bibfnamefont {M.}~\bibnamefont {W{\l}och}},\
  }\href@noop {} {\bibfield  {journal} {\bibinfo  {journal} {Int. J. Quantum
  Chem.}\ }\textbf {\bibinfo {volume} {109}},\ \bibinfo {pages} {3268}
  (\bibinfo {year} {2009})}\BibitemShut {NoStop}%
\bibitem [{\citenamefont {Fradelos}\ \emph {et~al.}(2011)\citenamefont
  {Fradelos}, \citenamefont {Lutz}, \citenamefont {Weso{\l}owski},
  \citenamefont {Piecuch},\ and\ \citenamefont {W{\l}och}}]{7hq}%
  \BibitemOpen
  \bibfield  {author} {\bibinfo {author} {\bibfnamefont {G.}~\bibnamefont
  {Fradelos}}, \bibinfo {author} {\bibfnamefont {J.~J.}\ \bibnamefont {Lutz}},
  \bibinfo {author} {\bibfnamefont {T.~A.}\ \bibnamefont {Weso{\l}owski}},
  \bibinfo {author} {\bibfnamefont {P.}~\bibnamefont {Piecuch}}, \ and\
  \bibinfo {author} {\bibfnamefont {M.}~\bibnamefont {W{\l}och}},\ }\href@noop
  {} {\bibfield  {journal} {\bibinfo  {journal} {J. Chem. Theory Comput.}\
  }\textbf {\bibinfo {volume} {7}},\ \bibinfo {pages} {1647} (\bibinfo {year}
  {2011})}\BibitemShut {NoStop}%
\bibitem [{\citenamefont {Jankowski}, \citenamefont {Paldus},\ and\
  \citenamefont {Piecuch}(1991)}]{moments}%
  \BibitemOpen
  \bibfield  {author} {\bibinfo {author} {\bibfnamefont {K.}~\bibnamefont
  {Jankowski}}, \bibinfo {author} {\bibfnamefont {J.}~\bibnamefont {Paldus}}, \
  and\ \bibinfo {author} {\bibfnamefont {P.}~\bibnamefont {Piecuch}},\
  }\href@noop {} {\bibfield  {journal} {\bibinfo  {journal} {Theor. Chim.
  Acta}\ }\textbf {\bibinfo {volume} {80}},\ \bibinfo {pages} {223} (\bibinfo
  {year} {1991})}\BibitemShut {NoStop}%
\bibitem [{\citenamefont {Dunning}(1989)}]{ccpvnz}%
  \BibitemOpen
  \bibfield  {author} {\bibinfo {author} {\bibfnamefont {T.~H.}\ \bibnamefont
  {Dunning}, \bibfnamefont {Jr.}},\ }\href@noop {} {\bibfield  {journal}
  {\bibinfo  {journal} {J. Chem. Phys.}\ }\textbf {\bibinfo {volume} {90}},\
  \bibinfo {pages} {1007} (\bibinfo {year} {1989})}\BibitemShut {NoStop}%
\bibitem [{\citenamefont {Kendall}, \citenamefont {Dunning},\ and\
  \citenamefont {Harrison}(1992)}]{augccpvnz}%
  \BibitemOpen
  \bibfield  {author} {\bibinfo {author} {\bibfnamefont {R.~A.}\ \bibnamefont
  {Kendall}}, \bibinfo {author} {\bibfnamefont {T.~H.}\ \bibnamefont {Dunning},
  \bibfnamefont {Jr.}}, \ and\ \bibinfo {author} {\bibfnamefont {R.~J.}\
  \bibnamefont {Harrison}},\ }\href@noop {} {\bibfield  {journal} {\bibinfo
  {journal} {J. Chem. Phys.}\ }\textbf {\bibinfo {volume} {96}},\ \bibinfo
  {pages} {6796} (\bibinfo {year} {1992})}\BibitemShut {NoStop}%
\bibitem [{\citenamefont {Hehre}, \citenamefont {Ditchfield},\ and\
  \citenamefont {Pople}(1972)}]{631gs}%
  \BibitemOpen
  \bibfield  {author} {\bibinfo {author} {\bibfnamefont {W.~J.}\ \bibnamefont
  {Hehre}}, \bibinfo {author} {\bibfnamefont {R.}~\bibnamefont {Ditchfield}}, \
  and\ \bibinfo {author} {\bibfnamefont {J.~A.}\ \bibnamefont {Pople}},\
  }\href@noop {} {\bibfield  {journal} {\bibinfo  {journal} {J. Chem. Phys.}\
  }\textbf {\bibinfo {volume} {56}},\ \bibinfo {pages} {2257} (\bibinfo {year}
  {1972})}\BibitemShut {NoStop}%
\bibitem [{\citenamefont {Hariharan}\ and\ \citenamefont {Pople}(1973)}]{ss}%
  \BibitemOpen
  \bibfield  {author} {\bibinfo {author} {\bibfnamefont {P.~C.}\ \bibnamefont
  {Hariharan}}\ and\ \bibinfo {author} {\bibfnamefont {J.~A.}\ \bibnamefont
  {Pople}},\ }\href@noop {} {\bibfield  {journal} {\bibinfo  {journal} {Theor.
  Chim. Acta}\ }\textbf {\bibinfo {volume} {28}},\ \bibinfo {pages} {213}
  (\bibinfo {year} {1973})}\BibitemShut {NoStop}%
\bibitem [{\citenamefont {Schmidt}\ \emph {et~al.}(1993)\citenamefont
  {Schmidt}, \citenamefont {Baldridge}, \citenamefont {Boatz}, \citenamefont
  {Elbert}, \citenamefont {Gordon}, \citenamefont {Jensen}, \citenamefont
  {Koseki}, \citenamefont {Matsunaga}, \citenamefont {Nguyen}, \citenamefont
  {Su}, \citenamefont {Windus}, \citenamefont {Dupuis},\ and\ \citenamefont
  {Montgomery}}]{gamess}%
  \BibitemOpen
  \bibfield  {author} {\bibinfo {author} {\bibfnamefont {M.~W.}\ \bibnamefont
  {Schmidt}}, \bibinfo {author} {\bibfnamefont {K.~K.}\ \bibnamefont
  {Baldridge}}, \bibinfo {author} {\bibfnamefont {J.~A.}\ \bibnamefont
  {Boatz}}, \bibinfo {author} {\bibfnamefont {S.~T.}\ \bibnamefont {Elbert}},
  \bibinfo {author} {\bibfnamefont {M.~S.}\ \bibnamefont {Gordon}}, \bibinfo
  {author} {\bibfnamefont {J.~H.}\ \bibnamefont {Jensen}}, \bibinfo {author}
  {\bibfnamefont {S.}~\bibnamefont {Koseki}}, \bibinfo {author} {\bibfnamefont
  {N.}~\bibnamefont {Matsunaga}}, \bibinfo {author} {\bibfnamefont {K.~A.}\
  \bibnamefont {Nguyen}}, \bibinfo {author} {\bibfnamefont {S.}~\bibnamefont
  {Su}}, \bibinfo {author} {\bibfnamefont {T.~L.}\ \bibnamefont {Windus}},
  \bibinfo {author} {\bibfnamefont {M.}~\bibnamefont {Dupuis}}, \ and\ \bibinfo
  {author} {\bibfnamefont {J.~A.}\ \bibnamefont {Montgomery}, \bibfnamefont
  {Jr.}},\ }\href@noop {} {\bibfield  {journal} {\bibinfo  {journal} {J.
  Comput. Chem.}\ }\textbf {\bibinfo {volume} {14}},\ \bibinfo {pages} {1347}
  (\bibinfo {year} {1993})}\BibitemShut {NoStop}%
\bibitem [{\citenamefont {Barca}\ \emph {et~al.}(2020)\citenamefont {Barca},
  \citenamefont {Bertoni}, \citenamefont {Carrington}, \citenamefont {Datta},
  \citenamefont {De~Silva}, \citenamefont {Deustua}, \citenamefont {Fedorov},
  \citenamefont {Gour}, \citenamefont {Gunina}, \citenamefont {Guidez},
  \citenamefont {Harville}, \citenamefont {Irle}, \citenamefont {Ivanic},
  \citenamefont {Kowalski}, \citenamefont {Leang}, \citenamefont {Li},
  \citenamefont {Li}, \citenamefont {Lutz}, \citenamefont {Magoulas},
  \citenamefont {Mato}, \citenamefont {Mironov}, \citenamefont {Nakata},
  \citenamefont {Pham}, \citenamefont {Piecuch}, \citenamefont {Poole},
  \citenamefont {Pruitt}, \citenamefont {Rendell}, \citenamefont {Roskop},
  \citenamefont {Ruedenberg}, \citenamefont {Sattasathuchana}, \citenamefont
  {Schmidt}, \citenamefont {Shen}, \citenamefont {Slipchenko}, \citenamefont
  {Sosonkina}, \citenamefont {Sundriyal}, \citenamefont {Tiwari}, \citenamefont
  {Vallejo}, \citenamefont {Westheimer}, \citenamefont {W{\l}och},
  \citenamefont {Xu}, \citenamefont {Zahariev},\ and\ \citenamefont
  {Gordon}}]{gamess2020}%
  \BibitemOpen
  \bibfield  {author} {\bibinfo {author} {\bibfnamefont {G.~M.~J.}\
  \bibnamefont {Barca}}, \bibinfo {author} {\bibfnamefont {C.}~\bibnamefont
  {Bertoni}}, \bibinfo {author} {\bibfnamefont {L.}~\bibnamefont {Carrington}},
  \bibinfo {author} {\bibfnamefont {D.}~\bibnamefont {Datta}}, \bibinfo
  {author} {\bibfnamefont {N.}~\bibnamefont {De~Silva}}, \bibinfo {author}
  {\bibfnamefont {J.~E.}\ \bibnamefont {Deustua}}, \bibinfo {author}
  {\bibfnamefont {D.~G.}\ \bibnamefont {Fedorov}}, \bibinfo {author}
  {\bibfnamefont {J.~R.}\ \bibnamefont {Gour}}, \bibinfo {author}
  {\bibfnamefont {A.~O.}\ \bibnamefont {Gunina}}, \bibinfo {author}
  {\bibfnamefont {E.}~\bibnamefont {Guidez}}, \bibinfo {author} {\bibfnamefont
  {T.}~\bibnamefont {Harville}}, \bibinfo {author} {\bibfnamefont
  {S.}~\bibnamefont {Irle}}, \bibinfo {author} {\bibfnamefont {J.}~\bibnamefont
  {Ivanic}}, \bibinfo {author} {\bibfnamefont {K.}~\bibnamefont {Kowalski}},
  \bibinfo {author} {\bibfnamefont {S.~S.}\ \bibnamefont {Leang}}, \bibinfo
  {author} {\bibfnamefont {H.}~\bibnamefont {Li}}, \bibinfo {author}
  {\bibfnamefont {W.}~\bibnamefont {Li}}, \bibinfo {author} {\bibfnamefont
  {J.~J.}\ \bibnamefont {Lutz}}, \bibinfo {author} {\bibfnamefont
  {I.}~\bibnamefont {Magoulas}}, \bibinfo {author} {\bibfnamefont
  {J.}~\bibnamefont {Mato}}, \bibinfo {author} {\bibfnamefont {V.}~\bibnamefont
  {Mironov}}, \bibinfo {author} {\bibfnamefont {H.}~\bibnamefont {Nakata}},
  \bibinfo {author} {\bibfnamefont {B.~Q.}\ \bibnamefont {Pham}}, \bibinfo
  {author} {\bibfnamefont {P.}~\bibnamefont {Piecuch}}, \bibinfo {author}
  {\bibfnamefont {D.}~\bibnamefont {Poole}}, \bibinfo {author} {\bibfnamefont
  {S.~R.}\ \bibnamefont {Pruitt}}, \bibinfo {author} {\bibfnamefont {A.~P.}\
  \bibnamefont {Rendell}}, \bibinfo {author} {\bibfnamefont {L.~B.}\
  \bibnamefont {Roskop}}, \bibinfo {author} {\bibfnamefont {K.}~\bibnamefont
  {Ruedenberg}}, \bibinfo {author} {\bibfnamefont {T.}~\bibnamefont
  {Sattasathuchana}}, \bibinfo {author} {\bibfnamefont {M.~W.}\ \bibnamefont
  {Schmidt}}, \bibinfo {author} {\bibfnamefont {J.}~\bibnamefont {Shen}},
  \bibinfo {author} {\bibfnamefont {L.}~\bibnamefont {Slipchenko}}, \bibinfo
  {author} {\bibfnamefont {M.}~\bibnamefont {Sosonkina}}, \bibinfo {author}
  {\bibfnamefont {V.}~\bibnamefont {Sundriyal}}, \bibinfo {author}
  {\bibfnamefont {A.}~\bibnamefont {Tiwari}}, \bibinfo {author} {\bibfnamefont
  {J.~L.~G.}\ \bibnamefont {Vallejo}}, \bibinfo {author} {\bibfnamefont
  {B.}~\bibnamefont {Westheimer}}, \bibinfo {author} {\bibfnamefont
  {M.}~\bibnamefont {W{\l}och}}, \bibinfo {author} {\bibfnamefont
  {P.}~\bibnamefont {Xu}}, \bibinfo {author} {\bibfnamefont {F.}~\bibnamefont
  {Zahariev}}, \ and\ \bibinfo {author} {\bibfnamefont {M.~S.}\ \bibnamefont
  {Gordon}},\ }\href@noop {} {\bibfield  {journal} {\bibinfo  {journal} {J.
  Chem. Phys.}\ }\textbf {\bibinfo {volume} {152}},\ \bibinfo {pages} {154102}
  (\bibinfo {year} {2020})}\BibitemShut {NoStop}%
\bibitem [{\citenamefont {Spencer}\ \emph {et~al.}(2015)\citenamefont
  {Spencer}, \citenamefont {Blunt}, \citenamefont {Vigor}, \citenamefont
  {Malone}, \citenamefont {Foulkes}, \citenamefont {Shepherd},\ and\
  \citenamefont {Thom}}]{hande-jors-2015}%
  \BibitemOpen
  \bibfield  {author} {\bibinfo {author} {\bibfnamefont {J.~S.}\ \bibnamefont
  {Spencer}}, \bibinfo {author} {\bibfnamefont {N.~S.}\ \bibnamefont {Blunt}},
  \bibinfo {author} {\bibfnamefont {W.~A.}\ \bibnamefont {Vigor}}, \bibinfo
  {author} {\bibfnamefont {F.~D.}\ \bibnamefont {Malone}}, \bibinfo {author}
  {\bibfnamefont {W.~M.~C.}\ \bibnamefont {Foulkes}}, \bibinfo {author}
  {\bibfnamefont {J.~J.}\ \bibnamefont {Shepherd}}, \ and\ \bibinfo {author}
  {\bibfnamefont {A.~J.~W.}\ \bibnamefont {Thom}},\ }\href@noop {} {\bibfield
  {journal} {\bibinfo  {journal} {J. Open Res. Softw.}\ }\textbf {\bibinfo
  {volume} {3}},\ \bibinfo {pages} {e9} (\bibinfo {year} {2015})}\BibitemShut
  {NoStop}%
\bibitem [{\citenamefont {Spencer}\ \emph {et~al.}(2019)\citenamefont
  {Spencer}, \citenamefont {Blunt}, \citenamefont {Choi}, \citenamefont
  {Etrych}, \citenamefont {Filip}, \citenamefont {Foulkes}, \citenamefont
  {Franklin}, \citenamefont {Handley}, \citenamefont {Malone}, \citenamefont
  {Neufeld}, \citenamefont {Di~Remigio}, \citenamefont {Rogers}, \citenamefont
  {Scott}, \citenamefont {Shepherd}, \citenamefont {Vigor}, \citenamefont
  {Weston}, \citenamefont {Xu},\ and\ \citenamefont {Thom}}]{hande-jctc-2019}%
  \BibitemOpen
  \bibfield  {author} {\bibinfo {author} {\bibfnamefont {J.~S.}\ \bibnamefont
  {Spencer}}, \bibinfo {author} {\bibfnamefont {N.~S.}\ \bibnamefont {Blunt}},
  \bibinfo {author} {\bibfnamefont {S.}~\bibnamefont {Choi}}, \bibinfo {author}
  {\bibfnamefont {J.}~\bibnamefont {Etrych}}, \bibinfo {author} {\bibfnamefont
  {M.-A.}\ \bibnamefont {Filip}}, \bibinfo {author} {\bibfnamefont {W.~M.~C.}\
  \bibnamefont {Foulkes}}, \bibinfo {author} {\bibfnamefont {R.~S.~T.}\
  \bibnamefont {Franklin}}, \bibinfo {author} {\bibfnamefont {W.~J.}\
  \bibnamefont {Handley}}, \bibinfo {author} {\bibfnamefont {F.~D.}\
  \bibnamefont {Malone}}, \bibinfo {author} {\bibfnamefont {V.~A.}\
  \bibnamefont {Neufeld}}, \bibinfo {author} {\bibfnamefont {R.}~\bibnamefont
  {Di~Remigio}}, \bibinfo {author} {\bibfnamefont {T.~W.}\ \bibnamefont
  {Rogers}}, \bibinfo {author} {\bibfnamefont {C.~J.~C.}\ \bibnamefont
  {Scott}}, \bibinfo {author} {\bibfnamefont {J.~J.}\ \bibnamefont {Shepherd}},
  \bibinfo {author} {\bibfnamefont {W.~A.}\ \bibnamefont {Vigor}}, \bibinfo
  {author} {\bibfnamefont {J.}~\bibnamefont {Weston}}, \bibinfo {author}
  {\bibfnamefont {R.}~\bibnamefont {Xu}}, \ and\ \bibinfo {author}
  {\bibfnamefont {A.~J.~W.}\ \bibnamefont {Thom}},\ }\href@noop {} {\bibfield
  {journal} {\bibinfo  {journal} {J. Chem. Theory Comput.}\ }\textbf {\bibinfo
  {volume} {15}},\ \bibinfo {pages} {1728} (\bibinfo {year}
  {2019})}\BibitemShut {NoStop}%
\bibitem [{\citenamefont {Dunning}(1971)}]{tz2p}%
  \BibitemOpen
  \bibfield  {author} {\bibinfo {author} {\bibfnamefont {T.~H.}\ \bibnamefont
  {Dunning}, \bibfnamefont {Jr.}},\ }\href@noop {} {\bibfield  {journal}
  {\bibinfo  {journal} {J. Chem. Phys.}\ }\textbf {\bibinfo {volume} {55}},\
  \bibinfo {pages} {716} (\bibinfo {year} {1971})}\BibitemShut {NoStop}%
\bibitem [{\citenamefont {Sherrill}\ \emph {et~al.}(1998)\citenamefont
  {Sherrill}, \citenamefont {Leininger}, \citenamefont {Van~Huis},\ and\
  \citenamefont {Schaefer}}]{ch2tz2p}%
  \BibitemOpen
  \bibfield  {author} {\bibinfo {author} {\bibfnamefont {C.~D.}\ \bibnamefont
  {Sherrill}}, \bibinfo {author} {\bibfnamefont {M.~L.}\ \bibnamefont
  {Leininger}}, \bibinfo {author} {\bibfnamefont {T.~J.}\ \bibnamefont
  {Van~Huis}}, \ and\ \bibinfo {author} {\bibfnamefont {H.~F.}\ \bibnamefont
  {Schaefer}, \bibfnamefont {III}},\ }\href@noop {} {\bibfield  {journal}
  {\bibinfo  {journal} {J. Chem. Phys.}\ }\textbf {\bibinfo {volume} {108}},\
  \bibinfo {pages} {1040} (\bibinfo {year} {1998})}\BibitemShut {NoStop}%
\bibitem [{\citenamefont {Bauschlicher}\ and\ \citenamefont
  {Taylor}(1986)}]{ch2dzp}%
  \BibitemOpen
  \bibfield  {author} {\bibinfo {author} {\bibfnamefont {C.~W.}\ \bibnamefont
  {Bauschlicher}, \bibfnamefont {Jr.}}\ and\ \bibinfo {author} {\bibfnamefont
  {P.~R.}\ \bibnamefont {Taylor}},\ }\href@noop {} {\bibfield  {journal}
  {\bibinfo  {journal} {J. Chem. Phys.}\ }\textbf {\bibinfo {volume} {85}},\
  \bibinfo {pages} {6510} (\bibinfo {year} {1986})}\BibitemShut {NoStop}%
\bibitem [{\citenamefont {Bauschlicher}, \citenamefont {Langhoff},\ and\
  \citenamefont {Taylor}(1987)}]{ch2-bausch}%
  \BibitemOpen
  \bibfield  {author} {\bibinfo {author} {\bibfnamefont {C.~W.}\ \bibnamefont
  {Bauschlicher}, \bibfnamefont {Jr.}}, \bibinfo {author} {\bibfnamefont
  {S.~R.}\ \bibnamefont {Langhoff}}, \ and\ \bibinfo {author} {\bibfnamefont
  {P.~R.}\ \bibnamefont {Taylor}},\ }\href@noop {} {\bibfield  {journal}
  {\bibinfo  {journal} {J. Chem. Phys.}\ }\textbf {\bibinfo {volume} {87}},\
  \bibinfo {pages} {387} (\bibinfo {year} {1987})}\BibitemShut {NoStop}%
\bibitem [{\citenamefont {McLean}\ \emph {et~al.}(1987)\citenamefont {McLean},
  \citenamefont {Bunker}, \citenamefont {Escribano},\ and\ \citenamefont
  {Jensen}}]{ch2-mclean}%
  \BibitemOpen
  \bibfield  {author} {\bibinfo {author} {\bibfnamefont {A.~D.}\ \bibnamefont
  {McLean}}, \bibinfo {author} {\bibfnamefont {P.~R.}\ \bibnamefont {Bunker}},
  \bibinfo {author} {\bibfnamefont {R.~M.}\ \bibnamefont {Escribano}}, \ and\
  \bibinfo {author} {\bibfnamefont {P.}~\bibnamefont {Jensen}},\ }\href@noop {}
  {\bibfield  {journal} {\bibinfo  {journal} {J. Chem. Phys.}\ }\textbf
  {\bibinfo {volume} {87}},\ \bibinfo {pages} {2166} (\bibinfo {year}
  {1987})}\BibitemShut {NoStop}%
\bibitem [{\citenamefont {Comeau}\ \emph {et~al.}(1989)\citenamefont {Comeau},
  \citenamefont {Shavitt}, \citenamefont {Jensen},\ and\ \citenamefont
  {Bunker}}]{ch2-shavitt}%
  \BibitemOpen
  \bibfield  {author} {\bibinfo {author} {\bibfnamefont {D.~C.}\ \bibnamefont
  {Comeau}}, \bibinfo {author} {\bibfnamefont {I.}~\bibnamefont {Shavitt}},
  \bibinfo {author} {\bibfnamefont {P.}~\bibnamefont {Jensen}}, \ and\ \bibinfo
  {author} {\bibfnamefont {P.~R.}\ \bibnamefont {Bunker}},\ }\href@noop {}
  {\bibfield  {journal} {\bibinfo  {journal} {J. Chem. Phys.}\ }\textbf
  {\bibinfo {volume} {90}},\ \bibinfo {pages} {6491} (\bibinfo {year}
  {1989})}\BibitemShut {NoStop}%
\bibitem [{\citenamefont {Knowles}\ \emph {et~al.}(1990)\citenamefont
  {Knowles}, \citenamefont {Alvarez-Collado}, \citenamefont {Hirsch},\ and\
  \citenamefont {Buenker}}]{ch2-buenker}%
  \BibitemOpen
  \bibfield  {author} {\bibinfo {author} {\bibfnamefont {D.~B.}\ \bibnamefont
  {Knowles}}, \bibinfo {author} {\bibfnamefont {J.~R.}\ \bibnamefont
  {Alvarez-Collado}}, \bibinfo {author} {\bibfnamefont {G.}~\bibnamefont
  {Hirsch}}, \ and\ \bibinfo {author} {\bibfnamefont {R.~J.}\ \bibnamefont
  {Buenker}},\ }\href@noop {} {\bibfield  {journal} {\bibinfo  {journal} {J.
  Chem. Phys.}\ }\textbf {\bibinfo {volume} {92}},\ \bibinfo {pages} {585}
  (\bibinfo {year} {1990})}\BibitemShut {NoStop}%
\bibitem [{\citenamefont {Li}, \citenamefont {Piecuch},\ and\ \citenamefont
  {Paldus}(1994)}]{ch2_ppjp-dzp}%
  \BibitemOpen
  \bibfield  {author} {\bibinfo {author} {\bibfnamefont {X.}~\bibnamefont
  {Li}}, \bibinfo {author} {\bibfnamefont {P.}~\bibnamefont {Piecuch}}, \ and\
  \bibinfo {author} {\bibfnamefont {J.}~\bibnamefont {Paldus}},\ }\href@noop {}
  {\bibfield  {journal} {\bibinfo  {journal} {Chem. Phys. Lett.}\ }\textbf
  {\bibinfo {volume} {224}},\ \bibinfo {pages} {267} (\bibinfo {year}
  {1994})}\BibitemShut {NoStop}%
\bibitem [{\citenamefont {Piecuch}, \citenamefont {Li},\ and\ \citenamefont
  {Paldus}(1994)}]{ch2_ppjp}%
  \BibitemOpen
  \bibfield  {author} {\bibinfo {author} {\bibfnamefont {P.}~\bibnamefont
  {Piecuch}}, \bibinfo {author} {\bibfnamefont {X.}~\bibnamefont {Li}}, \ and\
  \bibinfo {author} {\bibfnamefont {J.}~\bibnamefont {Paldus}},\ }\href@noop {}
  {\bibfield  {journal} {\bibinfo  {journal} {Chem. Phys. Lett.}\ }\textbf
  {\bibinfo {volume} {230}},\ \bibinfo {pages} {377} (\bibinfo {year}
  {1994})}\BibitemShut {NoStop}%
\bibitem [{\citenamefont {Balkov\'{a}}\ and\ \citenamefont
  {Bartlett}(1995)}]{succ-balkova-ch2}%
  \BibitemOpen
  \bibfield  {author} {\bibinfo {author} {\bibfnamefont {A.}~\bibnamefont
  {Balkov\'{a}}}\ and\ \bibinfo {author} {\bibfnamefont {R.~J.}\ \bibnamefont
  {Bartlett}},\ }\href@noop {} {\bibfield  {journal} {\bibinfo  {journal} {J.
  Chem. Phys.}\ }\textbf {\bibinfo {volume} {102}},\ \bibinfo {pages} {7116}
  (\bibinfo {year} {1995})}\BibitemShut {NoStop}%
\bibitem [{\citenamefont {Li}\ and\ \citenamefont {Paldus}(2006)}]{rmrcc9}%
  \BibitemOpen
  \bibfield  {author} {\bibinfo {author} {\bibfnamefont {X.}~\bibnamefont
  {Li}}\ and\ \bibinfo {author} {\bibfnamefont {J.}~\bibnamefont {Paldus}},\
  }\href@noop {} {\bibfield  {journal} {\bibinfo  {journal} {J. Chem. Phys.}\
  }\textbf {\bibinfo {volume} {124}},\ \bibinfo {pages} {174101} (\bibinfo
  {year} {2006})}\BibitemShut {NoStop}%
\bibitem [{\citenamefont {Gour}, \citenamefont {Piecuch},\ and\ \citenamefont
  {W{\l}och}(2010)}]{gour2010}%
  \BibitemOpen
  \bibfield  {author} {\bibinfo {author} {\bibfnamefont {J.~R.}\ \bibnamefont
  {Gour}}, \bibinfo {author} {\bibfnamefont {P.}~\bibnamefont {Piecuch}}, \
  and\ \bibinfo {author} {\bibfnamefont {M.}~\bibnamefont {W{\l}och}},\
  }\href@noop {} {\bibfield  {journal} {\bibinfo  {journal} {Mol. Phys.}\
  }\textbf {\bibinfo {volume} {108}},\ \bibinfo {pages} {2633} (\bibinfo {year}
  {2010})}\BibitemShut {NoStop}%
\bibitem [{\citenamefont {Jensen}\ and\ \citenamefont
  {Bunker}(1988)}]{methylene5}%
  \BibitemOpen
  \bibfield  {author} {\bibinfo {author} {\bibfnamefont {P.}~\bibnamefont
  {Jensen}}\ and\ \bibinfo {author} {\bibfnamefont {P.~R.}\ \bibnamefont
  {Bunker}},\ }\href@noop {} {\bibfield  {journal} {\bibinfo  {journal} {J.
  Chem. Phys.}\ }\textbf {\bibinfo {volume} {89}},\ \bibinfo {pages} {1327}
  (\bibinfo {year} {1988})}\BibitemShut {NoStop}%
\bibitem [{\citenamefont {Davidson}, \citenamefont {Feller},\ and\
  \citenamefont {Phillips}(1980)}]{ch2-rel}%
  \BibitemOpen
  \bibfield  {author} {\bibinfo {author} {\bibfnamefont {E.~R.}\ \bibnamefont
  {Davidson}}, \bibinfo {author} {\bibfnamefont {D.}~\bibnamefont {Feller}}, \
  and\ \bibinfo {author} {\bibfnamefont {P.}~\bibnamefont {Phillips}},\
  }\href@noop {} {\bibfield  {journal} {\bibinfo  {journal} {Chem. Phys.
  Lett.}\ }\textbf {\bibinfo {volume} {76}},\ \bibinfo {pages} {416} (\bibinfo
  {year} {1980})}\BibitemShut {NoStop}%
\bibitem [{\citenamefont {Handy}, \citenamefont {Yamaguchi},\ and\
  \citenamefont {Schaefer}(1986)}]{ch2-bodc}%
  \BibitemOpen
  \bibfield  {author} {\bibinfo {author} {\bibfnamefont {N.~C.}\ \bibnamefont
  {Handy}}, \bibinfo {author} {\bibfnamefont {Y.}~\bibnamefont {Yamaguchi}}, \
  and\ \bibinfo {author} {\bibfnamefont {H.~F.}\ \bibnamefont {Schaefer},
  \bibfnamefont {III}},\ }\href@noop {} {\bibfield  {journal} {\bibinfo
  {journal} {J. Chem. Phys.}\ }\textbf {\bibinfo {volume} {84}},\ \bibinfo
  {pages} {4481} (\bibinfo {year} {1986})}\BibitemShut {NoStop}%
\bibitem [{\citenamefont {Shen}\ \emph {et~al.}(2011)\citenamefont {Shen},
  \citenamefont {Kou}, \citenamefont {Xu},\ and\ \citenamefont {Li}}]{h3}%
  \BibitemOpen
  \bibfield  {author} {\bibinfo {author} {\bibfnamefont {J.}~\bibnamefont
  {Shen}}, \bibinfo {author} {\bibfnamefont {Z.}~\bibnamefont {Kou}}, \bibinfo
  {author} {\bibfnamefont {E.}~\bibnamefont {Xu}}, \ and\ \bibinfo {author}
  {\bibfnamefont {S.}~\bibnamefont {Li}},\ }\href@noop {} {\bibfield  {journal}
  {\bibinfo  {journal} {J. Chem. Phys.}\ }\textbf {\bibinfo {volume} {134}},\
  \bibinfo {pages} {044134} (\bibinfo {year} {2011})}\BibitemShut {NoStop}%
\bibitem [{\citenamefont {Schurkus}\ \emph {et~al.}(2020)\citenamefont
  {Schurkus}, \citenamefont {Chen}, \citenamefont {Cheng}, \citenamefont
  {Chan},\ and\ \citenamefont {Stanton}}]{hfh-2020-stanton}%
  \BibitemOpen
  \bibfield  {author} {\bibinfo {author} {\bibfnamefont {H.}~\bibnamefont
  {Schurkus}}, \bibinfo {author} {\bibfnamefont {D.-T.}\ \bibnamefont {Chen}},
  \bibinfo {author} {\bibfnamefont {H.-P.}\ \bibnamefont {Cheng}}, \bibinfo
  {author} {\bibfnamefont {G.}~\bibnamefont {Chan}}, \ and\ \bibinfo {author}
  {\bibfnamefont {J.}~\bibnamefont {Stanton}},\ }\href@noop {} {\bibfield
  {journal} {\bibinfo  {journal} {J. Chem. Phys.}\ }\textbf {\bibinfo {volume}
  {152}},\ \bibinfo {pages} {234115} (\bibinfo {year} {2020})}\BibitemShut
  {NoStop}%
\bibitem [{\citenamefont {Szalay}\ and\ \citenamefont
  {Bartlett}(1993)}]{mraqcc1}%
  \BibitemOpen
  \bibfield  {author} {\bibinfo {author} {\bibfnamefont {P.~G.}\ \bibnamefont
  {Szalay}}\ and\ \bibinfo {author} {\bibfnamefont {R.~J.}\ \bibnamefont
  {Bartlett}},\ }\href@noop {} {\bibfield  {journal} {\bibinfo  {journal}
  {Chem. Phys. Lett.}\ }\textbf {\bibinfo {volume} {214}},\ \bibinfo {pages}
  {481} (\bibinfo {year} {1993})}\BibitemShut {NoStop}%
\bibitem [{\citenamefont {Szalay}\ and\ \citenamefont
  {Bartlett}(1995)}]{mraqcc2}%
  \BibitemOpen
  \bibfield  {author} {\bibinfo {author} {\bibfnamefont {P.~G.}\ \bibnamefont
  {Szalay}}\ and\ \bibinfo {author} {\bibfnamefont {R.~J.}\ \bibnamefont
  {Bartlett}},\ }\href@noop {} {\bibfield  {journal} {\bibinfo  {journal} {J.
  Chem. Phys.}\ }\textbf {\bibinfo {volume} {103}},\ \bibinfo {pages} {3600}
  (\bibinfo {year} {1995})}\BibitemShut {NoStop}%
\bibitem [{\citenamefont {Eckert-Maksi\'{c}}\ \emph {et~al.}(2006)\citenamefont
  {Eckert-Maksi\'{c}}, \citenamefont {Vazdar}, \citenamefont {Barbatti},
  \citenamefont {Lischka},\ and\ \citenamefont {Maksi\'{c}}}]{MR-AQCC}%
  \BibitemOpen
  \bibfield  {author} {\bibinfo {author} {\bibfnamefont {M.}~\bibnamefont
  {Eckert-Maksi\'{c}}}, \bibinfo {author} {\bibfnamefont {M.}~\bibnamefont
  {Vazdar}}, \bibinfo {author} {\bibfnamefont {M.}~\bibnamefont {Barbatti}},
  \bibinfo {author} {\bibfnamefont {H.}~\bibnamefont {Lischka}}, \ and\
  \bibinfo {author} {\bibfnamefont {Z.~B.}\ \bibnamefont {Maksi\'{c}}},\
  }\href@noop {} {\bibfield  {journal} {\bibinfo  {journal} {J. Chem. Phys.}\
  }\textbf {\bibinfo {volume} {125}},\ \bibinfo {pages} {064310} (\bibinfo
  {year} {2006})}\BibitemShut {NoStop}%
\bibitem [{\citenamefont {Balkov\'{a}}\ and\ \citenamefont
  {Bartlett}(1994)}]{balkova1994}%
  \BibitemOpen
  \bibfield  {author} {\bibinfo {author} {\bibfnamefont {A.}~\bibnamefont
  {Balkov\'{a}}}\ and\ \bibinfo {author} {\bibfnamefont {R.~J.}\ \bibnamefont
  {Bartlett}},\ }\href@noop {} {\bibfield  {journal} {\bibinfo  {journal} {J.
  Chem. Phys.}\ }\textbf {\bibinfo {volume} {101}},\ \bibinfo {pages} {8972}
  (\bibinfo {year} {1994})}\BibitemShut {NoStop}%
\bibitem [{\citenamefont {Levchenko}\ and\ \citenamefont
  {Krylov}(2004)}]{krylov2004}%
  \BibitemOpen
  \bibfield  {author} {\bibinfo {author} {\bibfnamefont {S.~V.}\ \bibnamefont
  {Levchenko}}\ and\ \bibinfo {author} {\bibfnamefont {A.~I.}\ \bibnamefont
  {Krylov}},\ }\href@noop {} {\bibfield  {journal} {\bibinfo  {journal} {J.
  Chem. Phys.}\ }\textbf {\bibinfo {volume} {120}},\ \bibinfo {pages} {175}
  (\bibinfo {year} {2004})}\BibitemShut {NoStop}%
\bibitem [{\citenamefont {Zimmerman}(2017)}]{zimmerman2017}%
  \BibitemOpen
  \bibfield  {author} {\bibinfo {author} {\bibfnamefont {P.~M.}\ \bibnamefont
  {Zimmerman}},\ }\href@noop {} {\bibfield  {journal} {\bibinfo  {journal} {J.
  Phys. Chem. A}\ }\textbf {\bibinfo {volume} {121}},\ \bibinfo {pages} {4712}
  (\bibinfo {year} {2017})}\BibitemShut {NoStop}%
\bibitem [{\citenamefont {Boyn}\ and\ \citenamefont
  {Mazziotti}(2021)}]{mazziotti2021}%
  \BibitemOpen
  \bibfield  {author} {\bibinfo {author} {\bibfnamefont {J.-N.}\ \bibnamefont
  {Boyn}}\ and\ \bibinfo {author} {\bibfnamefont {D.~A.}\ \bibnamefont
  {Mazziotti}},\ }\href@noop {} {\bibfield  {journal} {\bibinfo  {journal} {J.
  Chem. Phys.}\ }\textbf {\bibinfo {volume} {154}},\ \bibinfo {pages} {134103}
  (\bibinfo {year} {2021})}\BibitemShut {NoStop}%
\bibitem [{\citenamefont {Monino}\ \emph {et~al.}()\citenamefont {Monino},
  \citenamefont {Boggio-Pasqua}, \citenamefont {Scemama}, \citenamefont
  {Jacquemin},\ and\ \citenamefont {Loos}}]{loos2022}%
  \BibitemOpen
  \bibfield  {author} {\bibinfo {author} {\bibfnamefont {E.}~\bibnamefont
  {Monino}}, \bibinfo {author} {\bibfnamefont {M.}~\bibnamefont
  {Boggio-Pasqua}}, \bibinfo {author} {\bibfnamefont {A.}~\bibnamefont
  {Scemama}}, \bibinfo {author} {\bibfnamefont {D.}~\bibnamefont {Jacquemin}},
  \ and\ \bibinfo {author} {\bibfnamefont {P.-F.}\ \bibnamefont {Loos}},\
  }\href@noop {} {}\bibinfo {note} {ArXiv:2204.05098 (2022)}\BibitemShut
  {NoStop}%
\bibitem [{\citenamefont {Coulson}(1948)}]{TMM-1}%
  \BibitemOpen
  \bibfield  {author} {\bibinfo {author} {\bibfnamefont {C.~A.}\ \bibnamefont
  {Coulson}},\ }\href@noop {} {\bibfield  {journal} {\bibinfo  {journal} {J.
  Chim. Phys. Phys.-Chim. Biol.}\ }\textbf {\bibinfo {volume} {45}},\ \bibinfo
  {pages} {243} (\bibinfo {year} {1948})}\BibitemShut {NoStop}%
\bibitem [{\citenamefont {Longuet-Higgins}(1950)}]{TMM-2}%
  \BibitemOpen
  \bibfield  {author} {\bibinfo {author} {\bibfnamefont {H.~C.}\ \bibnamefont
  {Longuet-Higgins}},\ }\href@noop {} {\bibfield  {journal} {\bibinfo
  {journal} {J. Chem. Phys.}\ }\textbf {\bibinfo {volume} {18}},\ \bibinfo
  {pages} {265} (\bibinfo {year} {1950})}\BibitemShut {NoStop}%
\bibitem [{\citenamefont {Dowd}(1966)}]{TMM-32}%
  \BibitemOpen
  \bibfield  {author} {\bibinfo {author} {\bibfnamefont {P.}~\bibnamefont
  {Dowd}},\ }\href@noop {} {\bibfield  {journal} {\bibinfo  {journal} {J. Am.
  Chem. Soc.}\ }\textbf {\bibinfo {volume} {88}},\ \bibinfo {pages} {2587}
  (\bibinfo {year} {1966})}\BibitemShut {NoStop}%
\bibitem [{\citenamefont {Baseman}\ \emph {et~al.}(1976)\citenamefont
  {Baseman}, \citenamefont {Pratt}, \citenamefont {Chow},\ and\ \citenamefont
  {Dowd}}]{TMM-33}%
  \BibitemOpen
  \bibfield  {author} {\bibinfo {author} {\bibfnamefont {R.~J.}\ \bibnamefont
  {Baseman}}, \bibinfo {author} {\bibfnamefont {D.~W.}\ \bibnamefont {Pratt}},
  \bibinfo {author} {\bibfnamefont {M.}~\bibnamefont {Chow}}, \ and\ \bibinfo
  {author} {\bibfnamefont {P.}~\bibnamefont {Dowd}},\ }\href@noop {} {\bibfield
   {journal} {\bibinfo  {journal} {J. Am. Chem. Soc.}\ }\textbf {\bibinfo
  {volume} {98}},\ \bibinfo {pages} {5726} (\bibinfo {year}
  {1976})}\BibitemShut {NoStop}%
\bibitem [{\citenamefont {Borden}\ and\ \citenamefont
  {Davidson}(1981)}]{TMM-8}%
  \BibitemOpen
  \bibfield  {author} {\bibinfo {author} {\bibfnamefont {W.~T.}\ \bibnamefont
  {Borden}}\ and\ \bibinfo {author} {\bibfnamefont {E.~R.}\ \bibnamefont
  {Davidson}},\ }\href@noop {} {\bibfield  {journal} {\bibinfo  {journal} {Acc.
  Chem. Res.}\ }\textbf {\bibinfo {volume} {14}},\ \bibinfo {pages} {69}
  (\bibinfo {year} {1981})}\BibitemShut {NoStop}%
\bibitem [{\citenamefont {Yarkony}\ and\ \citenamefont
  {Schaefer}(1974)}]{TMM-14}%
  \BibitemOpen
  \bibfield  {author} {\bibinfo {author} {\bibfnamefont {D.~R.}\ \bibnamefont
  {Yarkony}}\ and\ \bibinfo {author} {\bibfnamefont {H.~F.}\ \bibnamefont
  {Schaefer}, \bibfnamefont {III}},\ }\href@noop {} {\bibfield  {journal}
  {\bibinfo  {journal} {J. Am. Chem. Soc.}\ }\textbf {\bibinfo {volume} {96}},\
  \bibinfo {pages} {3754} (\bibinfo {year} {1974})}\BibitemShut {NoStop}%
\bibitem [{\citenamefont {Hehre}, \citenamefont {Salem},\ and\ \citenamefont
  {Willcott}(1974)}]{TMM-15}%
  \BibitemOpen
  \bibfield  {author} {\bibinfo {author} {\bibfnamefont {W.~J.}\ \bibnamefont
  {Hehre}}, \bibinfo {author} {\bibfnamefont {L.}~\bibnamefont {Salem}}, \ and\
  \bibinfo {author} {\bibfnamefont {M.~R.}\ \bibnamefont {Willcott}},\
  }\href@noop {} {\bibfield  {journal} {\bibinfo  {journal} {J. Am. Chem.
  Soc.}\ }\textbf {\bibinfo {volume} {96}},\ \bibinfo {pages} {4328} (\bibinfo
  {year} {1974})}\BibitemShut {NoStop}%
\bibitem [{\citenamefont {Davis}\ and\ \citenamefont {Goddard}(1977)}]{TMM-16}%
  \BibitemOpen
  \bibfield  {author} {\bibinfo {author} {\bibfnamefont {J.~H.}\ \bibnamefont
  {Davis}}\ and\ \bibinfo {author} {\bibfnamefont {W.~A.}\ \bibnamefont
  {Goddard}, \bibfnamefont {III}},\ }\href@noop {} {\bibfield  {journal}
  {\bibinfo  {journal} {J. Am. Chem. Soc.}\ }\textbf {\bibinfo {volume} {99}},\
  \bibinfo {pages} {4242} (\bibinfo {year} {1977})}\BibitemShut {NoStop}%
\bibitem [{\citenamefont {Hood}, \citenamefont {Pitzer},\ and\ \citenamefont
  {Schaefer}(1978)}]{TMM-17}%
  \BibitemOpen
  \bibfield  {author} {\bibinfo {author} {\bibfnamefont {D.~M.}\ \bibnamefont
  {Hood}}, \bibinfo {author} {\bibfnamefont {R.~M.}\ \bibnamefont {Pitzer}}, \
  and\ \bibinfo {author} {\bibfnamefont {H.~F.}\ \bibnamefont {Schaefer},
  \bibfnamefont {III}},\ }\href@noop {} {\bibfield  {journal} {\bibinfo
  {journal} {J. Am. Chem. Soc.}\ }\textbf {\bibinfo {volume} {100}},\ \bibinfo
  {pages} {2227} (\bibinfo {year} {1978})}\BibitemShut {NoStop}%
\bibitem [{\citenamefont {Hood}, \citenamefont {Schaefer},\ and\ \citenamefont
  {Pitzer}(1978)}]{TMM-18}%
  \BibitemOpen
  \bibfield  {author} {\bibinfo {author} {\bibfnamefont {D.~M.}\ \bibnamefont
  {Hood}}, \bibinfo {author} {\bibfnamefont {H.~F.}\ \bibnamefont {Schaefer},
  \bibfnamefont {III}}, \ and\ \bibinfo {author} {\bibfnamefont {R.~M.}\
  \bibnamefont {Pitzer}},\ }\href@noop {} {\bibfield  {journal} {\bibinfo
  {journal} {J. Am. Chem. Soc.}\ }\textbf {\bibinfo {volume} {100}},\ \bibinfo
  {pages} {8009} (\bibinfo {year} {1978})}\BibitemShut {NoStop}%
\bibitem [{\citenamefont {Dixon}\ \emph {et~al.}(1978)\citenamefont {Dixon},
  \citenamefont {Foster}, \citenamefont {Halgren},\ and\ \citenamefont
  {Lipscomb}}]{TMM-19}%
  \BibitemOpen
  \bibfield  {author} {\bibinfo {author} {\bibfnamefont {D.~A.}\ \bibnamefont
  {Dixon}}, \bibinfo {author} {\bibfnamefont {R.}~\bibnamefont {Foster}},
  \bibinfo {author} {\bibfnamefont {T.~A.}\ \bibnamefont {Halgren}}, \ and\
  \bibinfo {author} {\bibfnamefont {W.~N.}\ \bibnamefont {Lipscomb}},\
  }\href@noop {} {\bibfield  {journal} {\bibinfo  {journal} {J. Am. Chem.
  Soc.}\ }\textbf {\bibinfo {volume} {100}},\ \bibinfo {pages} {1359} (\bibinfo
  {year} {1978})}\BibitemShut {NoStop}%
\bibitem [{\citenamefont {Feller}, \citenamefont {Borden},\ and\ \citenamefont
  {Davidson}(1981)}]{TMM-20}%
  \BibitemOpen
  \bibfield  {author} {\bibinfo {author} {\bibfnamefont {D.}~\bibnamefont
  {Feller}}, \bibinfo {author} {\bibfnamefont {W.~T.}\ \bibnamefont {Borden}},
  \ and\ \bibinfo {author} {\bibfnamefont {E.~R.}\ \bibnamefont {Davidson}},\
  }\href@noop {} {\bibfield  {journal} {\bibinfo  {journal} {J. Chem. Phys.}\
  }\textbf {\bibinfo {volume} {74}},\ \bibinfo {pages} {2256} (\bibinfo {year}
  {1981})}\BibitemShut {NoStop}%
\bibitem [{\citenamefont {Auster}, \citenamefont {Pitzer},\ and\ \citenamefont
  {Platz}(1982)}]{TMM-21}%
  \BibitemOpen
  \bibfield  {author} {\bibinfo {author} {\bibfnamefont {S.~B.}\ \bibnamefont
  {Auster}}, \bibinfo {author} {\bibfnamefont {R.~M.}\ \bibnamefont {Pitzer}},
  \ and\ \bibinfo {author} {\bibfnamefont {M.~S.}\ \bibnamefont {Platz}},\
  }\href@noop {} {\bibfield  {journal} {\bibinfo  {journal} {J. Am. Chem.
  Soc.}\ }\textbf {\bibinfo {volume} {104}},\ \bibinfo {pages} {3812} (\bibinfo
  {year} {1982})}\BibitemShut {NoStop}%
\bibitem [{\citenamefont {Borden}, \citenamefont {Davidson},\ and\
  \citenamefont {Feller}(1982)}]{TMM-22}%
  \BibitemOpen
  \bibfield  {author} {\bibinfo {author} {\bibfnamefont {W.~T.}\ \bibnamefont
  {Borden}}, \bibinfo {author} {\bibfnamefont {E.~R.}\ \bibnamefont
  {Davidson}}, \ and\ \bibinfo {author} {\bibfnamefont {D.}~\bibnamefont
  {Feller}},\ }\href@noop {} {\bibfield  {journal} {\bibinfo  {journal}
  {Tetrahedron}\ }\textbf {\bibinfo {volume} {38}},\ \bibinfo {pages} {737}
  (\bibinfo {year} {1982})}\BibitemShut {NoStop}%
\bibitem [{\citenamefont {Lahti}, \citenamefont {Rossi},\ and\ \citenamefont
  {Berson}(1985)}]{TMM-23}%
  \BibitemOpen
  \bibfield  {author} {\bibinfo {author} {\bibfnamefont {P.~M.}\ \bibnamefont
  {Lahti}}, \bibinfo {author} {\bibfnamefont {A.~R.}\ \bibnamefont {Rossi}}, \
  and\ \bibinfo {author} {\bibfnamefont {J.~A.}\ \bibnamefont {Berson}},\
  }\href@noop {} {\bibfield  {journal} {\bibinfo  {journal} {J. Am. Chem.
  Soc.}\ }\textbf {\bibinfo {volume} {107}},\ \bibinfo {pages} {2273} (\bibinfo
  {year} {1985})}\BibitemShut {NoStop}%
\bibitem [{\citenamefont {Skanche}, \citenamefont {Schaad},\ and\ \citenamefont
  {Hess}(1988)}]{TMM-24}%
  \BibitemOpen
  \bibfield  {author} {\bibinfo {author} {\bibfnamefont {A.}~\bibnamefont
  {Skanche}}, \bibinfo {author} {\bibfnamefont {L.~J.}\ \bibnamefont {Schaad}},
  \ and\ \bibinfo {author} {\bibfnamefont {B.~A.}\ \bibnamefont {Hess},
  \bibfnamefont {Jr.}},\ }\href@noop {} {\bibfield  {journal} {\bibinfo
  {journal} {J. Am. Chem. Soc.}\ }\textbf {\bibinfo {volume} {110}},\ \bibinfo
  {pages} {5315} (\bibinfo {year} {1988})}\BibitemShut {NoStop}%
\bibitem [{\citenamefont {Olivella}\ and\ \citenamefont
  {Salvador}(1990)}]{TMM-25}%
  \BibitemOpen
  \bibfield  {author} {\bibinfo {author} {\bibfnamefont {S.}~\bibnamefont
  {Olivella}}\ and\ \bibinfo {author} {\bibfnamefont {J.}~\bibnamefont
  {Salvador}},\ }\href@noop {} {\bibfield  {journal} {\bibinfo  {journal} {Int.
  J. Quantum Chem.}\ }\textbf {\bibinfo {volume} {37}},\ \bibinfo {pages} {713}
  (\bibinfo {year} {1990})}\BibitemShut {NoStop}%
\bibitem [{\citenamefont {Radhakrishnan}(1991)}]{TMM-26}%
  \BibitemOpen
  \bibfield  {author} {\bibinfo {author} {\bibfnamefont {T.~P.}\ \bibnamefont
  {Radhakrishnan}},\ }\href@noop {} {\bibfield  {journal} {\bibinfo  {journal}
  {Tetrahedron Lett.}\ }\textbf {\bibinfo {volume} {32}},\ \bibinfo {pages}
  {4601} (\bibinfo {year} {1991})}\BibitemShut {NoStop}%
\bibitem [{\citenamefont {Ichimura}, \citenamefont {Koga},\ and\ \citenamefont
  {Iwamura}(1994)}]{TMM-27}%
  \BibitemOpen
  \bibfield  {author} {\bibinfo {author} {\bibfnamefont {A.~S.}\ \bibnamefont
  {Ichimura}}, \bibinfo {author} {\bibfnamefont {N.}~\bibnamefont {Koga}}, \
  and\ \bibinfo {author} {\bibfnamefont {H.}~\bibnamefont {Iwamura}},\
  }\href@noop {} {\bibfield  {journal} {\bibinfo  {journal} {J. Phys. Org.
  Chem.}\ }\textbf {\bibinfo {volume} {7}},\ \bibinfo {pages} {207} (\bibinfo
  {year} {1994})}\BibitemShut {NoStop}%
\bibitem [{\citenamefont {Borden}(1993)}]{TMM-28}%
  \BibitemOpen
  \bibfield  {author} {\bibinfo {author} {\bibfnamefont {W.~T.}\ \bibnamefont
  {Borden}},\ }\href@noop {} {\bibfield  {journal} {\bibinfo  {journal} {Mol.
  Cryst. Liq. Cryst.}\ }\textbf {\bibinfo {volume} {232}},\ \bibinfo {pages}
  {195} (\bibinfo {year} {1993})}\BibitemShut {NoStop}%
\bibitem [{\citenamefont {Cramer}\ and\ \citenamefont {Smith}(1996)}]{TMM-29}%
  \BibitemOpen
  \bibfield  {author} {\bibinfo {author} {\bibfnamefont {C.~J.}\ \bibnamefont
  {Cramer}}\ and\ \bibinfo {author} {\bibfnamefont {B.~A.}\ \bibnamefont
  {Smith}},\ }\href@noop {} {\bibfield  {journal} {\bibinfo  {journal} {J.
  Phys. Chem.}\ }\textbf {\bibinfo {volume} {100}},\ \bibinfo {pages} {9664}
  (\bibinfo {year} {1996})}\BibitemShut {NoStop}%
\bibitem [{\citenamefont {Ma}\ and\ \citenamefont {Schaefer}(1996)}]{TMM-30}%
  \BibitemOpen
  \bibfield  {author} {\bibinfo {author} {\bibfnamefont {B.}~\bibnamefont
  {Ma}}\ and\ \bibinfo {author} {\bibfnamefont {H.~F.}\ \bibnamefont
  {Schaefer}, \bibfnamefont {III}},\ }\href@noop {} {\bibfield  {journal}
  {\bibinfo  {journal} {Chem. Phys.}\ }\textbf {\bibinfo {volume} {207}},\
  \bibinfo {pages} {31} (\bibinfo {year} {1996})}\BibitemShut {NoStop}%
\bibitem [{\citenamefont {Slipchenko}\ and\ \citenamefont
  {Krylov}(2003)}]{TMM-krylov}%
  \BibitemOpen
  \bibfield  {author} {\bibinfo {author} {\bibfnamefont {L.~V.}\ \bibnamefont
  {Slipchenko}}\ and\ \bibinfo {author} {\bibfnamefont {A.~I.}\ \bibnamefont
  {Krylov}},\ }\href@noop {} {\bibfield  {journal} {\bibinfo  {journal} {J.
  Chem. Phys.}\ }\textbf {\bibinfo {volume} {118}},\ \bibinfo {pages} {6874}
  (\bibinfo {year} {2003})}\BibitemShut {NoStop}%
\bibitem [{\citenamefont {Slipchenko}\ and\ \citenamefont
  {Krylov}(2005)}]{krylov-little-t}%
  \BibitemOpen
  \bibfield  {author} {\bibinfo {author} {\bibfnamefont {L.~V.}\ \bibnamefont
  {Slipchenko}}\ and\ \bibinfo {author} {\bibfnamefont {A.~I.}\ \bibnamefont
  {Krylov}},\ }\href@noop {} {\bibfield  {journal} {\bibinfo  {journal} {J.
  Chem. Phys.}\ }\textbf {\bibinfo {volume} {123}},\ \bibinfo {pages} {084107}
  (\bibinfo {year} {2005})}\BibitemShut {NoStop}%
\bibitem [{\citenamefont {Brabec}\ and\ \citenamefont
  {Pittner}(2006)}]{TMM-BWCC}%
  \BibitemOpen
  \bibfield  {author} {\bibinfo {author} {\bibfnamefont {J.}~\bibnamefont
  {Brabec}}\ and\ \bibinfo {author} {\bibfnamefont {J.}~\bibnamefont
  {Pittner}},\ }\href@noop {} {\bibfield  {journal} {\bibinfo  {journal} {J.
  Phys. Chem. A}\ }\textbf {\bibinfo {volume} {110}},\ \bibinfo {pages} {11765}
  (\bibinfo {year} {2006})}\BibitemShut {NoStop}%
\bibitem [{\citenamefont {Shen}\ \emph {et~al.}(2008)\citenamefont {Shen},
  \citenamefont {Fang}, \citenamefont {Li},\ and\ \citenamefont
  {Jiang}}]{BCCC5}%
  \BibitemOpen
  \bibfield  {author} {\bibinfo {author} {\bibfnamefont {J.}~\bibnamefont
  {Shen}}, \bibinfo {author} {\bibfnamefont {T.}~\bibnamefont {Fang}}, \bibinfo
  {author} {\bibfnamefont {S.}~\bibnamefont {Li}}, \ and\ \bibinfo {author}
  {\bibfnamefont {Y.}~\bibnamefont {Jiang}},\ }\href@noop {} {\bibfield
  {journal} {\bibinfo  {journal} {J. Phys. Chem. A}\ }\textbf {\bibinfo
  {volume} {112}},\ \bibinfo {pages} {12518} (\bibinfo {year}
  {2008})}\BibitemShut {NoStop}%
\bibitem [{\citenamefont {Perera}\ \emph {et~al.}(2014)\citenamefont {Perera},
  \citenamefont {Molt}, \citenamefont {Lotrich},\ and\ \citenamefont
  {Bartlett}}]{TMM-EXTRA-1}%
  \BibitemOpen
  \bibfield  {author} {\bibinfo {author} {\bibfnamefont {A.}~\bibnamefont
  {Perera}}, \bibinfo {author} {\bibfnamefont {R.~W.}\ \bibnamefont {Molt},
  \bibfnamefont {Jr.}}, \bibinfo {author} {\bibfnamefont {V.~F.}\ \bibnamefont
  {Lotrich}}, \ and\ \bibinfo {author} {\bibfnamefont {R.~J.}\ \bibnamefont
  {Bartlett}},\ }\href@noop {} {\bibfield  {journal} {\bibinfo  {journal}
  {Theor. Chem. Acc.}\ }\textbf {\bibinfo {volume} {133}},\ \bibinfo {pages}
  {1514} (\bibinfo {year} {2014})}\BibitemShut {NoStop}%
\bibitem [{\citenamefont {Sinha~Ray}\ \emph {et~al.}(2019)\citenamefont
  {Sinha~Ray}, \citenamefont {Manna}, \citenamefont {Ghosh}, \citenamefont
  {Chaudhuri},\ and\ \citenamefont {Chattopadhyay}}]{ssmrpt2019}%
  \BibitemOpen
  \bibfield  {author} {\bibinfo {author} {\bibfnamefont {S.}~\bibnamefont
  {Sinha~Ray}}, \bibinfo {author} {\bibfnamefont {S.}~\bibnamefont {Manna}},
  \bibinfo {author} {\bibfnamefont {A.}~\bibnamefont {Ghosh}}, \bibinfo
  {author} {\bibfnamefont {R.~K.}\ \bibnamefont {Chaudhuri}}, \ and\ \bibinfo
  {author} {\bibfnamefont {S.}~\bibnamefont {Chattopadhyay}},\ }\href@noop {}
  {\bibfield  {journal} {\bibinfo  {journal} {Int. J. Quantum Chem.}\ }\textbf
  {\bibinfo {volume} {119}},\ \bibinfo {pages} {e25776} (\bibinfo {year}
  {2019})}\BibitemShut {NoStop}%
\bibitem [{\citenamefont {Chattopadhyay}(2021)}]{ssmrpt2021}%
  \BibitemOpen
  \bibfield  {author} {\bibinfo {author} {\bibfnamefont {S.}~\bibnamefont
  {Chattopadhyay}},\ }\href {\doibase 10.1021/acsomega.0c05714} {\bibfield
  {journal} {\bibinfo  {journal} {ACS Omega}\ }\textbf {\bibinfo {volume}
  {6}},\ \bibinfo {pages} {1668} (\bibinfo {year} {2021})}\BibitemShut
  {NoStop}%
\bibitem [{\citenamefont {Wenthold}\ \emph {et~al.}(1996)\citenamefont
  {Wenthold}, \citenamefont {Hu}, \citenamefont {Squires},\ and\ \citenamefont
  {Lineberger}}]{TMM-Expt-1}%
  \BibitemOpen
  \bibfield  {author} {\bibinfo {author} {\bibfnamefont {P.~G.}\ \bibnamefont
  {Wenthold}}, \bibinfo {author} {\bibfnamefont {J.}~\bibnamefont {Hu}},
  \bibinfo {author} {\bibfnamefont {R.~R.}\ \bibnamefont {Squires}}, \ and\
  \bibinfo {author} {\bibfnamefont {W.~C.}\ \bibnamefont {Lineberger}},\
  }\href@noop {} {\bibfield  {journal} {\bibinfo  {journal} {J. Am. Chem.
  Soc.}\ }\textbf {\bibinfo {volume} {118}},\ \bibinfo {pages} {475} (\bibinfo
  {year} {1996})}\BibitemShut {NoStop}%
\bibitem [{\citenamefont {Wenthold}\ \emph {et~al.}(1999)\citenamefont
  {Wenthold}, \citenamefont {Hu}, \citenamefont {Squires},\ and\ \citenamefont
  {Lineberger}}]{TMM-Expt-2}%
  \BibitemOpen
  \bibfield  {author} {\bibinfo {author} {\bibfnamefont {P.~G.}\ \bibnamefont
  {Wenthold}}, \bibinfo {author} {\bibfnamefont {J.}~\bibnamefont {Hu}},
  \bibinfo {author} {\bibfnamefont {R.~R.}\ \bibnamefont {Squires}}, \ and\
  \bibinfo {author} {\bibfnamefont {W.~C.}\ \bibnamefont {Lineberger}},\
  }\href@noop {} {\bibfield  {journal} {\bibinfo  {journal} {J. Am. Soc. Mass
  Spectrom.}\ }\textbf {\bibinfo {volume} {10}},\ \bibinfo {pages} {800}
  (\bibinfo {year} {1999})}\BibitemShut {NoStop}%
\end{thebibliography}
%

\clearpage


\onecolumngrid

\squeezetable
\begin{table*}[h!]
\caption{\label{table:table1}
Convergence of the CC($P$) and CC($P$;$Q$) energies of the ${\rm X \: ^3B_1}$
and ${\rm A \: ^1A_1}$ states of methylene, as described by the aug-cc-pVTZ
basis set, and of the corresponding adiabatic singlet--triplet gaps toward
their parent CCSDT values. The geometries of the ${\rm X \: ^3B_1}$ and
${\rm A \: ^1A_1}$ states, optimized in the FCI calculations using the TZ2P
basis set, were taken from Ref.\ \onlinecite{ch2tz2p}. The $P$ spaces used in
the CC($P$) and CC($P$;$Q$) calculations were defined as all singly and doubly
excited determinants and subsets of triply excited determinants extracted from
the $i$-FCIQMC propagations with $\delta\tau=0.0001$ a.u. The $Q$ spaces used
to determine the CC($P$;$Q$) corrections consisted of the triply excited
determinants not captured by the corresponding $i$-FCIQMC runs. The $i$-FCIQMC
calculations preceding the CC($P$) and CC($P$;$Q$) steps were initiated by
placing 1500 walkers on the ROHF (${\rm X \: ^3B_1}$ state) and RHF
(${\rm A \: ^1A_1}$ state) reference determinants and the $n_{a}$ parameter of
the initiator algorithm was set at 3. In all post-Hartree--Fock calculations, the lowest
core orbital was kept frozen and the spherical components of $d$ and $f$
orbitals were employed throughout.}
\begin{ruledtabular}
\begin{tabular}{lddddddrr}
& \multicolumn{3}{c}{${\rm X\: ^3B_1}$} & \multicolumn{3}{c}{${\rm A \: ^1A_1}$}
& \multicolumn{2}{c}{${\rm A\: ^1A_1 - X \: ^3B_1}$} \\
\cline{2-4} \cline{5-7} \cline{8-9}
MC Iterations 
& \multicolumn{1}{c}{$P$\protect\footnotemark[1]} & \multicolumn{1}{c}{$(P;Q)$\protect\footnotemark[1]}
& \multicolumn{1}{c}{\%T\protect\footnotemark[2]}
& \multicolumn{1}{c}{$P$\protect\footnotemark[1]} & \multicolumn{1}{c}{$(P;Q)$\protect\footnotemark[1]} 
& \multicolumn{1}{c}{\%T\protect\footnotemark[2]}
& \multicolumn{1}{c}{$P$\protect\footnotemark[3]} & \multicolumn{1}{c}{$(P;Q)$\protect\footnotemark[3]} \\
\hline
0        & 4.187\protect\footnotemark[4] & 0.177\protect\footnotemark[5] & 0  
         & 5.918\protect\footnotemark[4] & 0.656\protect\footnotemark[5] & 0  
         &   380\protect\footnotemark[4] &   105\protect\footnotemark[5] \\
2000     & 3.948 & 0.162 &  1.8 & 5.361 & 0.549 &  3.0 &   310  &  85 \\
4000     & 3.281 & 0.111 &  7.1 & 3.908 & 0.304 & 11.9 &   138  &  42 \\
6000     & 2.749 & 0.072 & 12.4 & 2.993 & 0.190 & 19.7 &    53  &  26 \\
8000     & 2.428 & 0.049 & 16.3 & 2.444 & 0.106 & 24.9 &     3  &  13 \\
10000    & 2.192 & 0.038 & 19.0 & 2.093 & 0.080 & 28.7 &  $-22$ &   9 \\
20000    & 1.703 & 0.018 & 26.3 & 1.358 & 0.025 & 37.7 &  $-76$ &   2 \\
50000    & 1.133 & 0.005 & 39.1 & 0.644 & 0.004 & 54.8 & $-107$ &   0 \\
100000   & 0.532 & 0.000 & 59.5 & 0.171 & 0.000 & 76.5 &  $-79$ &   0 \\
150000   & 0.218 & 0.000 & 76.8 & 0.037 & 0.000 & 90.7 &  $-40$ &   0 \\
200000   & 0.076 & 0.000 & 88.7 & 0.006 & 0.000 & 97.2 &  $-15$ &   0 \\
$\infty$ & \multicolumn{3}{c}{$-39.080575$\protect\footnotemark[6]} 
         & \multicolumn{3}{c}{$-39.065411$\protect\footnotemark[6]} 
         & \multicolumn{2}{c}{$3328$\protect\footnotemark[7]} \\
\end{tabular}
\end{ruledtabular}
\footnotetext[1]{
\setlength{\baselineskip}{1em}Unless otherwise stated, all energies are
reported as errors relative to CCSDT in millihartree.
}
\footnotetext[2]{
\setlength{\baselineskip}{1em}The \%T values are the percentages of triples
captured during the $i$-FCIQMC propagations (the $S_{z} = 1$ triply excited
determinants of the ${\rm B}_{1}$ symmetry in the case of the ${\rm X \: ^3B_1}$
state and the $S_{z} = 0$ triply excited determinants of the ${\rm A}_{1}$
symmetry in the case of the ${\rm A \: ^1A_1}$ state).}
\footnotetext[3]{
\setlength{\baselineskip}{1em}Unless otherwise specified, the values of the
singlet--triplet gap are reported as errors relative to CCSDT in ${\rm cm^{-1}}$.}
\footnotetext[4]{
\setlength{\baselineskip}{1em}Equivalent to CCSD.
}
\footnotetext[5]{
\setlength{\baselineskip}{1em}Equivalent to CR-CC(2,3) [the most complete
variant of CR-CC(2,3) abbreviated sometimes as CR-CC(2,3),D or
CR-CC(2,3)$_{\rm D}$].}
\footnotetext[6]{
\setlength{\baselineskip}{1em}Total CCSDT energy in hartree.}
\footnotetext[7]{
\setlength{\baselineskip}{1em}The CCSDT singlet--triplet gap in
${\rm cm^{-1}}$.}
\end{table*}

\squeezetable
\begin{table*}[h!]
\centering

\caption{\label{table:table3}
Convergence of the CC($P$) and CC($P$;$Q$) energies of the
${\rm X \: ^1\Sigma}^+_g$ state of ${\rm (HFH)^-}$, as described by the
6-31G(d,p) basis set, at selected H--F distances $R_{\rm H\mbox{-}F}$ toward
their parent CCSDT values. The $P$ spaces used in the CC($P$) and CC($P$;$Q$)
calculations were defined as all singly and doubly excited determinants and
subsets of triply excited determinants extracted from the $i$-FCIQMC
propagations with $\delta\tau=0.0001$ a.u. The $Q$ spaces used to determine the
CC($P$;$Q$) corrections consisted of the triply excited determinants not
captured by the corresponding $i$-FCIQMC runs. The $i$-FCIQMC calculations
preceding the CC($P$) and CC($P$;$Q$) steps were initiated by placing 1500 
walkers on the RHF reference determinant and the $n_{a}$ parameter of the
initiator algorithm was set at 3. In all post-Hartree--Fock calculations, the lowest core
orbital was kept frozen and the spherical components of $d$ orbitals were
employed throughout.}
\begin{ruledtabular}
\begin{tabular}{lrrcrrcrrcrrcrrc}
& \multicolumn{3}{c}{$R_{\rm H\mbox{-}F} = 1.50$ \AA} & \multicolumn{3}{c}{$R_{\rm H\mbox{-}F} = 1.75$ \AA}
& \multicolumn{3}{c}{$R_{\rm H\mbox{-}F} = 2.00$ \AA} 
& \multicolumn{3}{c}{$R_{\rm H\mbox{-}F} = 2.50$ \AA} & \multicolumn{3}{c}{$R_{\rm H\mbox{-}F} = 4.00$ \AA} \\
\cline{2-4} \cline{5-7} \cline{8-10} \cline{11-13} \cline{14-16}
MC Iterations & $P$\protect\footnotemark[1] & $(P;Q)$\protect\footnotemark[1] & \%T\protect\footnotemark[2]
              & $P$\protect\footnotemark[1] & $(P;Q)$\protect\footnotemark[1] & \%T\protect\footnotemark[2]
              & $P$\protect\footnotemark[1] & $(P;Q)$\protect\footnotemark[1] & \%T\protect\footnotemark[2] 
              & $P$\protect\footnotemark[1] & $(P;Q)$\protect\footnotemark[1] & \%T\protect\footnotemark[2] 
              & $P$\protect\footnotemark[1] & $(P;Q)$\protect\footnotemark[1] & \%T\protect\footnotemark[2] \\
\hline
0      & 11.412\protect\footnotemark[3] & $-0.343$\protect\footnotemark[4] &  0    
       & 14.738\protect\footnotemark[3] & $-0.686$\protect\footnotemark[4] &  0    
       & 17.453\protect\footnotemark[3] & $-1.455$\protect\footnotemark[4] &  0    
       & 17.051\protect\footnotemark[3] & $-2.800$\protect\footnotemark[4] &  0
       &  1.907\protect\footnotemark[3] & $-0.291$\protect\footnotemark[4] &  0    \\
2000   &  2.601 & $-0.035$ & 34.2 &  3.998 & $-0.056$ &  30.5 &  3.511 & $-0.110$ & 22.6 
       &  6.586 & $-0.583$ & 15.2 &  0.412 & $-0.025$ &   7.6 \\
4000   &  0.843 & $-0.028$ & 56.1 &  1.078 & $-0.009$ &  49.6 &  1.979 & $-0.017$ & 40.5 
       &  0.973 & $-0.050$ & 25.6 &  0.141 & $-0.004$ &  11.7 \\
6000   &  0.595 & $-0.004$ & 63.9 &  0.434 & $-0.003$ &  58.1 &  0.432 & $-0.010$ & 46.7 
       &  0.459 & $-0.012$ & 30.2 &  0.076 & $-0.003$ &  13.2 \\
8000   &  0.225 & $-0.003$ & 68.6 &  0.477 & $-0.007$ &  61.4 &  0.187 & $-0.003$ & 50.2 
       &  0.225 & $-0.003$ & 33.9 &  0.037 & $-0.001$ &  14.4 \\
10000  &  0.258 & $-0.003$ & 70.9 &  0.161 & $-0.002$ &  63.3 &  0.136 & $-0.003$ & 54.5 
       &  0.167 &   0.000  & 35.4 &  0.025 & $-0.001$ &  15.3 \\
20000  &  0.112 &   0.000  & 77.2 &  0.056 & $-0.001$ &  71.0 &  0.079 & $-0.002$ & 61.1 
       &  0.042 & $-0.001$ & 41.8 &  0.026 & $-0.001$ &  19.0 \\
50000  &  0.017 &   0.000  & 88.4 &  0.019 &   0.000  &  85.8 &  0.005 &   0.000  & 77.5 
       &  0.009 &   0.000  & 58.8 &  0.002 & $-0.001$ &  28.6 \\
100000 &  0.002 &  0.000   & 97.7 &  0.001 &   0.000  &  96.3 &  0.001 &   0.000  & 94.4 
       &  0.000 &  0.000   & 81.8 &  0.000 &   0.000  &  54.8 \\
150000 &  0.000 &  0.000   & 99.5 &  0.000 &   0.000  &  99.4 &  0.000 &   0.000  & 99.2 
       &  0.000 &  0.000   & 94.1 &  0.000 &   0.000  &  73.7 \\ 
200000 &  0.000 &  0.000   & 99.9 &  0.000 &   0.000  & 100.0 &  0.000 &   0.000  & 99.9 
       &  0.000 &  0.000   & 99.2 &  0.000 &   0.000  &  86.9 \\
$\infty$ & \multicolumn{3}{c}{$-100.588130$\protect\footnotemark[5]} 
         & \multicolumn{3}{c}{$-100.576056$\protect\footnotemark[5]} 
         & \multicolumn{3}{c}{$-100.561110$\protect\footnotemark[5]} 
         & \multicolumn{3}{c}{$-100.539783$\protect\footnotemark[5]} 
         & \multicolumn{3}{c}{$-100.525901$\protect\footnotemark[5]} \\
\end{tabular}
\end{ruledtabular}

\footnotetext[1]{
\setlength{\baselineskip}{1em}Unless otherwise stated, all energies are
reported as errors relative to CCSDT in millihartree.}
\footnotetext[2]{
\setlength{\baselineskip}{1em}The \%T values are the percentages of triples
captured during the $i$-FCIQMC propagations [the $S_{z} = 0$ triply excited
determinants of the ${\rm A}_{g}$(${D_{2h}}$) symmetry].}
\footnotetext[3]{
\setlength{\baselineskip}{1em}Equivalent to CCSD.}
\footnotetext[4]{
\setlength{\baselineskip}{1em}Equivalent to CR-CC(2,3) [the most complete
variant of CR-CC(2,3) abbreviated sometimes as CR-CC(2,3),D or
CR-CC(2,3)$_{\rm D}$].}
\footnotetext[5]{
\setlength{\baselineskip}{1em}Total CCSDT energy in hartree.}
\end{table*}

\begin{table*}[!htbp]
\centering
\caption{\label{table:table4}
Convergence of the CC($P$) and CC($P$;$Q$) energies of the
${\rm A \: ^3\Sigma}^+_u$ state of ${\rm (HFH)^-}$, as described by the
6-31G(d,p) basis set, at selected H--F distances $R_{\rm H\mbox{-}F}$ toward
their parent CCSDT values. The $P$ spaces used in the CC($P$) and CC($P$;$Q$)
calculations were defined as all singly and doubly excited determinants and
subsets of triply excited determinants extracted from the $i$-FCIQMC
propagations with $\delta\tau=0.0001$ a.u. The $Q$ spaces used to determine the
CC($P$;$Q$) corrections consisted of the triply excited determinants not
captured by the corresponding $i$-FCIQMC runs. The $i$-FCIQMC calculations
preceding the CC($P$) and CC($P$;$Q$) steps were initiated by placing
1500 walkers on the ROHF reference determinant and the $n_{a}$ parameter of the
initiator algorithm was set at 3. In all post-Hartree--Fock calculations, the lowest core
orbital was kept frozen and the spherical components of $d$ orbitals were
employed throughout.}
\begin{ruledtabular}
\begin{tabular}{lrrcrrcrrcrrcrrc}
& \multicolumn{3}{c}{$R_{\rm H\mbox{-}F} = 1.50$ \AA} & \multicolumn{3}{c}{$R_{\rm H\mbox{-}F} = 1.75$ \AA}
& \multicolumn{3}{c}{$R_{\rm H\mbox{-}F} = 2.00$ \AA}
& \multicolumn{3}{c}{$R_{\rm H\mbox{-}F} = 2.50$ \AA} & \multicolumn{3}{c}{$R_{\rm H\mbox{-}F} = 4.00$ \AA} \\
\cline{2-4} \cline{5-7} \cline{8-10} \cline{11-13} \cline{14-16}
MC Iterations & $P$\protect\footnotemark[1] & $(P;Q)$\protect\footnotemark[1] & \%T\protect\footnotemark[2]
              & $P$\protect\footnotemark[1] & $(P;Q)$\protect\footnotemark[1] & \%T\protect\footnotemark[2]   
              & $P$\protect\footnotemark[1] & $(P;Q)$\protect\footnotemark[1] & \%T\protect\footnotemark[2]
              & $P$\protect\footnotemark[1] & $(P;Q)$\protect\footnotemark[1] & \%T\protect\footnotemark[2]
              & $P$\protect\footnotemark[1] & $(P;Q)$\protect\footnotemark[1] & \%T\protect\footnotemark[2] \\
\hline
\hline
0      & 2.268\protect\footnotemark[3] & $-0.217$\protect\footnotemark[4] & 0
       & 1.967\protect\footnotemark[3] & $-0.181$\protect\footnotemark[4] & 0
       & 1.678\protect\footnotemark[3] & $-0.172$\protect\footnotemark[4] & 0
       & 1.277\protect\footnotemark[3] & $-0.167$\protect\footnotemark[4] & 0
       & 1.123\protect\footnotemark[3] & $-0.180$\protect\footnotemark[4] & 0 \\
2000   & 0.995 & $-0.040$ & 27.8 & 0.826 & $-0.024$ & 24.2 & 0.834 & $-0.038$ & 19.1 
       & 0.502 & $-0.029$ & 10.7 & 0.239 & $-0.014$ & 4.5 \\
4000   & 0.456 & $-0.010$ & 49.4 & 0.477 & $-0.010$ & 41.7 & 0.475 & $-0.012$ & 33.7  
       & 0.236 & $-0.009$ & 17.2 & 0.079 & $-0.002$ &  5.4 \\
6000   & 0.338 & $-0.005$ & 56.4 & 0.266 & $-0.001$ & 50.5 & 0.321 & $-0.005$ & 41.2 
       & 0.174 & $-0.003$ & 21.5 & 0.070 & $-0.003$ &  5.9 \\
8000   & 0.290 & $-0.003$ & 60.1 & 0.254 & $-0.003$ & 54.2 & 0.225 & $-0.003$ & 44.7 
       & 0.195 & $-0.006$ & 23.9 & 0.064 & $-0.002$ &  6.0\\
10000  & 0.271 & $-0.003$ & 61.1 & 0.267 & $-0.004$ & 56.6 & 0.201 & $-0.002$ & 45.7 
       & 0.064 & $-0.003$ & 25.0 & 0.056 & $-0.002$ &  6.4 \\
20000  & 0.201 & $-0.002$ & 67.9 & 0.151 & $-0.001$ & 62.1 & 0.157 & $-0.002$ & 52.2 
       & 0.078 & $-0.003$ & 28.6 & 0.025 & $-0.001$ &  7.4 \\
50000  & 0.082 &   0.000  & 80.0 & 0.056 &   0.000  & 76.3 & 0.069 & $-0.001$ & 66.1 
       & 0.049 & $-0.001$ & 37.4 & 0.012 &   0.000  &  8.4 \\
100000 & 0.021 &   0.000  & 91.8 & 0.016 &   0.000  & 89.4 & 0.015 &   0.000  & 82.8 
       & 0.014 &   0.000  & 52.9 & 0.002 &   0.000  & 11.7\\
150000 & 0.007 &   0.000  & 96.7 & 0.003 &   0.000  & 95.8 & 0.003 &   0.000  & 92.9 
       & 0.002 &   0.000  & 68.6 & 0.001 &   0.000  & 16.8 \\
200000 & 0.001 &   0.000  & 98.8 & 0.001 &   0.000  & 98.4 & 0.001 &   0.000  & 97.1 
       & 0.000 &   0.000  & 81.8 & 0.000 &   0.000  & 23.8 \\
%

$\infty$ & \multicolumn{3}{c}{$-100.545633$\protect\footnotemark[5]} 
         & \multicolumn{3}{c}{$-100.554908$\protect\footnotemark[5]} 
         & \multicolumn{3}{c}{$-100.552882$\protect\footnotemark[5]} 
         & \multicolumn{3}{c}{$-100.540435$\protect\footnotemark[5]} 
         & \multicolumn{3}{c}{$-100.526164$\protect\footnotemark[5]} \\
\end{tabular}
\end{ruledtabular}

\footnotetext[1]{
\setlength{\baselineskip}{1em}Unless otherwise stated, all energies are
reported as errors relative to CCSDT in millihartree.}

\footnotetext[2]{
\setlength{\baselineskip}{1em}The \%T values are the percentages of triples
captured during the $i$-FCIQMC propagations [the $S_{z} = 1$ triply excited
determinants of the ${\rm B}_{1u}$(${D_{2h}}$) symmetry].}
 
\footnotetext[3]{
\setlength{\baselineskip}{1em}Equivalent to CCSD.}
 
\footnotetext[4]{
\setlength{\baselineskip}{1em}Equivalent to CR-CC(2,3) [the most complete
variant of CR-CC(2,3) abbreviated sometimes as CR-CC(2,3),D or
CR-CC(2,3)$_{\rm D}$].}
 
\footnotetext[5]{
\setlength{\baselineskip}{1em}Total CCSDT energy in hartree.}

\end{table*}

\squeezetable
\begin{table*}[!htbp]
\centering
\caption{\label{table:table5} 
Convergence of the CC($P$) and CC($P$;$Q$) singlet--triplet gaps of
${\rm (HFH)^-}$, as described by the 6-31G(d,p) basis set, at selected H--F
distances $R_{\rm H\mbox{-}F}$ toward their parent CCSDT values. The $P$ spaces
used in the CC($P$) and CC($P$;$Q$) calculations were defined as all singly and
doubly excited determinants and subsets of triply excited determinants
extracted from the $i$-FCIQMC propagations with $\delta\tau=0.0001$ a.u. The $Q$
spaces used to determine the CC($P$;$Q$) corrections consisted of the triply
excited determinants not captured by the corresponding $i$-FCIQMC runs. The
$i$-FCIQMC calculations preceding the CC($P$) and CC($P$;$Q$) steps were
initiated by placing 1500 walkers on the RHF (${\rm X \: ^1\Sigma}^+_g$ state)
and ROHF (${\rm A \: ^3\Sigma}^+_u$ state) reference determinants and the
$n_{a}$ parameter of the initiator algorithm was set at 3. In all post-Hartree--Fock
calculations, the lowest core orbital was kept frozen and the spherical
components of $d$ orbitals were employed throughout.}
\begin{ruledtabular}
\begin{tabular}{lrrrrrrrrrr}
& \multicolumn{2}{c}{$R_{\rm H\mbox{-}F} = 1.50$ \AA} & \multicolumn{2}{c}{$R_{\rm H\mbox{-}F} = 1.75$ \AA}
& \multicolumn{2}{c}{$R_{\rm H\mbox{-}F} = 2.00$ \AA}
& \multicolumn{2}{c}{$R_{\rm H\mbox{-}F} = 2.50$ \AA} & \multicolumn{2}{c}{$R_{\rm H\mbox{-}F} = 4.00$ \AA} \\
\cline{2-3} \cline{4-5} \cline{6-7} \cline{8-9} \cline{10-11}
MC Iterations & $P$\protect\footnotemark[1] & $(P;Q)$\protect\footnotemark[1] 
              & $P$\protect\footnotemark[1] & $(P;Q)$\protect\footnotemark[1]
              & $P$\protect\footnotemark[1] & $(P;Q)$\protect\footnotemark[1]
              & $P$\protect\footnotemark[1] & $(P;Q)$\protect\footnotemark[1]
              & $P$\protect\footnotemark[1] & $(P;Q)$\protect\footnotemark[1] \\

\hline
0      & 2007\protect\footnotemark[2] &  $-28$\protect\footnotemark[3]
       & 2803\protect\footnotemark[2] & $-111$\protect\footnotemark[3] 
       & 3462\protect\footnotemark[2] & $-282$\protect\footnotemark[3]  
       & 3462\protect\footnotemark[2] & $-578$\protect\footnotemark[3] 
       &  172\protect\footnotemark[2] &  $-24$\protect\footnotemark[3] \\
2000   &  353 &  1 &  696 & $-7$ &   588 & $-16$ & 1335 & $-122$ &  38  & $-2$\\
4000   &   85 & $-4$ & 132 &    0 &  330 &  $-1$ &  162 &   $-9$ &  14  &   0 \\
6000   &   56 &  0 &   37 &    0 &    24 &  $-1$ &   62 &   $-2$ &   1  &   0 \\
8000   & $-14$ & 0 &   49 & $-1$ &  $-8$ &    0  &    7 &     1  & $-6$ &   0 \\
10000  &  $-3$ & 0 & $-23$ &   0 & $-14$ &    0  &   23 &     0  & $-7$ &   0 \\
20000  & $-20$ & 0 & $-21$ &   0 & $-17$ &    0  & $-8$ &     1  &   0  &   0 \\
50000  & $-14$ & 0 & $-8$ &    0 & $-14$ &    0  & $-9$ &     0  & $-2$ &   0 \\
100000 &  $-4$ & 0 & $-3$ &    0 &  $-3$ &    0  & $-3$ &     0  & $-1$ &   0 \\
150000 &  $-2$ & 0 & $-1$ &    0 &  $-1$ &    0  &    0 &     0  &   0  &   0 \\
200000 &    0  & 0 &    0 &    0 &     0 &    0  &    0 &     0  &   0  &   0 \\


$\infty$ & \multicolumn{2}{c}{$-9327$\protect\footnotemark[4]} & \multicolumn{2}{c}{$-4641$\protect\footnotemark[4]} 
         & \multicolumn{2}{c}{$-1806$\protect\footnotemark[4]} & \multicolumn{2}{c}{143\protect\footnotemark[4]}     
         & \multicolumn{2}{c}{58\protect\footnotemark[4]} \\

\end{tabular}
\end{ruledtabular}

\footnotetext[1]{
\setlength{\baselineskip}{1em}Unless otherwise stated, all singlet--triplet
gaps are reported as errors relative to CCSDT in ${\rm cm^{-1}}$.}

\footnotetext[2]{
\setlength{\baselineskip}{1em}Equivalent to CCSD.}

\footnotetext[3]{
\setlength{\baselineskip}{1em}Equivalent to CR-CC(2,3) [the most complete
variant of CR-CC(2,3) abbreviated sometimes as CR-CC(2,3),D or
CR-CC(2,3)$_{\rm D}$].}

\footnotetext[4]{
\setlength{\baselineskip}{1em}The CCSDT singlet--triplet gap in
${\rm cm^{-1}}$.}

\end{table*}

\begin{table*}[!htbp]
\centering
\caption{\label{table:table7} 
Convergence of the CC($P$) and CC($P$;$Q$) energies of the ${\rm X \: ^1B}_{1g}$
and ${\rm A \: ^3A}_{2g}$ states of cyclobutadiene, as described by the cc-pVDZ
basis set, and of the corresponding vertical singlet--triplet gaps toward their
parent CCSDT values. All calculations were performed at the $D_{4h}$-symmetric
transition-state geometry of the ${\rm X\: ^1B}_{1g}$ state optimized in the MR-AQCC
calculations in Ref.\ \onlinecite{MR-AQCC}. The $P$ spaces used in the CC($P$)
and CC($P$;$Q$) calculations were defined as all singly and doubly excited
determinants and subsets of triply excited determinants extracted from the
$i$-FCIQMC propagations with $\delta\tau=0.0001$ a.u. The $Q$ spaces used to
determine the CC($P$;$Q$) corrections consisted of the triply excited
determinants not captured by the corresponding $i$-FCIQMC runs. The $i$-FCIQMC
calculations preceding the CC($P$) and CC($P$;$Q$) steps were initiated by
placing 1500 walkers on the RHF (${\rm X \: ^1B}_{1g}$ state) and ROHF
(${\rm A \: ^3A}_{2g}$ state) reference determinants and the $n_{a}$ parameter
of the initiator algorithm was set at 3. In all post-Hartree--Fock calculations, the four
lowest core orbitals were kept frozen and the spherical components of $d$
orbitals were employed throughout.}
\begin{ruledtabular}
\begin{tabular}{ldddddddd}
& \multicolumn{3}{c}{${\rm X\: ^1B}_{1g}$} & \multicolumn{3}{c}{${\rm A\: ^3A}_{2g}$} 
& \multicolumn{2}{c}{${\rm X\: ^1B}_{1g}-{\rm A\: ^3A}_{2g}$} \\
\cline{2-4} \cline{5-7} \cline{8-9}

MC Iterations
& \multicolumn{1}{c}{$P$\protect\footnotemark[1]} & \multicolumn{1}{c}{$(P;Q)$\protect\footnotemark[1]}
& \multicolumn{1}{c}{\%T\protect\footnotemark[2]}
& \multicolumn{1}{c}{$P$\protect\footnotemark[1]} & \multicolumn{1}{c}{$(P;Q)$\protect\footnotemark[1]}
& \multicolumn{1}{c}{\%T\protect\footnotemark[2]}
& \multicolumn{1}{c}{$P$\protect\footnotemark[3]} & \multicolumn{1}{c}{$(P;Q)$\protect\footnotemark[3]} \\
\hline

0     & 47.979\protect\footnotemark[4] & 14.636\protect\footnotemark[5] &  0 
      & 23.884\protect\footnotemark[4] & -0.060\protect\footnotemark[5] &  0
      & 15.1\protect\footnotemark[4]   &  9.2\protect\footnotemark[5] \\
2000  &	40.663 & 11.059 &  3.5 & 21.004 &  0.031 &  3.0 & 12.3 & 6.9 \\
4000  &	27.235 &  5.921 & 16.6 & 14.317 &  0.068 & 14.2 &  8.1 & 3.7 \\
6000  & 17.188 &  2.223 & 29.5 & 10.016 &  0.051 & 25.5 &  4.6 & 1.4 \\
8000  & 11.207 &  0.835 & 39.2 &  7.463 &  0.031 & 34.3 &  3.3 & 0.5 \\
10000 &	 8.299 &  0.429 & 46.6 &  5.865 &  0.020 & 41.0 &  1.5 & 0.3 \\
20000 &	 2.030 &  0.013 & 70.0 &  2.461 &  0.005 & 62.8 & -0.3 & 0.0 \\
50000 &	 0.049 &  0.000 & 96.9 &  0.166 &  0.000 & 94.2 & -0.1 & 0.0 \\
80000 &  0.001 &  0.000 & 99.9 &  0.009 &  0.000 & 99.6 &  0.0 & 0.0 \\

$\infty$ & \multicolumn{3}{c}{$-154.232002$\protect\footnotemark[6]} 
         & \multicolumn{3}{c}{$-154.224380$\protect\footnotemark[6]}
         & \multicolumn{2}{c}{$-4.8$\protect\footnotemark[7]} \\

\end{tabular}
\end{ruledtabular}

\footnotetext[1]{
\setlength{\baselineskip}{1em}Unless otherwise stated, all energies are reported as
errors relative to CCSDT in millihartree.
}
\footnotetext[2]{
\setlength{\baselineskip}{1em}The \%T values are the percentages of triples
captured during the $i$-FCIQMC propagations [the $S_{z} = 0$ triply excited
determinants of the ${\rm A}_{g}(D_{2h}$) symmetry in the case of the
${\rm X \: ^1B}_{1g}$ state and the $S_{z} = 1$ triply excited determinants of
the ${\rm B}_{1g}(D_{2h}$) symmetry in the case of the ${\rm A \: ^3A}_{2g}$
state].}
\footnotetext[3]{
\setlength{\baselineskip}{1em}Unless otherwise specified, the values of the
singlet--triplet gaps are reported as errors relative to CCSDT in kcal/mol.}
\footnotetext[4]{
\setlength{\baselineskip}{1em}Equivalent to CCSD.}
\footnotetext[5]{
\setlength{\baselineskip}{1em}Equivalent to CR-CC(2,3) [the most complete
variant of CR-CC(2,3) abbreviated sometimes as CR-CC(2,3),D or
CR-CC(2,3)$_{\rm D}$].}
\footnotetext[6]{
\setlength{\baselineskip}{1em}Total CCSDT energy in hartree.}
\footnotetext[7]{
\setlength{\baselineskip}{1em}The CCSDT singlet--triplet gap in kcal/mol.}

\end{table*}


\begin{table*}[!htbp]
\caption{\label{table:table9}Convergence of the CC($P$) and CC($P$;$Q$)
energies of the ${\rm X \: ^3A^\prime_2}$ and ${\rm A \: ^1E^\prime_2}$ states
of cyclopentadienyl cation, as described by the cc-pVDZ basis set,
and of the corresponding vertical singlet--triplet gaps toward their parent
CCSDT values. All calculations were performed at the $D_{5h}$-symmetric
geometry of the ${\rm X\: ^3A^\prime_2}$ state optimized using the unrestricted
CCSD/cc-pVDZ approach in Ref.\ \onlinecite{BiradicalGeom}. The $P$ spaces used
in the CC($P$) and CC($P$;$Q$) calculations were defined as all singly
and doubly excited determinants and subsets of triply excited determinants
extracted from the $i$-CISDTQ-MC propagations with $\delta\tau=0.0001$ a.u. The
$Q$ spaces used to determine the CC($P$;$Q$) corrections consisted of the
triply excited determinants not captured by the corresponding $i$-CISDTQ-MC
runs. The $i$-CISDTQ-MC calculations preceding the CC($P$) and CC($P$;$Q$)
steps were initiated by placing 1500 walkers on the ROHF
(${\rm X \: ^3A^\prime_2}$ state) and RHF (${\rm A \: ^1E^\prime_2}$ state)
reference determinants and the $n_{a}$ parameter of the initiator algorithm was
set at 3. In all post-Hartree--Fock calculations, the five lowest core orbitals were kept
frozen and the spherical components of $d$ orbitals were employed throughout.}
\begin{ruledtabular}
\begin{tabular}{ldddddddd}
            
& \multicolumn{3}{c}{${\rm X\: ^3A^\prime_2}$} & \multicolumn{3}{c}{${\rm A\: ^1E^\prime_2}$}
& \multicolumn{2}{c}{${\rm A\: ^1E^\prime_2}-{\rm X\: ^3A^\prime_2}$} \\
\cline{2-4} \cline{5-7} \cline{8-9}
            
MC Iterations
& \multicolumn{1}{c}{$P$\protect\footnotemark[1]} & \multicolumn{1}{c}{$(P;Q)$\protect\footnotemark[1]}
& \multicolumn{1}{c}{\%T\protect\footnotemark[2]}
& \multicolumn{1}{c}{$P$\protect\footnotemark[1]} & \multicolumn{1}{c}{$(P;Q)$\protect\footnotemark[1]}
& \multicolumn{1}{c}{\%T\protect\footnotemark[2]}
& \multicolumn{1}{c}{$P$\protect\footnotemark[3]} & \multicolumn{1}{c}{$(P;Q)$\protect\footnotemark[3]} \\
\hline
0     & 28.840\protect\footnotemark[4] & 0.245\protect\footnotemark[5] &  0 
& 38.572\protect\footnotemark[4] & 6.245\protect\footnotemark[5] &  0 
&  6.1\protect\footnotemark[4]   & 3.8\protect\footnotemark[5] \\
2000  & 27.396 & 0.272 &  0.8 & 35.598 & 5.948 &   1.0 &  5.1 & 3.6 \\
4000  & 22.253 & 0.267 &  5.1 & 27.946 & 5.078 &   6.5 &  3.6 & 3.0 \\
6000  & 17.394 & 0.212 & 11.6 & 21.124 & 3.971 &  14.7 &  2.3 & 2.4 \\
8000  & 13.743 & 0.152 & 18.3 & 16.042 & 2.756 &  23.0 &  1.4 & 1.6 \\
10000 & 11.027 & 0.108 & 24.8 & 12.947 & 2.248 &  30.9 &  1.2 & 1.3 \\
20000 &  4.250 & 0.026 & 52.1 &  3.964 & 0.217 &  61.4 & -0.2 & 0.1 \\
50000 &  0.155 & 0.001 & 95.3 &  0.060 & 0.001 &  98.3 & -0.1 & 0.0 \\
80000 &  0.007 & 0.000 & 99.8 &  0.001 & 0.000 & 100.0 &  0.0 & 0.0 \\
            
$\infty$ & \multicolumn{3}{c}{$-192.615924$\protect\footnotemark[6]} 
         & \multicolumn{3}{c}{$-192.589235$\protect\footnotemark[6]} 
         & \multicolumn{2}{c}{$16.7$\protect\footnotemark[7]} \\
            
\end{tabular}
\end{ruledtabular}
    
\footnotetext[1]{
\setlength{\baselineskip}{1em}Unless otherwise stated, all energies are
reported as errors relative to CCSDT in millihartree.}
\footnotetext[2]{
\setlength{\baselineskip}{1em}The \%T values are the percentages of triples
captured during the $i$-CISDTQ-MC propagations [the $S_{z} = 1$ triply excited
determinants of the ${\rm B}_{2}(C_{2v}$) symmetry in the case of the
${\rm X \: ^3A^\prime_2}$ state and the $S_{z} = 0$ triply excited determinants
of the ${\rm A}_{1}(C_{2v}$) symmetry in the case of the ${\rm A \:
^1E^\prime_2}$ state].}
\footnotetext[3]{
\setlength{\baselineskip}{1em}Unless otherwise specified, the values of the
singlet--triplet gaps are reported as errors relative to CCSDT in kcal/mol.}
\footnotetext[4]{
\setlength{\baselineskip}{1em}Equivalent to CCSD.
}
\footnotetext[5]{
\setlength{\baselineskip}{1em}Equivalent to CR-CC(2,3) [the most complete
variantof CR-CC(2,3) abbreviated sometimes as CR-CC(2,3),D or
CR-CC(2,3)$_{\rm D}$].}
\footnotetext[6]{
\setlength{\baselineskip}{1em}Total CCSDT energy in hartree.}
\footnotetext[7]{
\setlength{\baselineskip}{1em}The CCSDT singlet--triplet gap in kcal/mol.}
    
\end{table*}

\begin{table*}[!htbp]
\caption{\label{table:table11} 
Convergence of the CC($P$) and CC($P$;$Q$) energies of the ${\rm X \:
^3A^\prime_2}$ and ${\rm B \: ^1A_1}$ states of trimethylenemethane, as
described by the cc-pVDZ basis set, and of the corresponding adiabatic
singlet--triplet gaps toward their parent CCSDT values. The $D_{3h}$- and
$C_{2v}$-symmetric geometries of the ${\rm X \: ^3A^\prime_2}$ and ${\rm B \:
^1A_1}$ states, respectively, optimized in the SF-DFT/6-31G(d) calculations,
were taken from Ref.\ \onlinecite{ch2_krylov}. The $P$ spaces
used in the CC($P$) and CC($P$;$Q$) calculations were defined as all singly and
doubly excited determinants and subsets of triply excited determinants
extracted from the $i$-CISDTQ-MC propagations with $\delta\tau=0.0001$ a.u. The
$Q$ spaces used to determine the CC($P$;$Q$) corrections consisted of the
triply excited determinants not captured by the corresponding $i$-CISDTQ-MC
runs. The $i$-CISDTQ-MC calculations preceding the CC($P$) and CC($P$;$Q$)
steps were initiated by placing 1500 walkers on the ROHF (${\rm X \:
^3A^\prime_2}$ state) and RHF (${\rm B \: ^1A_1}$ state) reference determinants
and the $n_{a}$ parameter of the initiator algorithm was set at 3.
In all post-Hartree--Fock calculations, the four lowest core orbitals were kept frozen
and the spherical components of $d$ orbitals were employed throughout.}
\begin{ruledtabular}
\begin{tabular}{ldddddddd}

& \multicolumn{3}{c}{${\rm X\: ^3A^\prime_2}$} & \multicolumn{3}{c}{${\rm B\: ^1A_1}$}
& \multicolumn{2}{c}{${\rm B\: ^1A_1}-{\rm X\: ^3A^\prime_2}$} \\
\cline{2-4} \cline{5-7} \cline{8-9}

MC Iterations
& \multicolumn{1}{c}{$P$\protect\footnotemark[1]} & \multicolumn{1}{c}{$(P;Q)$\protect\footnotemark[1]}
& \multicolumn{1}{c}{\%T\protect\footnotemark[2]}
& \multicolumn{1}{c}{$P$\protect\footnotemark[1]} & \multicolumn{1}{c}{$(P;Q)$\protect\footnotemark[1]}
& \multicolumn{1}{c}{\%T\protect\footnotemark[2]}
& \multicolumn{1}{c}{$P$\protect\footnotemark[3]} & \multicolumn{1}{c}{$(P;Q)$\protect\footnotemark[3]} \\
\hline

0     & 19.202\protect\footnotemark[4] &  0.418\protect\footnotemark[5] &  0 
      & 58.051\protect\footnotemark[4] & 13.370\protect\footnotemark[5] &  0 
      & 24.4\protect\footnotemark[4] & 8.1\protect\footnotemark[5] \\
2000  & 17.975 & 0.422 &  1.1 & 50.012 &  9.362 &  1.2 & 20.1 & 5.6 \\
4000  & 14.462 & 0.357 &  6.6 & 32.925 &  3.236 &  7.7 & 11.6 & 1.8 \\
6000  & 11.319 & 0.253 & 14.1 & 20.628 &  1.260 & 16.8 &  5.8 & 0.6 \\
8000  &  9.066 & 0.173 & 21.3 & 14.601 &  0.649 & 25.5 &  3.5 & 0.3 \\
10000 &  7.429 & 0.123 & 27.9 & 10.680 &  0.314 & 33.1 &  2.0 & 0.1 \\
20000 &  3.294 & 0.031 & 52.3 &  2.675 &  0.028 & 61.1 & -0.4 & 0.0 \\
50000 &  0.213 & 0.001 & 92.8 &  0.061 &  0.000 & 97.1 & -0.1 & 0.0 \\
80000 &  0.012 & 0.000 & 99.5 &  0.002 &  0.000 & 99.9 &  0.0 & 0.0 \\
$\infty$ & \multicolumn{3}{c}{$-155.466242$\protect\footnotemark[6]} 
         & \multicolumn{3}{c}{$-155.431596$\protect\footnotemark[6]} 
         & \multicolumn{2}{c}{$21.7$\protect\footnotemark[7]} \\

\end{tabular}
\end{ruledtabular}

\footnotetext[1]{
\setlength{\baselineskip}{1em}Unless otherwise stated, all energies are
reported as errors relative to CCSDT in millihartree.}
\footnotetext[2]{
\setlength{\baselineskip}{1em}The \%T values are the percentages of triples
captured during the $i$-CISDTQ-MC propagations [the $S_{z} = 1$ triply excited
determinants of the ${\rm B}_{2}(C_{2v}$) symmetry in the case of the ${\rm X \: ^3A^\prime_2}$ state and
the $S_{z} = 0$ triply excited determinants of the ${\rm A}_{1}$ symmetry in the case of the ${\rm B \: ^1A_1}$ state].
}
\footnotetext[3]{
\setlength{\baselineskip}{1em}Unless otherwise specified, the values of the
singlet--triplet gaps are reported as errors relative to CCSDT in kcal/mol.
}
\footnotetext[4]{
\setlength{\baselineskip}{1em}Equivalent to CCSD.
}
\footnotetext[5]{
\setlength{\baselineskip}{1em}Equivalent to CR-CC(2,3) [the most complete variant
of CR-CC(2,3) abbreviated sometimes as CR-CC(2,3),D or CR-CC(2,3)$_{\rm D}$].
}
\footnotetext[6]{
\setlength{\baselineskip}{1em}Total CCSDT energy in hartree.
}
\footnotetext[7]{
\setlength{\baselineskip}{1em}The CCSDT singlet--triplet gap in kcal/mol.
}
 
\end{table*}


\clearpage

\end{document}